\documentclass[useAMS,usenatbib,usegraphicx]{mn2e}

\usepackage{natbib}
\usepackage{amsmath}
\usepackage{amssymb}
\usepackage{rotating}

\newcommand{\HI}{H~\texttt{I}~}
\newcommand{\HII}{H~\texttt{II}~}

\title[Deep surface photometry on $24$ BCGs]{Deep multiband surface photometry on star forming galaxies: \emph{I}. A sample of $24$ blue compact galaxies}
\author[Micheva et al.]{Genoveva Micheva$^{1}$\thanks{E-mail: genoveva@astro.su.se (GM)}, G\"oran \"Ostlin$^{2}$, Nils Bergvall$^{3}$, Erik Zackrisson$^{2}$,
\newauthor Josefa Masegosa$^{4}$, Isabel Marquez$^{4}$, Thomas Marquart$^{1,3}$, Florence Durret$^{5}$\\
$^{1}$Stockholm Observatory, Department of Astronomy, Stockholm University, 106\,91 Stockholm, Sweden\\
$^{2}$Oskar Klein Centre for Cosmoparticle Physics, Department of Astronomy, Stockholm University, 106\,91 Stockholm, Sweden\\
$^{3}$Division of Astronomy \& Space Physics, Uppsala university, 751\,20 Uppsala, Sweden\\
$^{4}$Instituto Astrofisica Andalucia (IAA), CSIC, Spain\\
$^{5}$UPMC-CNRS, UMR7095, Institut d'Astrophysique de Paris, F-75014, Paris, France\\
}
\begin{document}
\date{Accepted .... Received ...; in original form ...}

\pagerange{\pageref{firstpage}--\pageref{lastpage}} \pubyear{2011}

\maketitle

\label{firstpage}

\begin{abstract}

\noindent We present deep optical and near--infrared (NIR) $UBVRIHKs$ imaging data for $24$ blue compact galaxies (BCGs). The individual exposure times are on average $\sim40$ minutes in the optical ($B$) and $\sim90$ minutes in the NIR, but on occasion up to $\sim5$ hours for a single target and filter, observed on $2.5,~3.5,~8.2$--m telescopes. The sample contains luminous dwarf and intermediate--mass BCGs which are predominantly metal--poor, although a few have near--solar metallicities. We have analyzed isophotal and elliptical integration surface brightness and color profiles, extremely deep ($\mu_B\lesssim29$ mag arcsec${}^{-2}$) contour maps and RGB images for each galaxy in the sample, and provide a morphological classification where such is missing. We have measured the total galaxy colors, the colors of the underlying host galaxy, and the colors of the burst, and compare these to the predictions of new state--of--the--art spectral evolutionary models (SEMs) both with and without contribution by nebular emission. Separating the burst from the underlying host we find that regardless of the total luminosity the host galaxy has the properties of a low surface brightness (LSB) dwarf with $M_B\gtrsim-18$. For a number of galaxies we discover a distinct LSB component dominant around and beyond the Holmberg radius. For the specific case of ESO400--43A\&B we detect an optical bridge between the two companion galaxies at the $\mu_V\sim28$th mag arcsec${}^{-2}$ isophotal level. Synthetic disk tests are performed to verify that we can trace such faint components with negligible errors down to $\mu_B=28$ mag arcsec${}^{-2}$ and $\mu_K=23$ mag arcsec${}^{-2}$. By examining the structural parameters (central surface brightness $\mu_0$ and scale length $h_r$) derived from two radial ranges typically assumed to be dominated by the underlying host galaxy, we demonstrate the importance of sampling the host well away from the effects of the burst. We find that $\mu_0$ and $h_r$ of the BCGs host deviate from those of dwarf ellipticals (dE) and dwarf irregulars (dI) solely due to a strong burst contribution to the surface brightness profile almost down to the Holmberg radius. Structural parameters obtained from a fainter region, $\mu_B=26$--$28$ mag arcsec${}^{-2}$, are consistent with those of true LSB galaxies for the starbursting BCGs in our sample, and with dEs and dIs for the BCGs with less vigorous star formation. 
\end{abstract}


\section{Introduction}

\noindent Blue compact galaxies (BCGs) are gas--rich star--forming low redshift galaxies with low metallicities~\citep[for a review see][]{2000A&ARv..10....1K}. Their star formation rates (SFRs) can be as high as $\sim20~M_\odot~yr^{-1}$~\citep[e.g. Haro 11,][]{2007MNRAS.382.1465H} and their \HI gas masses range from $M(H\texttt{I})\sim10^7$ to a few tens of $10^8 M_\odot$~\citep[e.g.][]{1995ApJS...99..427T,1999A&AS..139....1T,2000A&A...361...19S}. BCGs are said to be starbursting when they will exhaust their gas supply in less than a Hubble time. The metallicities measured in these galaxies are low~\citep[e.g.][]{1994ApJ...420..576M,1994MNRAS.270...35M,1999ApJ...511..639I}, to extremely low~\citep{2005ApJ...632..210I}, once leading researchers to propose that BCGs are truly young galaxies~\citep{1972ApJ...173...25S}, experiencing their first episode of star formation in a primordial gas environment. There is mounting evidence to the contrary, however, and ``first--burst'' claims have been challenged using broadband colors and morphology~\citep[e.g.][]{1996A&A...314...59P,1997MNRAS.286..183T,2001ApJS..133..321C,2001ApJS..136..393C,2002A&A...390..891B}, absorption features in the spectra~\citep[e.g.][]{2004A&A...423..133W}, through the direct detection of globular clusters associated with the BCG~\citep{1998A&A...335...85O}, and for near--by targets through the detection of luminous red giants and asymptotic giant branch (AGB) stars in color--magnitude diagrams~\citep[e.g.][]{1999AJ....118..302A,2000ApJ...535L..99O}. In most cases the relative strength of the starburst can be so high that it completely dominates the light output of the galaxy, an obstacle which has been countered by deeper optical imaging data and observations in the near infra-red (NIR) regime~\citep[e.g.][]{2003A&A...410..481N,2003ApJ...593..312C} and an older population has been revealed. In other cases, e.g. SBS0335--052 E\&W, the burst can be spatially dominating over the galaxy even for extremely deep imaging, making it difficult to directly detect a possible old population. Hence, the youth hypothesis persists for these targets~\citep[e.g.][]{1997ApJ...489..623T,2005ApJ...632..210I} but has been challenged~\citep{2001A&A...371..429O}.\\

\noindent Nevertheless, the emerging consensus seems to be that BCGs are galaxies with an underlying old stellar population, often referred to as 'host', on which the current starburst is superposed. The host population is distinctly different in terms of optical and NIR colors~\cite[e.g.][]{2002A&A...390..891B,2003A&A...410..481N,2003ApJ...593..312C,2005mmgf.conf..355B,2005ApJS..157..218C,2009A&A...501...75A}, and appears to be smoothly distributed in a physical structure reminiscent of a disk with regular isophotes~\citep{1983MitAG..58..108L,1983ApJ...268..667T,1996A&A...314...59P,1999A&A...341..725S}. The discovery of this host population invariably led to a relaxation in the definition of this group of galaxies. The ``dwarf'' qualifier in the original naming convention of ``blue compact dwarf'' is often dropped by some authors to the benefit of the more inclusive ``galaxy'' term. There is a large fraction of BCGs which are too massive and bright to be dwarfs, with $M_B$ as high as $-20$, but they too exhibit starbursts in a gas--rich, metal--poor environment, and a similar morphological structure as the dwarfs -- an extended, often regular, disk--like structure hosts knots of star formation (SF), either centrally located or dispersed in seemingly random fashion around the galaxy. The name is still misleading since, as pointed out in~\citet{2003ApJS..147...29G}, some blue compact galaxies are neither blue, nor compact. Including the host, the requirements of~\citet{1981ApJ...247..823T} that a BCG must have an apparent optical size $r_{25}\lesssim1$ kpc in diameter can now only be met for some very compact sources, like HL293B. Further, the ``blue'' condition of $B-V\lesssim0.3^{\textrm{m}}$~\citep{1986A&AS...64..469B} can now only be applied to the starbursting region since inclusion of the host shifts the total $B-V$ color redwards of this (somewhat arbitrary) limit.\\

\noindent Despite such diversity all BCGs have at least one thing in common. They are a subgroup of emission line galaxies, well separated from Seyferts and quasars, with spectra reminiscent of \HII regions, only with a strong stellar continuum from the composite populations of the galaxy~\citep[e.g.][]{1983ApJ...273...81K,1994ApJ...420..576M,1989ApJS...70..447S}. Because of their \HII region--like spectrum they are often referred to as \HII galaxies. The SFR can drastically vary from BCG to BCG, with some extreme outliers like \emph{Haro 11} having a $SFR\gtrsim18~M_\odot~yr^{-1}$, which it can only sustain for $\lesssim50$ Myr~\citep{2000A&A...359...41B}. While not equally vigorous, the SF activity in most BCGs must similarly proceed in short--lived bursts, lasting from a few tens of Myr to $1$ Gyr. The duration of the typical burst has been constrained both from observed \HI masses, which are generally too low to sustain the current SFRs for a Hubble time, and from measured metallicities, which are also too low for the interstellar medium to have undergone enrichment by constant star formation. The latter argument can be partially circumvented if one allows for the outflow and loss of metals to the intergalactic medium due to supernovae winds, or for the inflow of pristine gas, which would dilute the metallicity. The argument of too low metallicity cannot be completely discarded, however, as there are indications that gas--rich galaxies are more likely to be closed systems~\citep{2006ApJ...636..214V}.\\

\noindent Some BCGs are clearly in various stages of merging, or undergoing tidal interaction. They can have very irregular morphologies, with complicated, almost chaotic kinematics, making it difficult to quantify the masses and ages of the current burst. Other BCGs show an otherwise regular elliptical disk with a central nuclear starburst, or sporadic star--forming knots offset from the center, a.k.a. \HII hotspots~\citep{1989ApJS...70..447S}, and it is not clear whether galaxies with such morphologies also are subject to the same SF trigger. \HI observations have shown that while the majority of BCGs are associated with massive \HI clouds, most likely fueling the ongoing star formation, they are generally isolated in the sense that they lack massive companion galaxies~\citep{1991A&A...241..358C,1993AJ....106.1784C,1995ApJS...99..427T,1999A&AS..139....1T}. However, \HI companions without a confirmed optical counterpart do appear to be common~\citep{1997ApJ...480..524T}, indicating that perhaps the companion galaxy is a faint gas--rich dwarf. This makes mergers the leading candidates for SF triggers since tidal interactions are unlikely to be the primary SF triggers in dwarf galaxies~\citep{2004MNRAS.349..357B}. For a starburst to take place the merger would not have to be major but it would have to be wet, meaning at least one of the merging galaxies would have to be gas--rich.\\

\noindent Ideally, one would like to study the SF as it occurs for the first time in the protogalaxies in the distant Universe. It is thought that observing such pristine SF activity would allow us to determine what the necessary trigger conditions for the onset of star formation are. Presently, we are unable to directly study the stellar populations of high--redshift galaxies in any great detail since we are constrained by cosmological effects and small intrinsic sizes of the targets, and the comparatively poor resolution of our currently available instrumentation. Because of their low metallicities, high SFRs, and their lack of spiral density waves, BCGs are thought to be reminiscent of the young galaxies in the universe, in which the star formation must have proceeded under similar metal--poor gas--rich conditions. Being much closer, BCGs are available for detailed photometric and spectroscopic studies.\\

\noindent Considering the degree of complication inherent to the analysis of such a heterogeneous group of galaxies, one needs to examine the properties of the BCGs from several angles: the behavior of the spectrum in the optical and NIR regimes on a large scale (broadband imaging), the individual line emissions (typically $H_\alpha$ and $[OIII]$), and the kinematics of the gas and the stars. We have undertaken a study of BCGs that combines optical and NIR broadband and narrowband imaging~\citep[this work,][]{Paper2,Paper3}, Fabry--Perot interferometry (\"Ostlin et al. 2012 in prep., Marquart et al. 2012 in prep), and grism spectroscopy of two distinct types of star--forming galaxies - high and low luminosity BCGs. In this paper, which is the first in a series, we present broadband $UBVRIHKs$ data for a sample of $24$ luminous galaxies classified as BCGs. This sample was hand--picked to contain interesting and representative cases, and is biased towards relatively luminous (median $M_B\sim-18$ mag) galaxies. Some of the targets come from the~\citet{2001A&A...374..800O} sample, others from the~\citet{1989ApJS...70..447S} UM survey of \HII galaxies. The latter were selected based on their emission line spectra to be actively star--forming if not always strongly starbursting, and to have metallicities and \HI masses available in the literature. Additionally, a number of famous and strongly starbursting BCGs (e.g. IIZw40, MK930, Haro11) were included in the project because they lacked equally deep observations and yet presented interesting targets to have in our analysis. Haro 11 was already presented in~\citet{2010MNRAS.405.1203M}, so it is not presented here but will be included in our future analysis. In a consecutive paper~\citep{Paper2} we present $UBVRIHKs$ broadband imaging data for a volume--limited sample of $21$ low luminosity emission line galaxies. These constitute a well defined complete sample of star forming galaxies, representative of the galaxy population in the local Universe. In total we have $46$ BCG and BCG--like targets in $7$ broadband and $2$ narrowband filters. In the optical, imaging observations of depth comparable to ours can be found in the literature for a number of galaxies~\citep{1996A&AS..120..207P,1999A&AS..138..213D,2001ApJS..133..321C,2002A&A...390..891B}. In the NIR, however, there is no sample of BCGs with as deep individual exposure times as ours~\citep[compare e.g.][]{2003A&A...410..481N,2003ApJ...593..312C}. In~\citet{Paper3} we combine both samples to juxtapose the high and low luminosity BCGs and analyse any differences in their host populations. In the latter paper we also present the narrowband data from both samples as well as derived estimates of the ages, masses and star formation rates through SED fitting. We have assumed $H_0=73~km^{-1}s^{-1}Mpc^{-1}$ throughout this paper. \\

\noindent The layout of this paper is as follows: \S~\ref{data} introduces the observations, reduction and calibration of the data. \S~\ref{sbsection}, \S~\ref{contourplots}, and \S~\ref{rgbimages} deal with the generation of surface brightness profiles, contour plots and three--color RGB images, respectively. \S~\ref{integrsurfphot} and \S~\ref{colors} explain our approach to obtaining total luminosities and total colors for the galaxies. \S~\ref{strucparam} illustrates how we obtain structural parameters for population separation. Characteristics of individual galaxies and comparison of integrated colors to stellar evolutionary models can be found in \S~\ref{individ}. Finally, discussion and concluding remarks are in \S~\ref{discuss} and \S~\ref{conclude}.

\section[]{Observations}\protect\label{data}

\noindent The data consist of optical and near infra-red (NIR) broadband imaging in the $UBVRIHK$ filters. They were obtained during the period $2001$--$2007$ with ALFOSC (Nordic Optical Telescope, NOT), EMMI (European Southern Observatory New Technology Telescope, ESO NTT), and FORS (ESO Very Large Telescope, VLT) in the optical, and with NOTCAM (NOT) and SOFI (ESO NTT) in the NIR. A log of observations is shown in Table~\ref{nedtable} and the individual exposure times for each filter in Table~\ref{exptable}. Since both the sample presented here and the volume limited sample of~\citet{Paper2} were reduced with the same pipelines, the description of the reductions and calibration that follows here applies to both. We will give the details in this paper and will not repeat them in~\citet{Paper2}.

\setcounter{table}{0}
\begin{table}
  \begin{minipage}{70mm}
    \tiny
    \caption{Total integration times for the sample. All times are given in minutes and converted to the framework of a 2.56 meter telescope where needed. The values are for observations in a single filter, e.g. only $SOFI~Ks$, and not $SOFI~Ks+NOTCAM~Ks$.}\protect\label{exptable}
    \begin{tabular}{|lccccccc|} 
      \hline
      &U&B&V&R&I&H&Ks\\\hline\hline
      ESO185--13&30&38&9&7&38&67&137\\\hline
      ESO249--31&&38&19&19&19&58&117\\\hline
      ESO338--04&&9&19&29&125&61&352\\\hline
      ESO400--43&&&77&&148&61&314\\\hline
      ESO421--02&30&38&9&19&19&19&114\\\hline
      ESO462--20&30&38&9&&&73&139\\\hline
      HE2--10&51&38&19&&&9&121\\\hline
      HL293B&26&45&25&20&40&25&64\\\hline
      IIZW40&10&40&20&&40&12&$96^\dagger$\\\hline
      MK600&20&40&23&20&40&24&119\\\hline
      MK900&20&40&25&20&40&21&32\\\hline
      MK930&20&40&25&30&40&37&127\\\hline
      MK996&20&40&20&20&26&15&75\\\hline
      SBS0335--052E\&W&&103&&&45&94&$298^\ddagger$\\\hline
      TOL0341--407&30&29&9&&38&67&117\\\hline
      TOL1457--262\emph{I}\&\emph{II}&&66&76&&&&125\\\hline
      UM133&20&53&23&20&40&22&53\\\hline
      UM160&20&40&26&20&40&96&56\\\hline
      UM238&20&40&23&20&40&21&96\\\hline
      UM417&20&40&23&20&40&18&58\\\hline
      UM448&&40&40&9&58&&32\\\hline
      UM619&40&40&40&9&58&32&$186^\dagger$\\\hline
    \end{tabular}
    \medskip
    ~\\
    $\dagger$ -- SOFI, $\ddagger$ -- NOTCAM
  \end{minipage}
\end{table}
\begin{center}
\begin{table*}
  \begin{minipage}{150mm}
    \tiny
    \caption{Log of the observations. Heliocentric redshift and cosmology--corrected luminosity distances from \emph{NED}.}\protect\label{nedtable}
    \begin{tabular}{@{}|lllllll|@{}}
      \hline
      Galaxy&Ra~Dec~(J2000)&Redshift&$\textrm{D}~[\textrm{Mpc}]$&Year&Instrument&Filters\\\hline
      ESO185--13&19h45m00.5s&0.018633&76.3&2001&EMMI&Bb\#605,V\#606,I\#610\\
      &-54d15m03s&&&&SOFI&Ks\#13\\
      &&&&2002&FORS1&U\_BESS,R\_BESS\\
      &&&&&SOFI&H\#12\\\hline
      ESO249--31&03h55m46.1s &0.002829&10.8&2001&EMMI&Bb\#605,V\#606,R\#608,I\#610\\
      &-42d22m05s&&&&SOFI &Ks\#13\\
      &&&&2002 &SOFI&H\#12\\\hline
      ESO338--I04&19h27m58.2s &0.009453&37.4&2001&EMMI&R\#608\\
      &-41d34m32s&&&&EMMI&Bb\#605,V\#606\\
      &&&&2005&EMMI&I\#610\\
      &&&&2005&SOFI&Ks\#13\\
      &&&&2006&SOFI&Ks\#13,H\#12\\\hline
      ESO400--G043&20h37m41.9s&0.019680&79.2&2005&EMMI&V\#606,I\#610\\
      &-35d29m08s&&&2006&SOFI&KS\#13,H\#12\\\hline
      ESO421--02&04h29m40.1s&0.003149&12.4&2001&EMMI&Bb\#605,V\#606,R\#608,I\#610\\
      &-27d24m31s&&&&SOFI&Ks\#13\\
      &&&&2002&FORS1&U\_BESS\\
      &&&&&SOFI&H\#12\\\hline
      ESO462--IG020&20h26m56.8s&0.019782&79.5&2001&EMMI&Bb\#605,V\#606\\
      &-29d07m01s&&&&SOFI&Ks\#13\\
      &&&&2002&FORS1&U\_BESS\\
      &&&&&SOFI&H\#12\\\hline
      HE2--10&08h36m15.1s&0.002912&15.8&2001&EMMI&Bb\#605,V\#606\\
      &-26d24m34s&&&2002&FORS1&U\_BESS\\
      &&&&&SOFI&H\#12,Ks\#13\\\hline
      HL293B&22h30m36.798s&0.005170&16.3&2001&ALFOSC--FASU&B\#74,V\#75,R\#76,I\#12\\
      &-00d06m36.99s&&&2002&ALFOSC--FASU&U\#7\\
      &&&&2003&NOTCAM&Ks\#207,H\#204\\\hline
      IIZW40&05h55m42.6s&0.002632&11.8&2001&ALFOSC--FASU&B\#74,I\#12\\
      &+03d23m32s&&&2002&ALFOSC--FASU&U\#7\\
      &&&&&NOTCAM&Ks\#207\\
      &&&&&SOFI&Ks\#13\\
      &&&&2003&ALFOSC--FASU&V\#75\\
      &&&&&NOTCAM&H\#204\\\hline
      MK600&02h51m04.6s&0.003362&10.9&2001&ALFOSC--FASU&B\#74,I\#12\\
      &+04d27m14s&&&2002&ALFOSC--FASU&U\#7,V\#75,R\#76\\
      &&&&&NOTCAM&Ks\#207\\
      &&&&2003&NOTCAM&H\#204\\\hline
      MK900&21h29m59.6s&0.003843&11.2&2001&ALFOSC--FASU&B\#74,V\#75,R\#76,I\#12\\
      &+02d24m51s&&&2002&ALFOSC--FASU&U\#7\\
      &&&&2003&NOTCAM&Ks\#207,H\#204\\\hline
      MK930&23h31m58.3s&0.018296&71.4&2001&ALFOSC--FASU&B\#74,V\#75,I\#12\\
      &+28d56m50s&&&2002&ALFOSC--FASU&U\#7,R\#76\\
      &&&&2003&NOTCAM&Ks\#207,H\#204\\\hline
      MK996&01h27m35.5s&0.005410&18.2&2001&ALFOSC--FASU&I\#12\\
      &-06d19m36s&&&2002&ALFOSC--FASU&U\#7\\
      &&&&&SOFI&Ks\#13\\
      &&&&2003&ALFOSC--FASU&B\#74,V\#75,R\#76\\
      &&&&&NOTCAM&H\#204\\\hline
      SBS0335--052E\&W&03h37m44.0s&0.013486&54&2001&ALFOSC--FASU&B\#74,I\#12\\
      &-05d02m40s&&&2002&SOFI&Ks\#13\\
      &&&&2003&NOTCAM&KS\#207,H\#204\\\hline
      TOL0341&03h42m49.4s&0.014827&60.6&2001&EMMI&Bb\#605,V\#606,I\#610\\
      &-40d35m56s&&&2002&FORS1&U\_BESS\\
      &&&&&SOFI&H\#12,Ks\#13\\\hline
      TOL1457--262\emph{I}\&\emph{II}&15h00m27.9s&0.016812&72.8&2004&ALFOSC--FASU&B\#74,V\#75\\
      &-26d27m02s&&&2005&SOFI&Ks\#13\\\hline
      UM133&01h44m41.3s&0.005414&18.3&2001&ALFOSC--FASU&B\#74,I\#12\\
      &+04d53m26s&&&2002&ALFOSC--FASU&U\#7,V\#75,R\#76\\
      &&&&2003&NOTCAM&Ks\#207,H\#204\\\hline
      UM160&23h24m19.8s&0.008006&28&2001&ALFOSC--FASU&B\#74,V\#75,I\#12\\
      &-00d07m01s&&&2002&ALFOSC--FASU&U\#7,R\#76\\
      &&&&&SOFI&H\#12,Ks\#13\\\hline
      UM238&00h24m42.3s&0.014230&54.2&2001&ALFOSC--FASU&B\#74,I\#12\\
      &+01d44m02s&&&2002&ALFOSC--FASU&U\#7,V\#75,R\#76\\
      &&&&&SOFI&Ks\#13\\
      &&&&2003&NOTCAM&H\#204\\\hline
      UM417&02h19m30.2s&0.009000&33.8&2001&ALFOSC--FASU&B\#74,I\#12\\
      &-00d59m11s&&&2002&ALFOSC--FASU&U\#7,V\#75,R\#76\\
      &&&&&NOTCAM&Ks\#207\\
      &&&&2003&NOTCAM&H\#204\\\hline
      UM448&11h42m12.4s&0.018559&74.6&2004&ALFOSC--FASU&B\#74,V\#75\\
      &+00d20m03s&&&2005&ALFOSC--FASU&U\#7\\
      &&&&&EMMI&R\#608,I\#610\\
      &&&&2007&NOTCAM&Ks\#207\\\hline
      UM619&13h52m44.7s&0.015494&63.1&2004&ALFOSC--FASU&B\#74,V\#75\\
      &+00d07m53s&&&2005&ALFOSC--FASU&U\#7\\
      &&&&&EMMI&R\#608,I\#610\\
      &&&&&NOTCAM&H\#204\\
      &&&&&SOFI&Ks\#13\\
      &&&&2006&SOFI&Ks\#13\\
      &&&&2007&NOTCAM&Ks\#207\\\hline
      \hline
    \end{tabular}
  \end{minipage}
\end{table*}
\end{center}
\begin{center}
\begin{figure*}
\begin{minipage}{150mm}
\centering
\includegraphics[width=15cm,height=18cm]{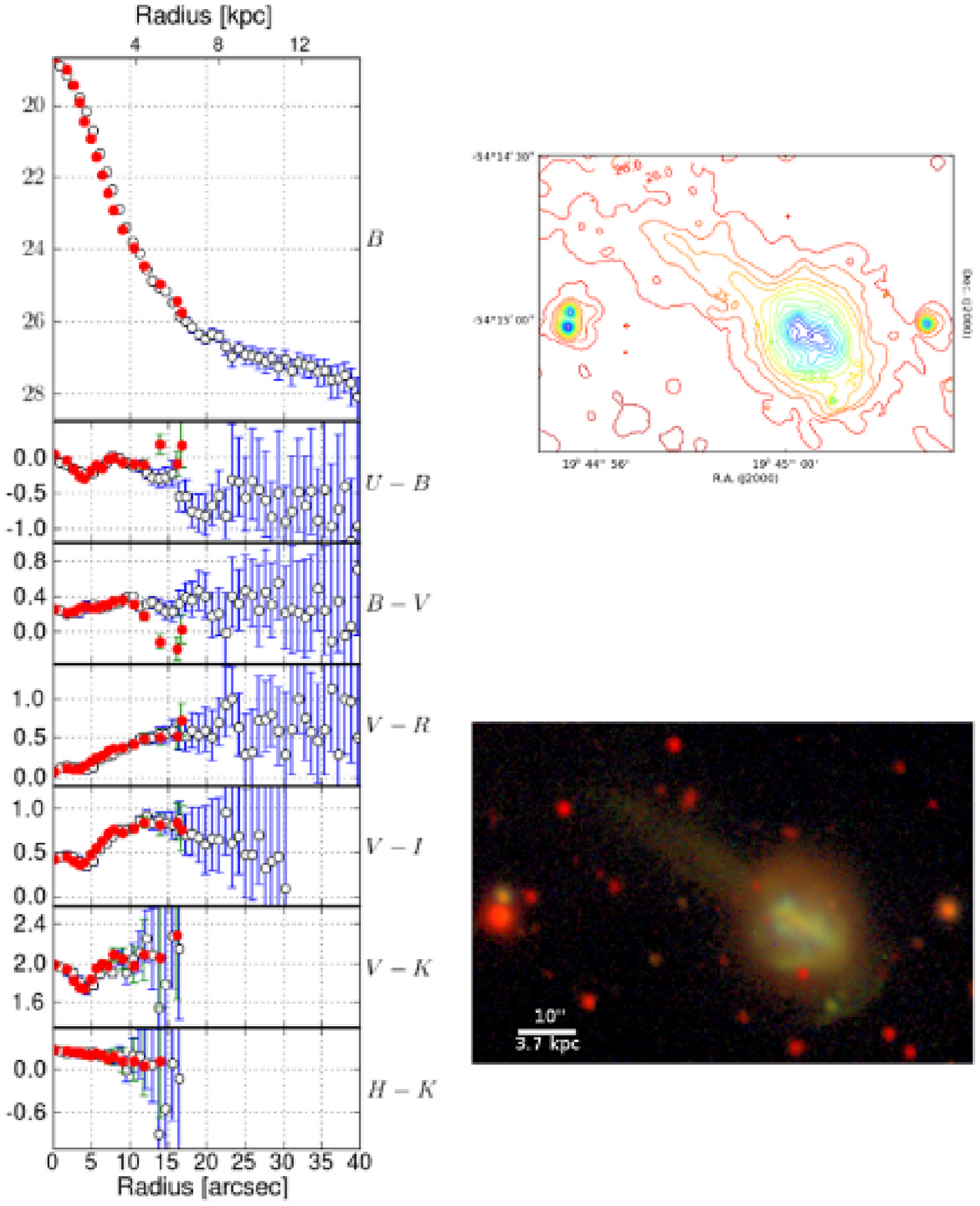}
\caption{\textbf{ESO185-13}. \textit{Left panel}: Surface brightness and color radial profiles for elliptical (open circles) and isophotal (red circles) integration. \textit{Upper right panel}: contour plot based on the $B$ band. Isophotes fainter than $22.5$, $25.5$ are iteratively smoothed with a boxcar median filter of size $5$, $15$ pixels respectively. \textit{Lower right panel}: A true color RGB composite image using the $U,B,I$ filters. Each channel has been corrected for Galactic extinction following \citet{1998ApJ...500..525S} and converted to the AB photometric system. The RGB composite was created by implementing the \citet{2004PASP..116..133L} algorithm.}
\protect\label{datafig}
\end{minipage}
\end{figure*}

\clearpage

\begin{figure*}
\begin{minipage}{150mm}
\includegraphics[width=15cm,height=18cm]{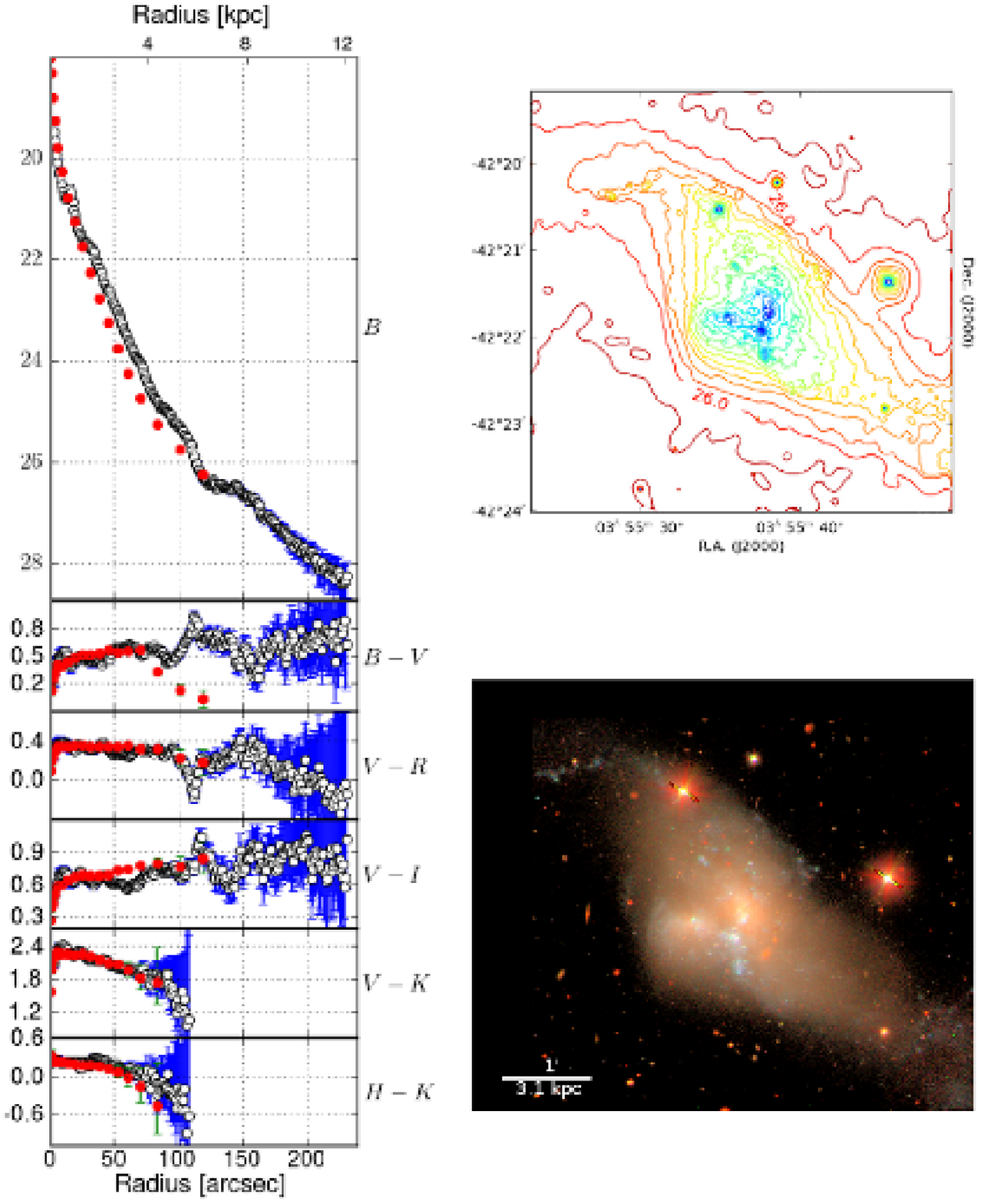}
\contcaption{\textbf{ESO249-31}. \textit{Left panel}: Surface brightness and color radial profiles for elliptical (open circles) and isophotal (red circles) integration. \textit{Upper right panel}: contour plot based on the $B$ band. Isophotes fainter than $22.0$, $24.5$, $25.5$ are iteratively smoothed with a boxcar median filter of size $5$, $15$, $25$ pixels respectively. \textit{Lower right panel}: A true color RGB composite image using the $B,V,I$ filters. Each channel has been corrected for Galactic extinction following \citet{1998ApJ...500..525S} and converted to the AB photometric system. The RGB composite was created by implementing the \citet{2004PASP..116..133L} algorithm.}
\end{minipage}
\end{figure*}

\clearpage

\begin{figure*}
\begin{minipage}{150mm}
\includegraphics[width=15cm,height=18cm]{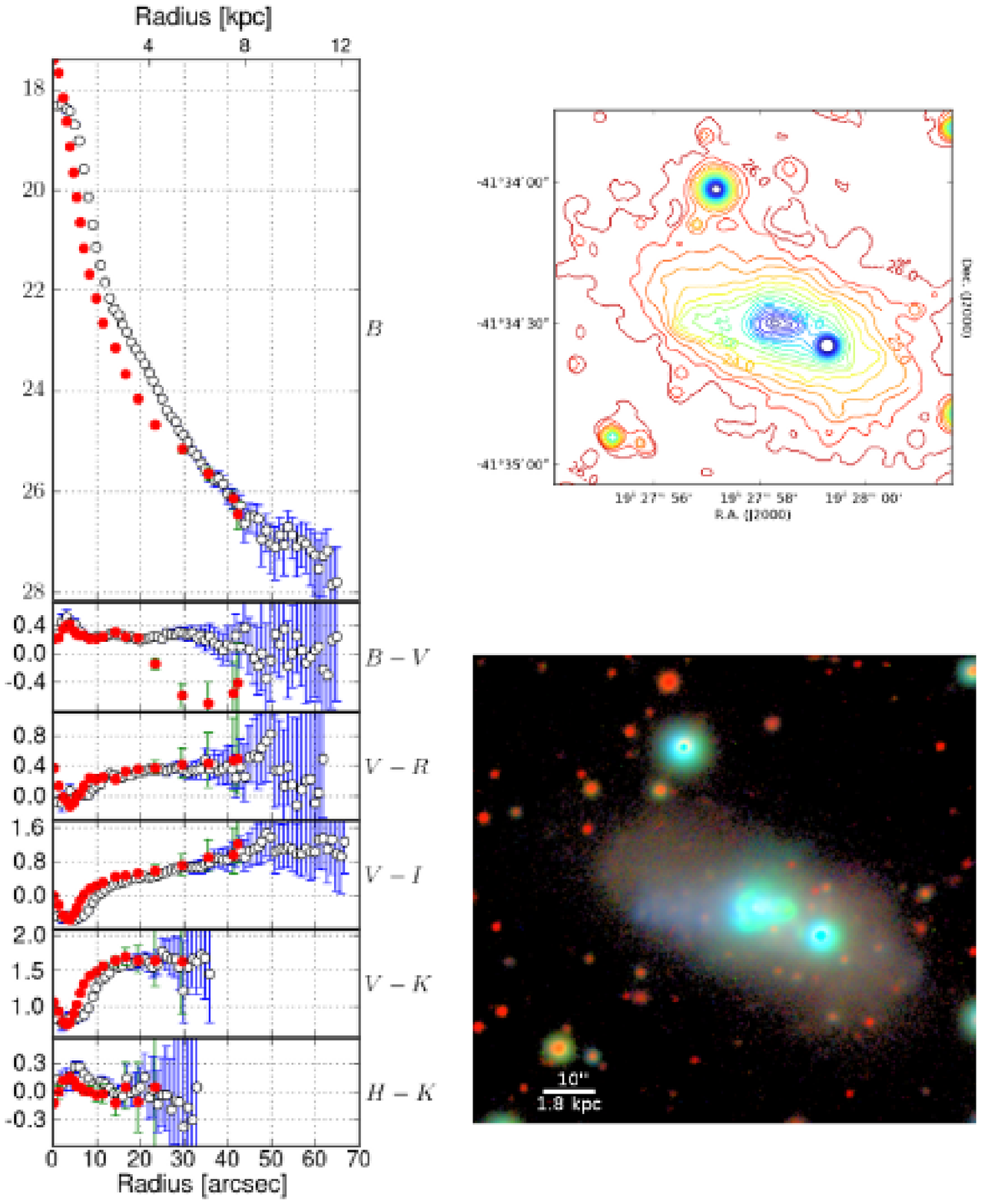}
\contcaption{\textbf{ESO338-04}. \textit{Left panel}: Surface brightness and color radial profiles for elliptical (open circles) and isophotal (red circles) integration. \textit{Upper right panel}: contour plot based on the $B$ band. Isophotes fainter than $22.5$, $25.5$ are iteratively smoothed with a boxcar median filter of size $5$, $15$ pixels respectively. \textit{Lower right panel}: A true color RGB composite image using the $B,V,I$ filters. Each channel has been corrected for Galactic extinction following \citet{1998ApJ...500..525S} and converted to the AB photometric system. The RGB composite was created by implementing the \citet{2004PASP..116..133L} algorithm.}
\end{minipage}
\end{figure*}

\clearpage

\begin{figure*}
\begin{minipage}{150mm}
\includegraphics[width=15cm,height=18cm]{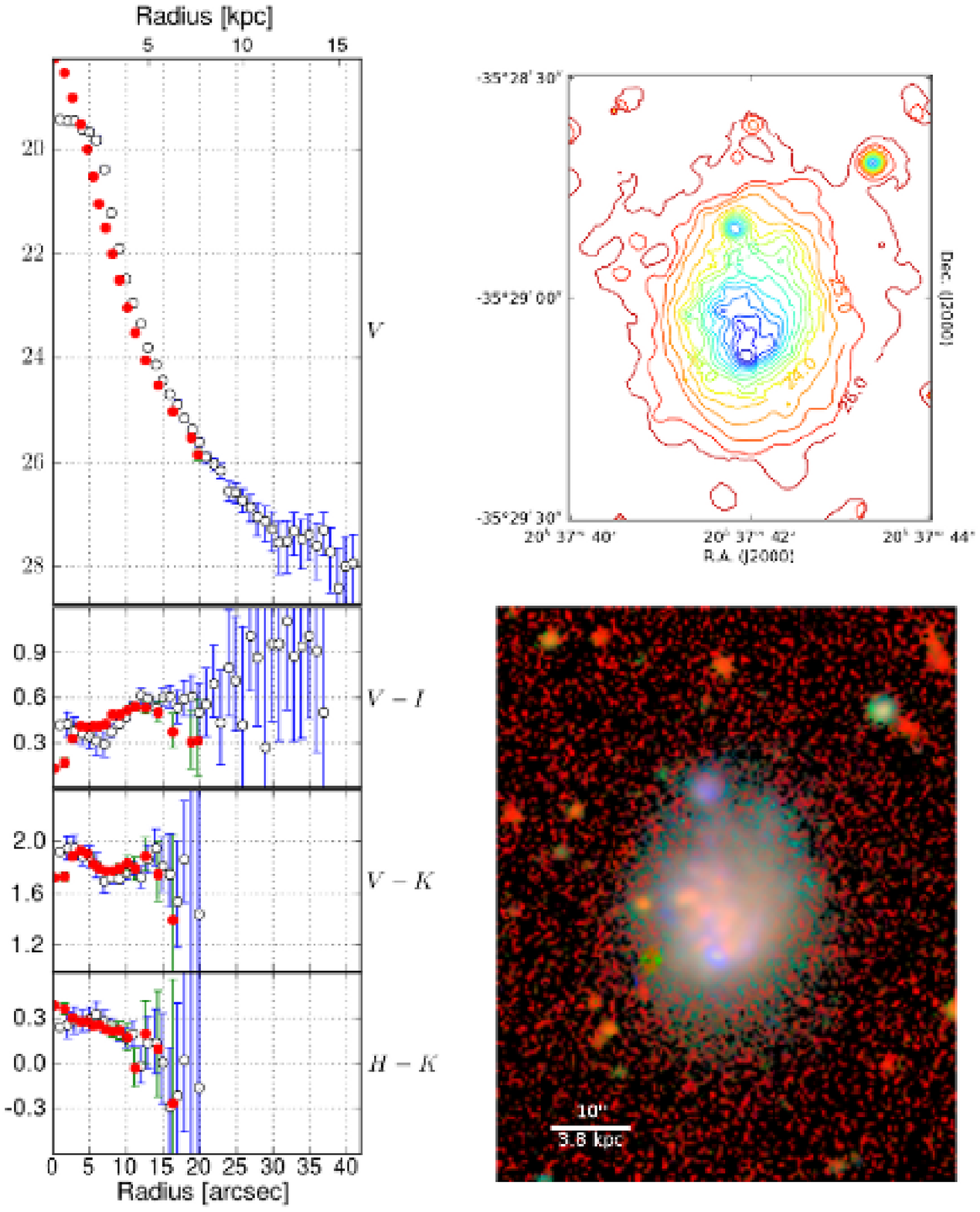}
\contcaption{\textbf{ESO400-43}. \textit{Left panel}: Surface brightness and color radial profiles for elliptical (open circles) and isophotal (red circles) integration. \textit{Upper right panel}: contour plot based on the $V$ band. Isophotes fainter than $23.5$, $25.5$ are iteratively smoothed with a boxcar median filter of size $5$, $15$ pixels respectively. \textit{Lower right panel}: A true color RGB composite image using the $V,I,H$ filters. Each channel has been corrected for Galactic extinction following \citet{1998ApJ...500..525S} and converted to the AB photometric system. The RGB composite was created by implementing the \citet{2004PASP..116..133L} algorithm.}
\end{minipage}
\end{figure*}

\clearpage

\begin{figure*}
\begin{minipage}{150mm}
\includegraphics[width=15cm,height=18cm]{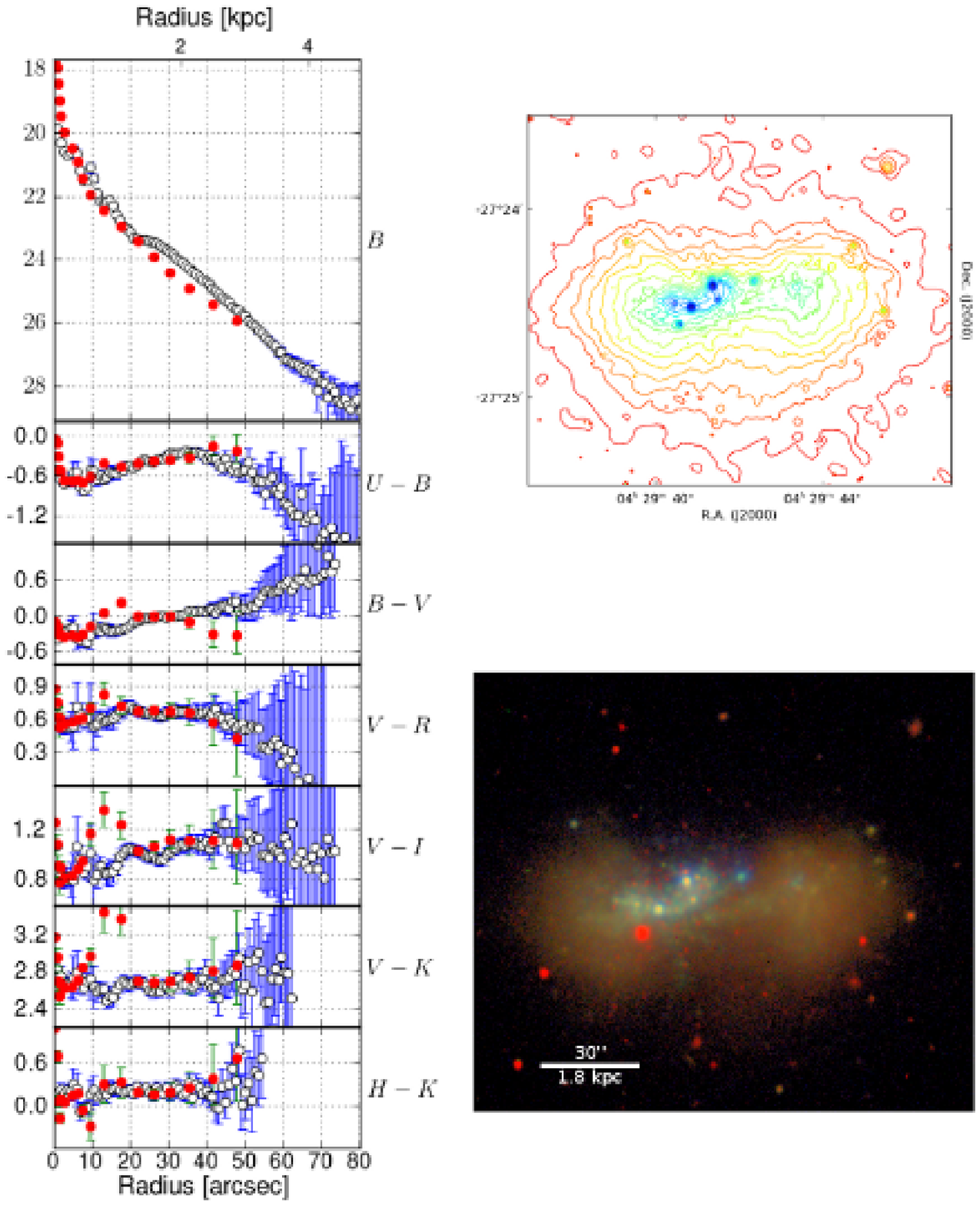}
\contcaption{\textbf{ESO421-02}. \textit{Left panel}: Surface brightness and color radial profiles for elliptical (open circles) and isophotal (red circles) integration. \textit{Upper right panel}: contour plot based on the $B$ band. Isophotes fainter than $23.5$, $25.5$ are iteratively smoothed with a boxcar median filter of size $5$, $15$ pixels respectively. \textit{Lower right panel}: A true color RGB composite image using the $U,B,I$ filters. Each channel has been corrected for Galactic extinction following \citet{1998ApJ...500..525S} and converted to the AB photometric system. The RGB composite was created by implementing the \citet{2004PASP..116..133L} algorithm.}
\end{minipage}
\end{figure*}

\clearpage

\begin{figure*}
\begin{minipage}{150mm}
\includegraphics[width=15cm,height=18cm]{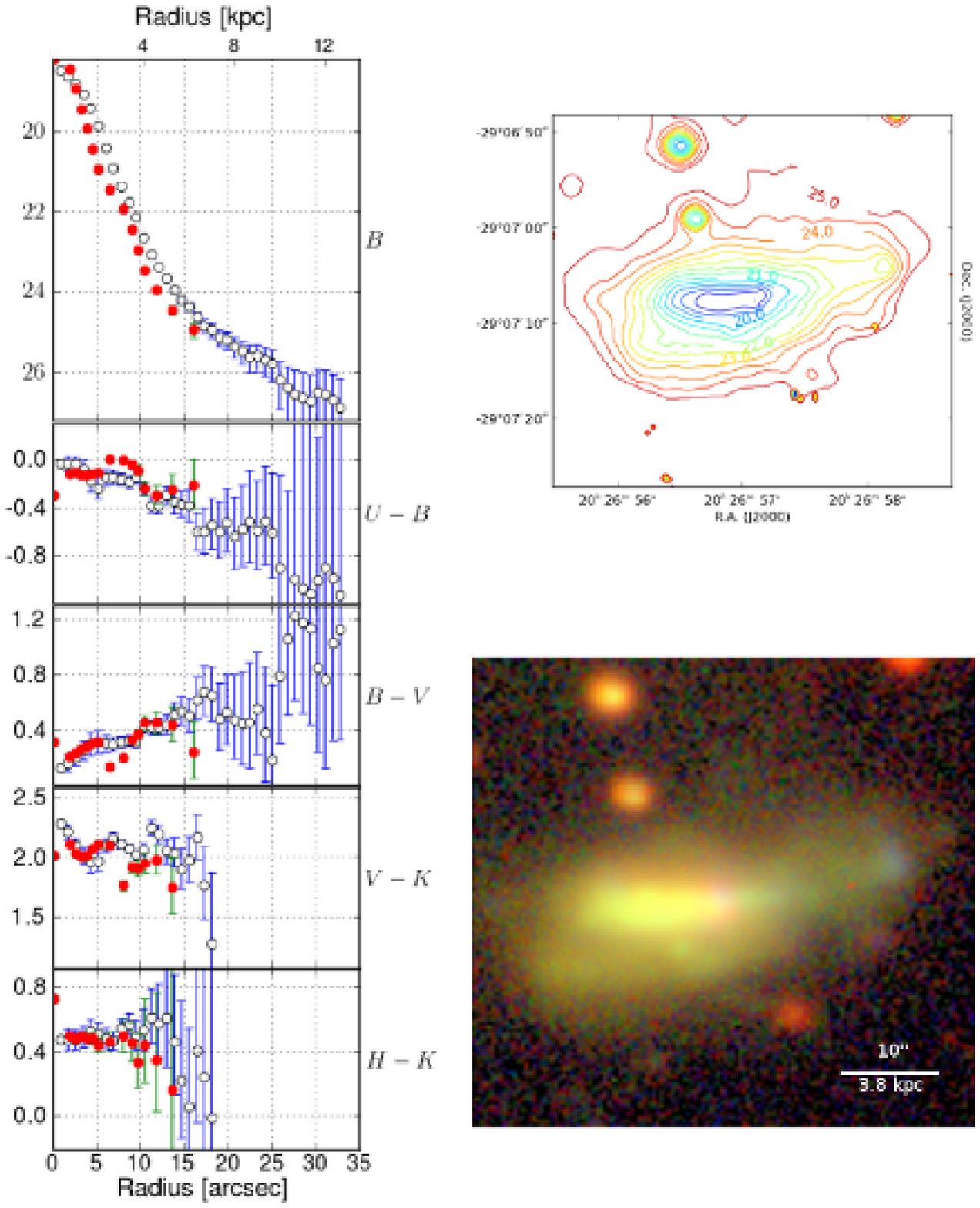}
\contcaption{\textbf{ESO462-20}. \textit{Left panel}: Surface brightness and color radial profiles for elliptical (open circles) and isophotal (red circles) integration. \textit{Upper right panel}: contour plot based on the $B$ band. Isophotes fainter than $22.5$ are smoothed with a boxcar median filter of size $5$ pixels. \textit{Lower right panel}: A true color RGB composite image using the $U,B,V$ filters. Each channel has been corrected for Galactic extinction following \citet{1998ApJ...500..525S} and converted to the AB photometric system. The RGB composite was created by implementing the \citet{2004PASP..116..133L} algorithm.}
\end{minipage}
\end{figure*}

\clearpage

\begin{figure*}
\begin{minipage}{150mm}
\includegraphics[width=15cm,height=18cm]{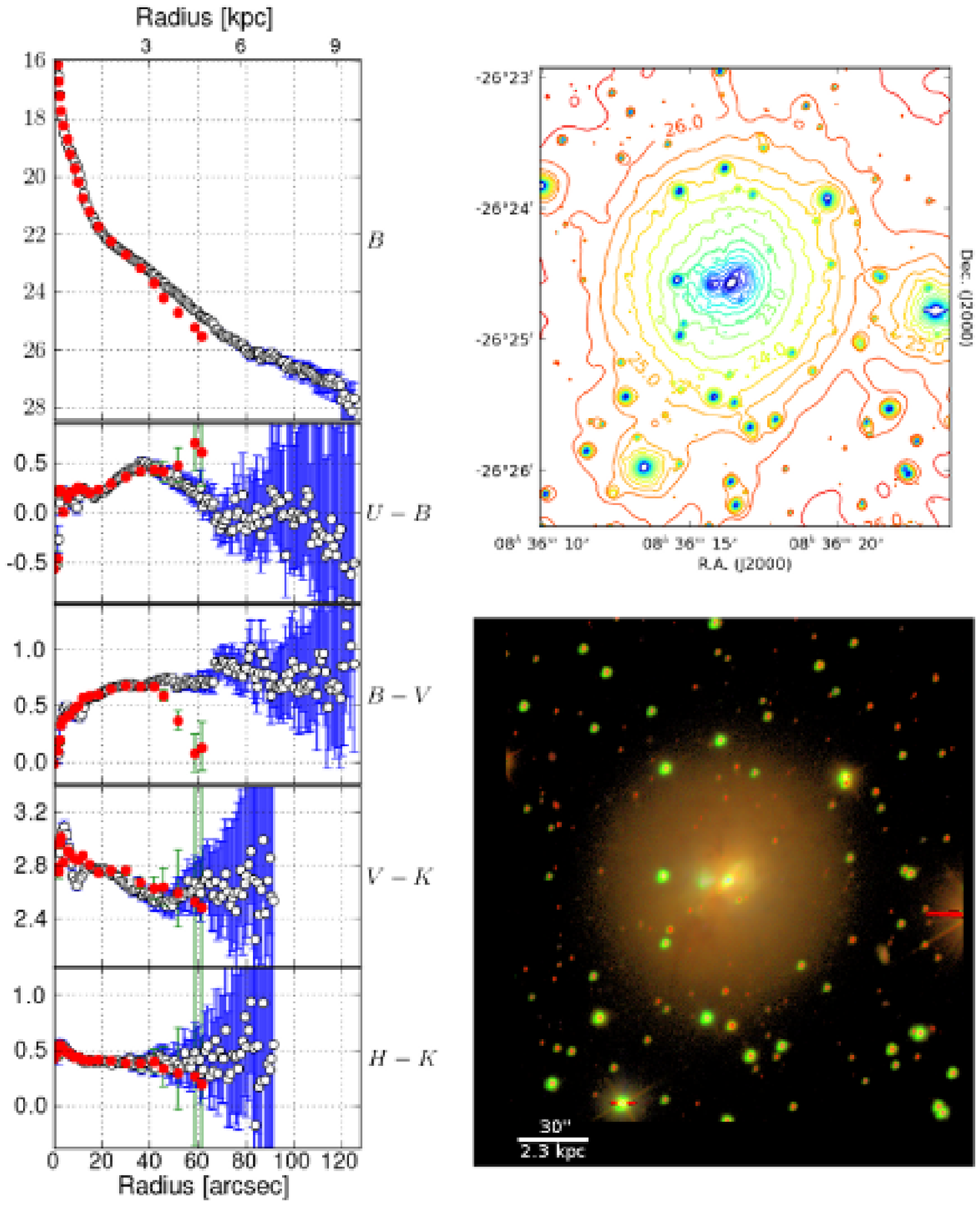}
\contcaption{\textbf{HE2-10}. \textit{Left panel}: Surface brightness and color radial profiles for elliptical (open circles) and isophotal (red circles) integration. \textit{Upper right panel}: contour plot based on the $B$ band. Isophotes fainter than $22.5$, $24.5$, $25.5$ are iteratively smoothed with a boxcar median filter of size $5$, $15$, $25$ pixels respectively. \textit{Lower right panel}: A true color RGB composite image using the $U,B,V$ filters. Each channel has been corrected for Galactic extinction following \citet{1998ApJ...500..525S} and converted to the AB photometric system. The RGB composite was created by implementing the \citet{2004PASP..116..133L} algorithm.}
\end{minipage}
\end{figure*}

\clearpage

\begin{figure*}
\begin{minipage}{150mm}
\includegraphics[width=15cm,height=18cm]{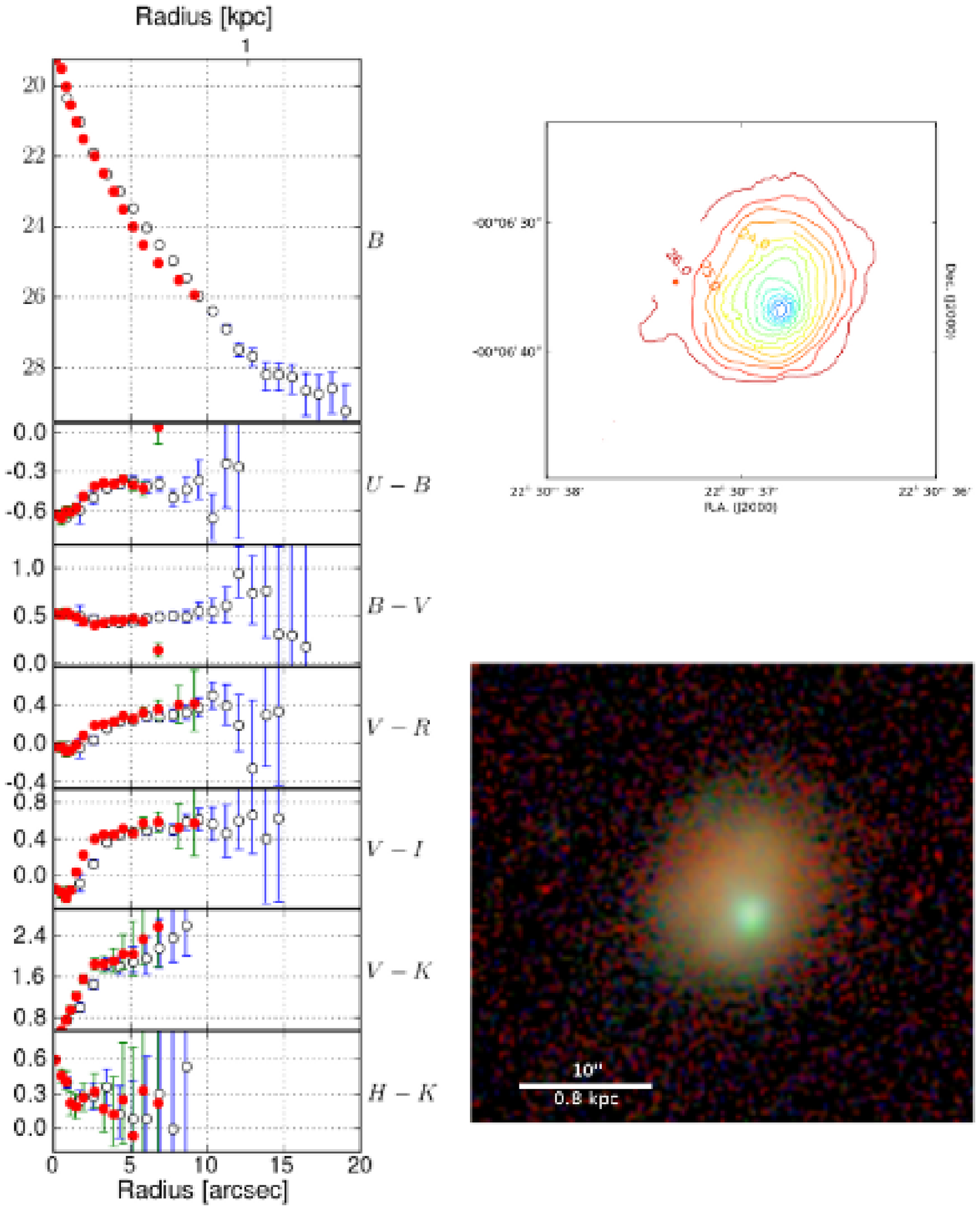}
\contcaption{\textbf{HL293B}. \textit{Left panel}: Surface brightness and color radial profiles for elliptical (open circles) and isophotal (red circles) integration. \textit{Upper right panel}: contour plot based on the $B$ band. Isophotes fainter than $23.5$, $25.5$ are iteratively smoothed with a boxcar median filter of size $5$, $15$ pixels respectively. \textit{Lower right panel}: A true color RGB composite image using the $U,B,I$ filters. Each channel has been corrected for Galactic extinction following \citet{1998ApJ...500..525S} and converted to the AB photometric system. The RGB composite was created by implementing the \citet{2004PASP..116..133L} algorithm.}
\end{minipage}
\end{figure*}

\clearpage

\begin{figure*}
\begin{minipage}{150mm}
\includegraphics[width=15cm,height=18cm]{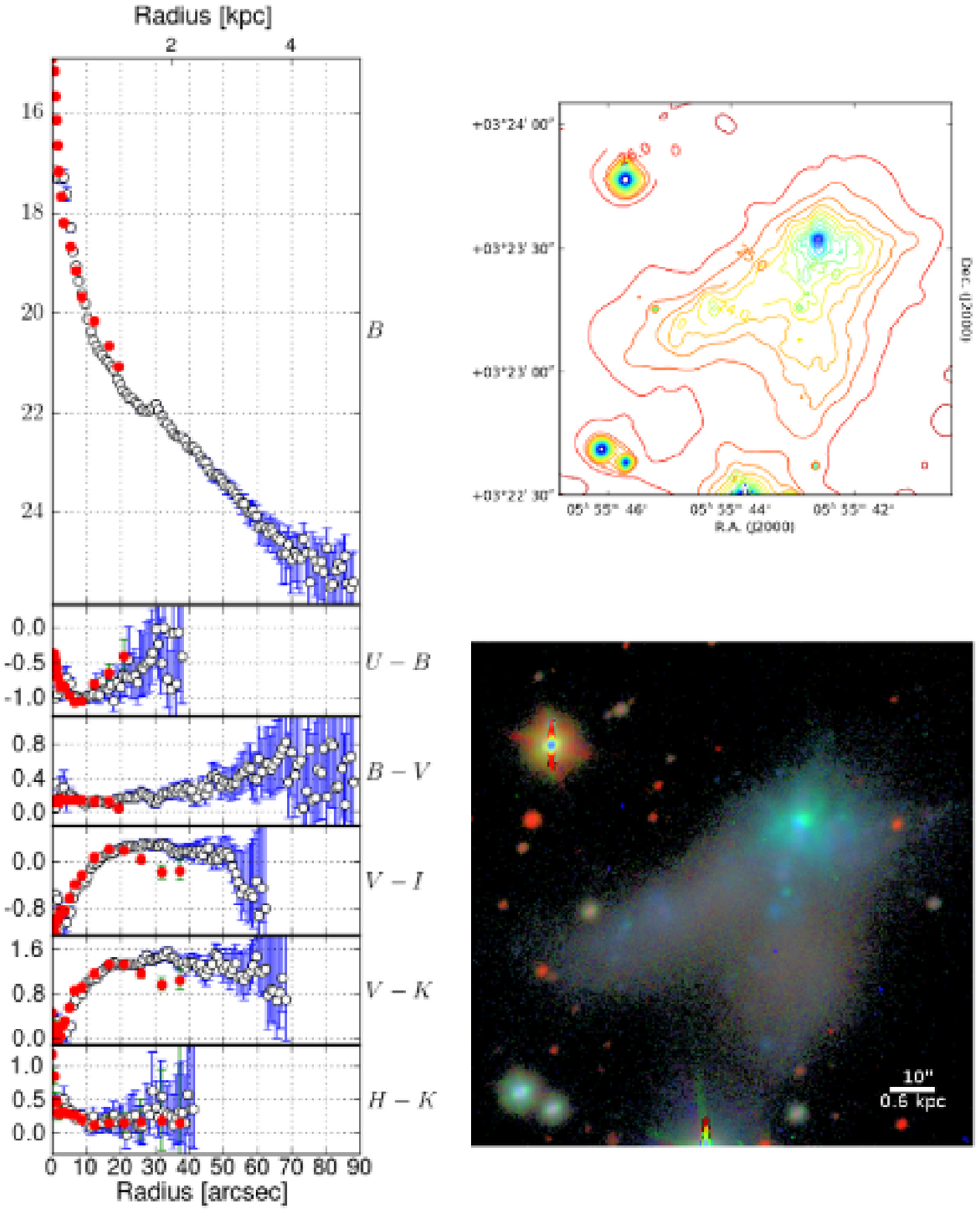}
\contcaption{\textbf{IIZw40}. \textit{Left panel}: Surface brightness and color radial profiles for elliptical (open circles) and isophotal (red circles) integration. \textit{Upper right panel}: contour plot based on the $B$ band. Isophotes fainter than $22.5$, $25.0$ are iteratively smoothed with a boxcar median filter of size $5$, $15$ pixels respectively. \textit{Lower right panel}: A true color RGB composite image using the $B,V,I$ filters. Each channel has been corrected for Galactic extinction following \citet{1998ApJ...500..525S} and converted to the AB photometric system. The RGB composite was created by implementing the \citet{2004PASP..116..133L} algorithm.}
\end{minipage}
\end{figure*}

\clearpage

%

\begin{figure*}
\begin{minipage}{150mm}
\includegraphics[width=15cm,height=18cm]{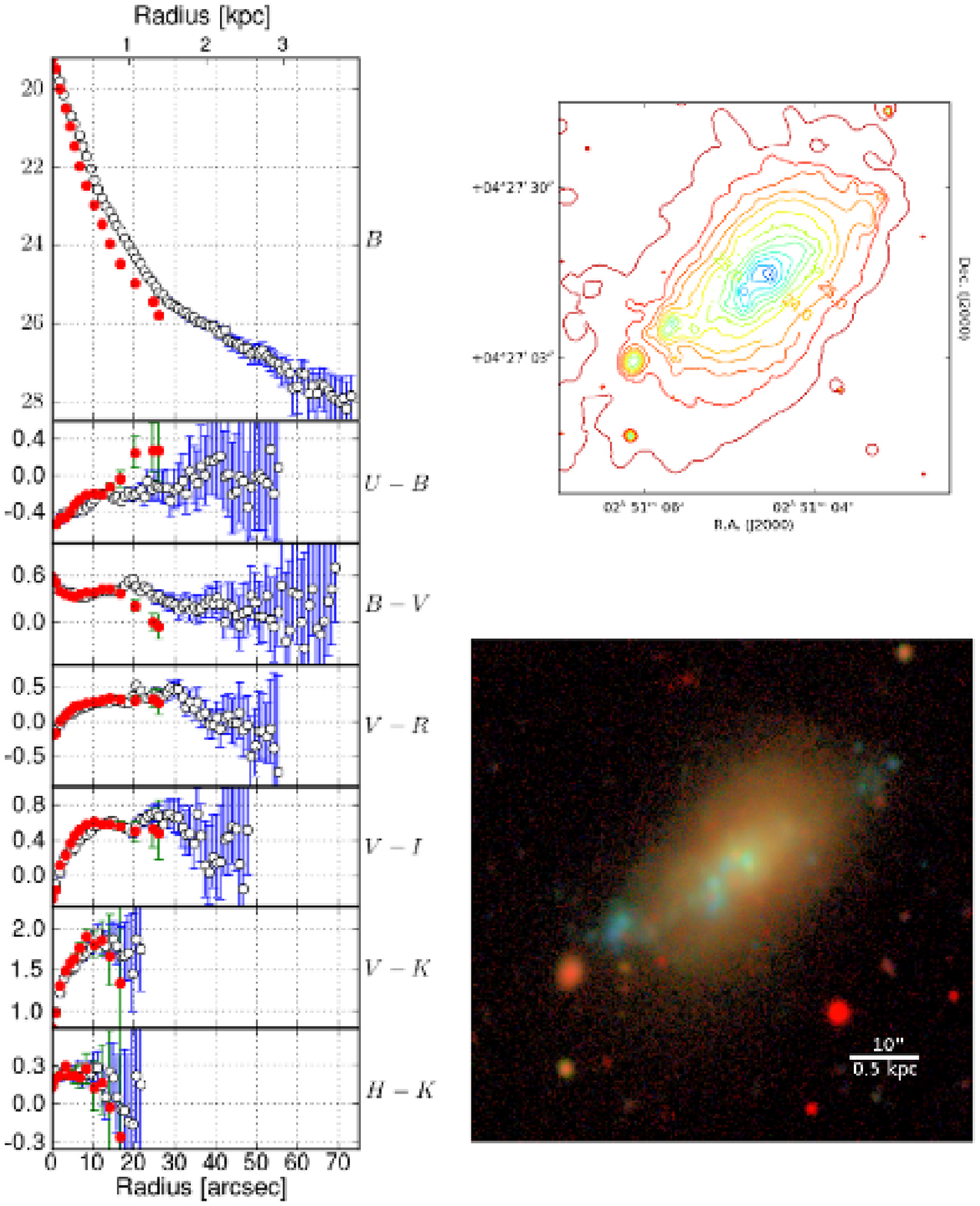}
\contcaption{\textbf{MK600}. \textit{Left panel}: Surface brightness and color radial profiles for elliptical (open circles) and isophotal (red circles) integration. \textit{Upper right panel}: contour plot based on the $B$ band. Isophotes fainter than $23.0$, $25.0$ are iteratively smoothed with a boxcar median filter of size $5$, $15$ pixels respectively. \textit{Lower right panel}: A true color RGB composite image using the $U,B,I$ filters. Each channel has been corrected for Galactic extinction following \citet{1998ApJ...500..525S} and converted to the AB photometric system. The RGB composite was created by implementing the \citet{2004PASP..116..133L} algorithm.}
\end{minipage}
\end{figure*}

\clearpage

\begin{figure*}
\begin{minipage}{150mm}
\includegraphics[width=15cm,height=18cm]{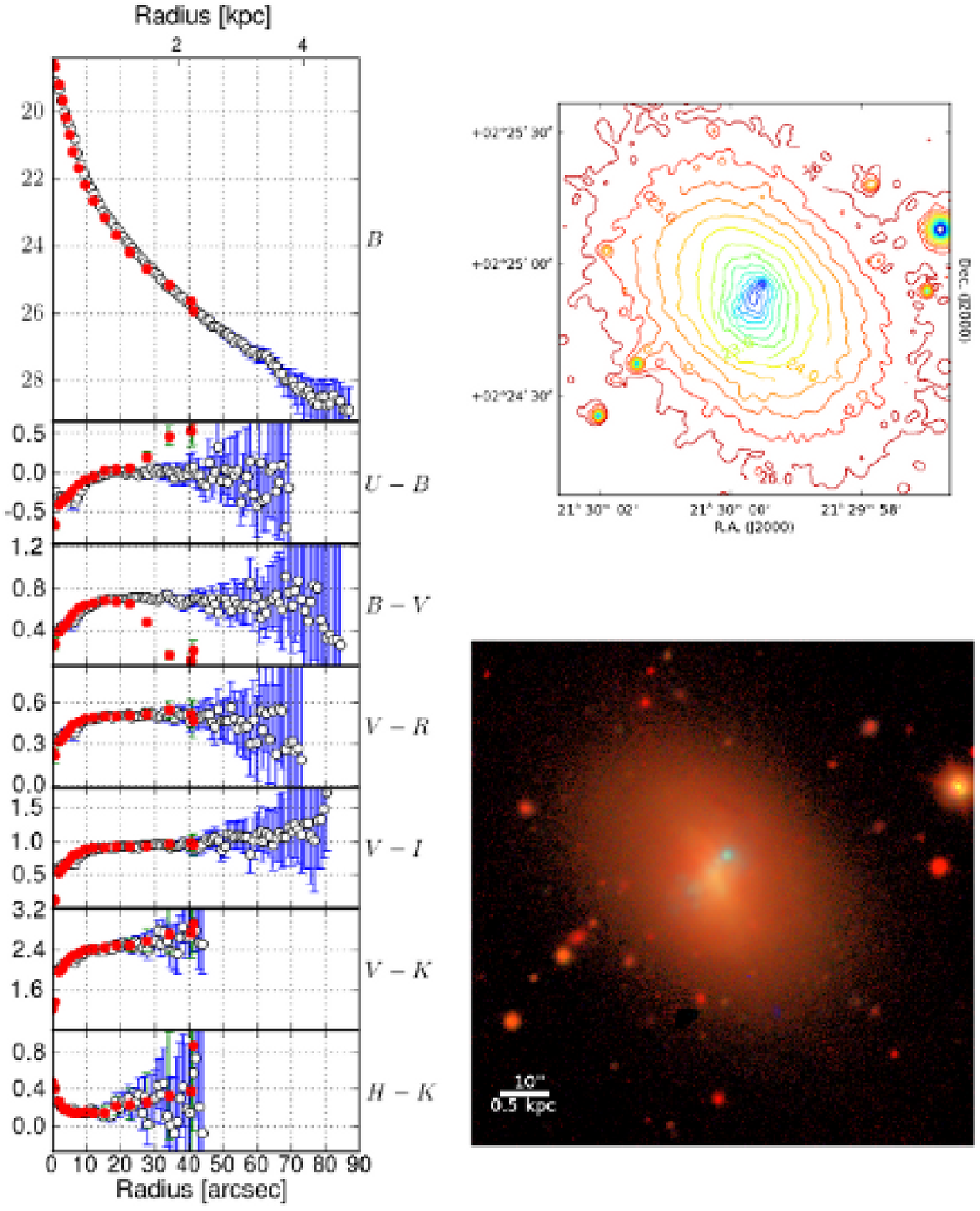}
\contcaption{\textbf{MK900}. \textit{Left panel}: Surface brightness and color radial profiles for elliptical (open circles) and isophotal (red circles) integration. \textit{Upper right panel}: contour plot based on the $B$ band. Isophotes fainter than $22.5$, $25.5$ are iteratively smoothed with a boxcar median filter of size $5$, $15$ pixels respectively. \textit{Lower right panel}: A true color RGB composite image using the $U,B,I$ filters. Each channel has been corrected for Galactic extinction following \citet{1998ApJ...500..525S} and converted to the AB photometric system. The RGB composite was created by implementing the \citet{2004PASP..116..133L} algorithm.}
\end{minipage}
\end{figure*}

\clearpage

\begin{figure*}
\begin{minipage}{150mm}
\includegraphics[width=15cm,height=18cm]{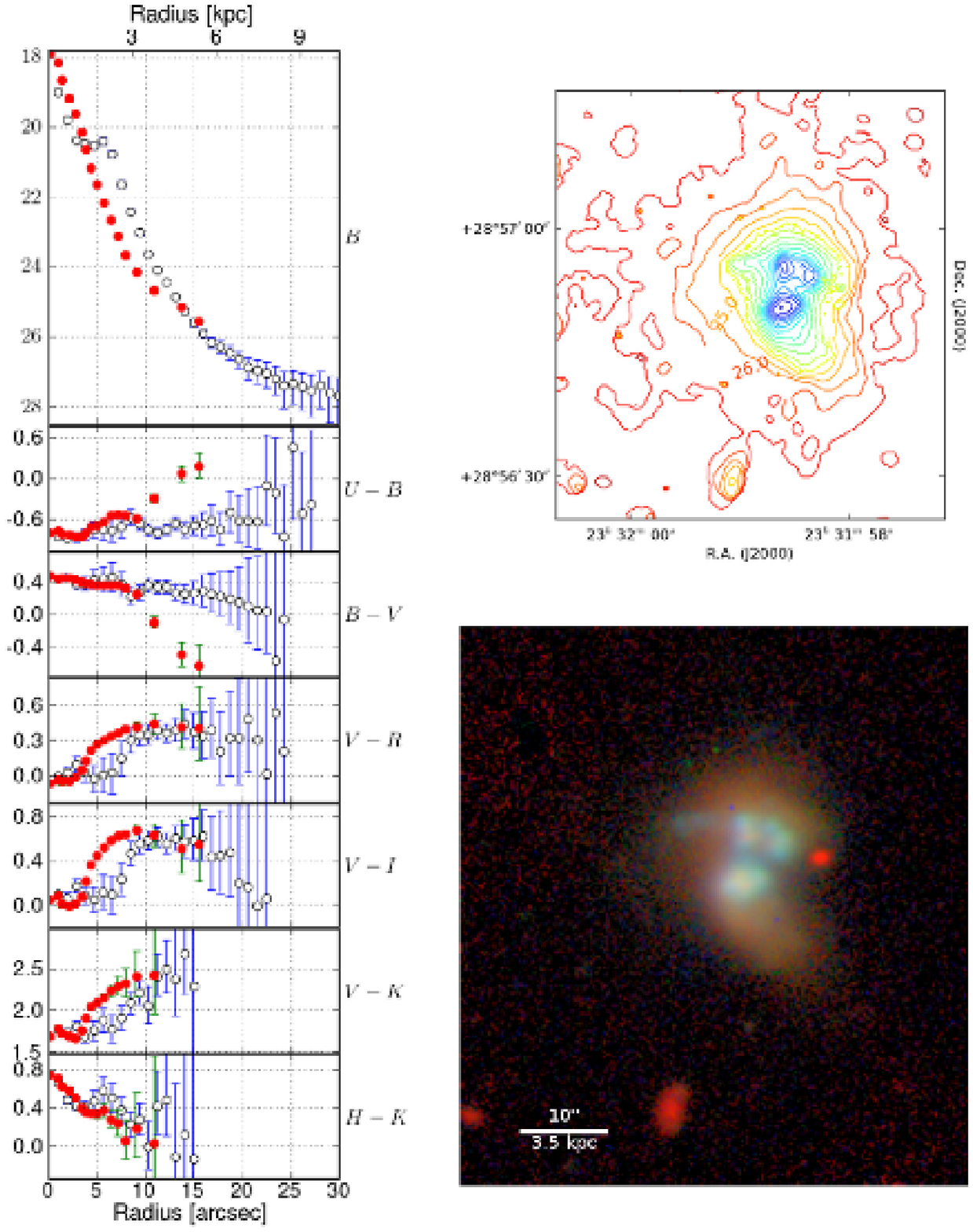}
\contcaption{\textbf{MK930}. \textit{Left panel}: Surface brightness and color radial profiles for elliptical (open circles) and isophotal (red circles) integration. \textit{Upper right panel}: contour plot based on the $B$ band. Isophotes fainter than $24.0$, $26.0$ are iteratively smoothed with a boxcar median filter of size $5$, $15$ pixels respectively. \textit{Lower right panel}: A true color RGB composite image using the $U,B,I$ filters. Each channel has been corrected for Galactic extinction following \citet{1998ApJ...500..525S} and converted to the AB photometric system. The RGB composite was created by implementing the \citet{2004PASP..116..133L} algorithm.}
\end{minipage}
\end{figure*}

\clearpage

\begin{figure*}
\begin{minipage}{150mm}
\includegraphics[width=15cm,height=18cm]{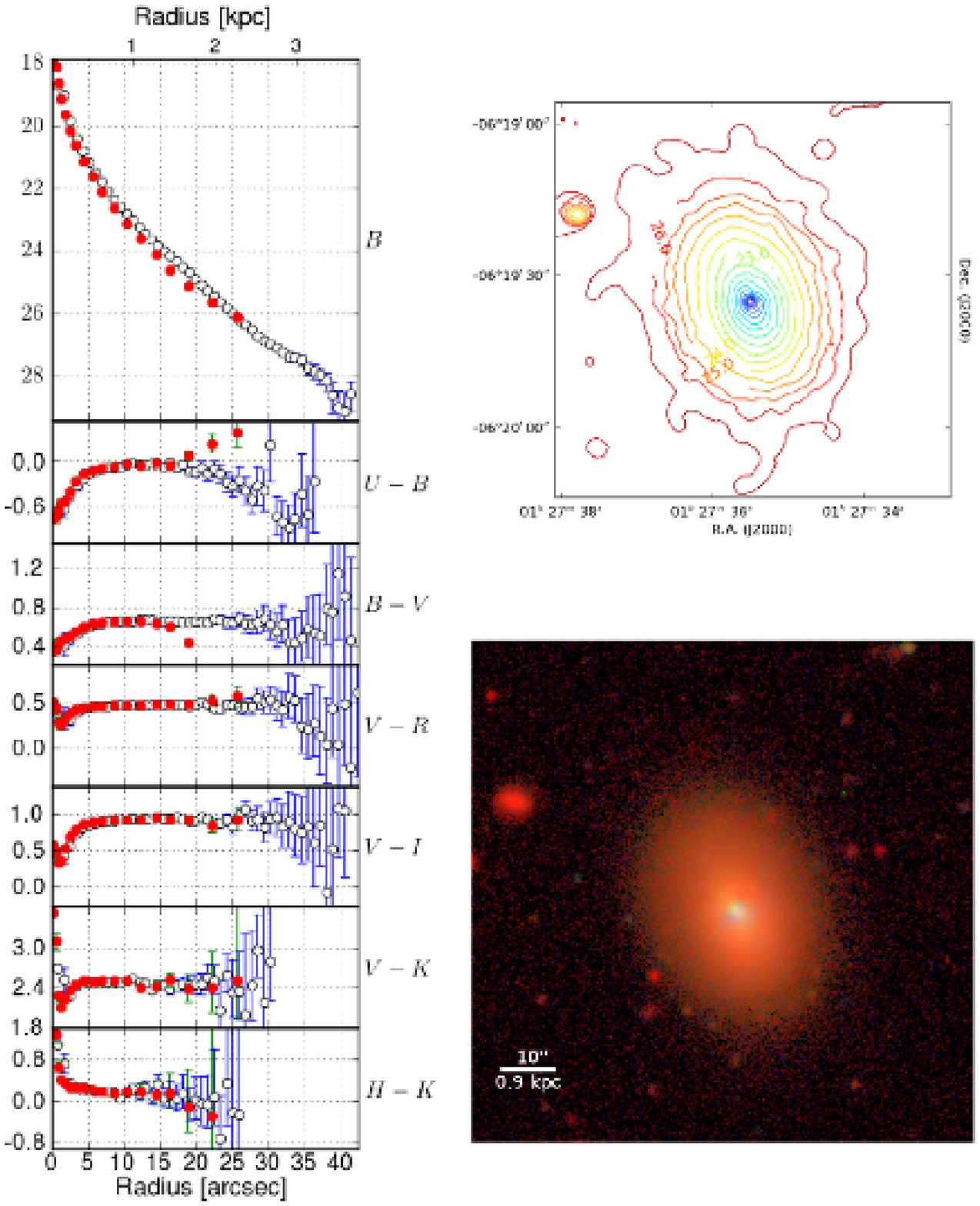}
\contcaption{\textbf{MK996}. \textit{Left panel}: Surface brightness and color radial profiles for elliptical (open circles) and isophotal (red circles) integration. \textit{Upper right panel}: contour plot based on the $B$ band. Isophotes fainter than $23.0$, $25.5$ are iteratively smoothed with a boxcar median filter of size $5$, $15$ pixels respectively. \textit{Lower right panel}: A true color RGB composite image using the $U,B,I$ filters. Each channel has been corrected for Galactic extinction following \citet{1998ApJ...500..525S} and converted to the AB photometric system. The RGB composite was created by implementing the \citet{2004PASP..116..133L} algorithm.}
\end{minipage}
\end{figure*}

\clearpage

%

\begin{figure*}
\begin{minipage}{150mm}
\includegraphics[width=15cm,height=18cm]{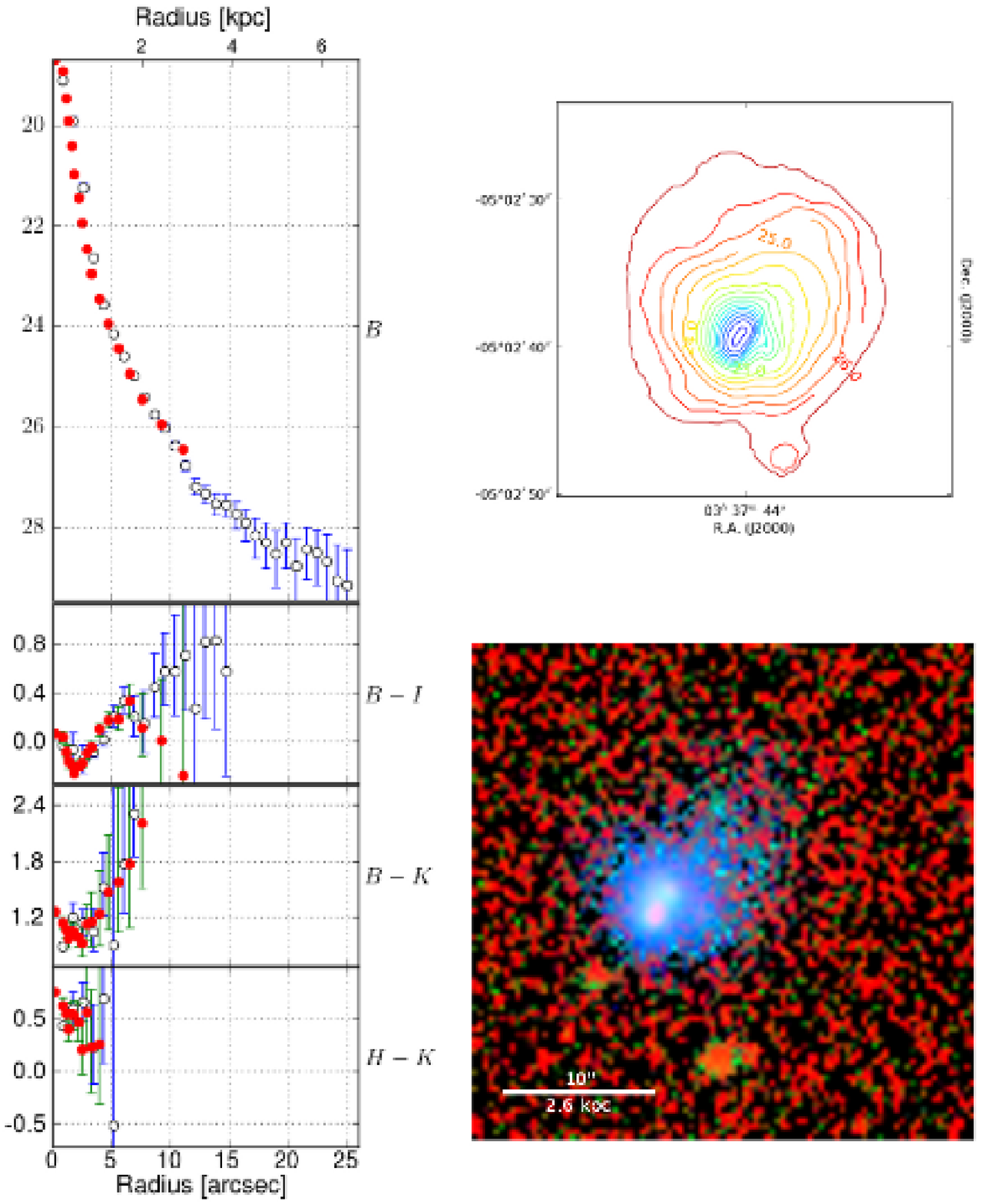}
\contcaption{\textbf{SBS0335-052 EAST}. \textit{Left panel}: Surface brightness and color radial profiles for elliptical (open circles) and isophotal (red circles) integration. \textit{Upper right panel}: contour plot based on the $B$ band. Isophotes fainter than $23.5$, $26.0$ are iteratively smoothed with a boxcar median filter of size $5$, $15$ pixels respectively. \textit{Lower right panel}: A true color RGB composite image using the $B,I,K$ filters. Each channel has been corrected for Galactic extinction following \citet{1998ApJ...500..525S} and converted to the AB photometric system. The RGB composite was created by implementing the \citet{2004PASP..116..133L} algorithm.}
\end{minipage}
\end{figure*}
\clearpage

\begin{figure*}
\begin{minipage}{150mm}
\includegraphics[width=15cm,height=18cm]{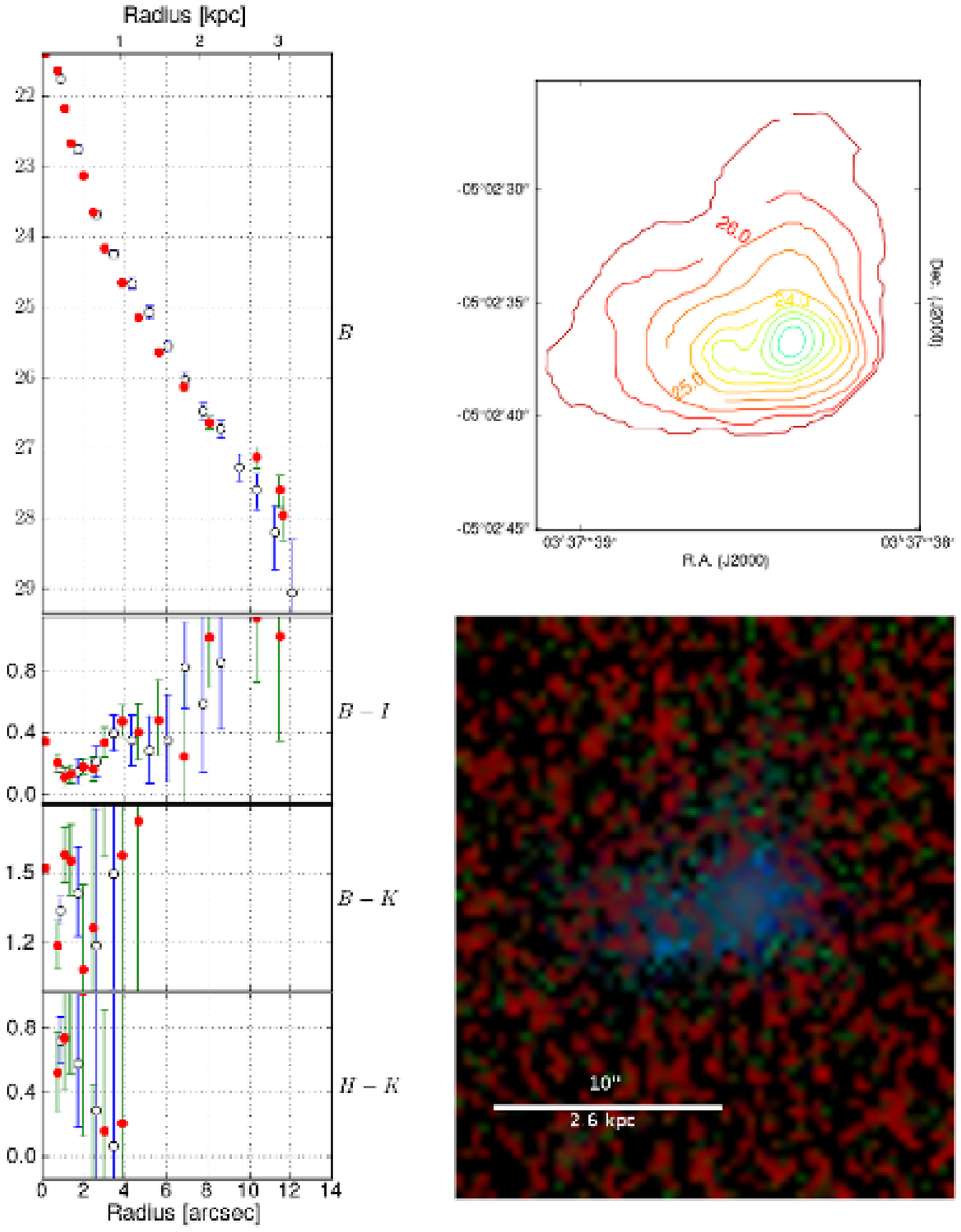}
\contcaption{\textbf{SBS0335-052 WEST}. \textit{Left panel}: Surface brightness and color radial profiles for elliptical (open circles) and isophotal (red circles) integration. \textit{Upper right panel}: contour plot based on the $B$ band. Isophotes fainter than $23.5$, $26.0$ are iteratively smoothed with a boxcar median filter of size $5$, $15$ pixels respectively. \textit{Lower right panel}: A true color RGB composite image using the $B,I,K$ filters. Each channel has been corrected for Galactic extinction following \citet{1998ApJ...500..525S} and converted to the AB photometric system. The RGB composite was created by implementing the \citet{2004PASP..116..133L} algorithm.}
\end{minipage}
\end{figure*}

\clearpage

\begin{figure*}
\begin{minipage}{150mm}
\includegraphics[width=15cm,height=18cm]{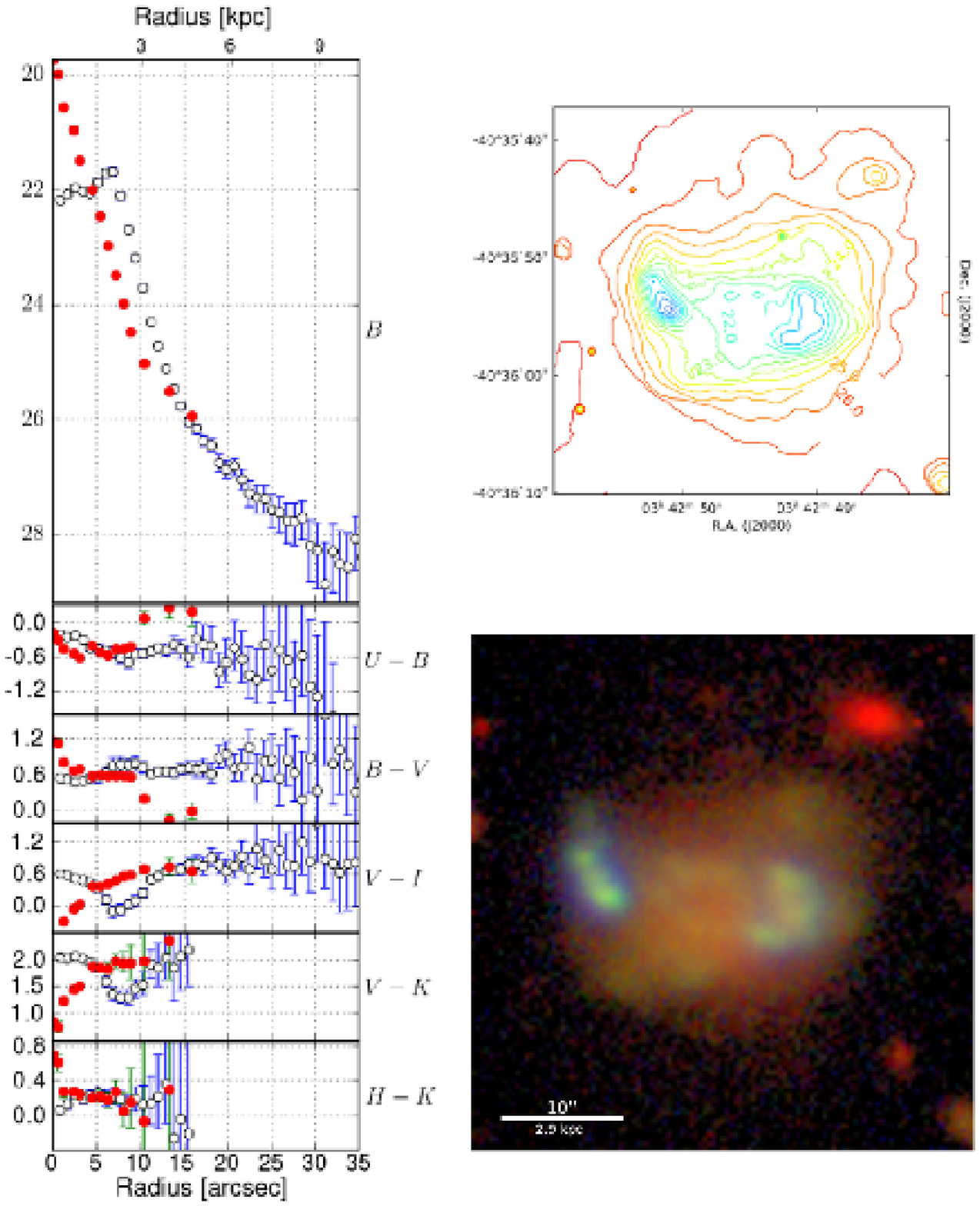}
\contcaption{\textbf{TOL0341-407}. \textit{Left panel}: Surface brightness and color radial profiles for elliptical (open circles) and isophotal (red circles) integration. \textit{Upper right panel}: contour plot based on the $B$ band. Isophotes fainter than $23.5$, $25.5$ are iteratively smoothed with a boxcar median filter of size $5$, $15$ pixels respectively. \textit{Lower right panel}: A true color RGB composite image using the $U,B,I$ filters. Each channel has been corrected for Galactic extinction following \citet{1998ApJ...500..525S} and converted to the AB photometric system. The RGB composite was created by implementing the \citet{2004PASP..116..133L} algorithm.}
\end{minipage}
\end{figure*}

\clearpage

\begin{figure*}
\begin{minipage}{150mm}
\includegraphics[width=15cm,height=18cm]{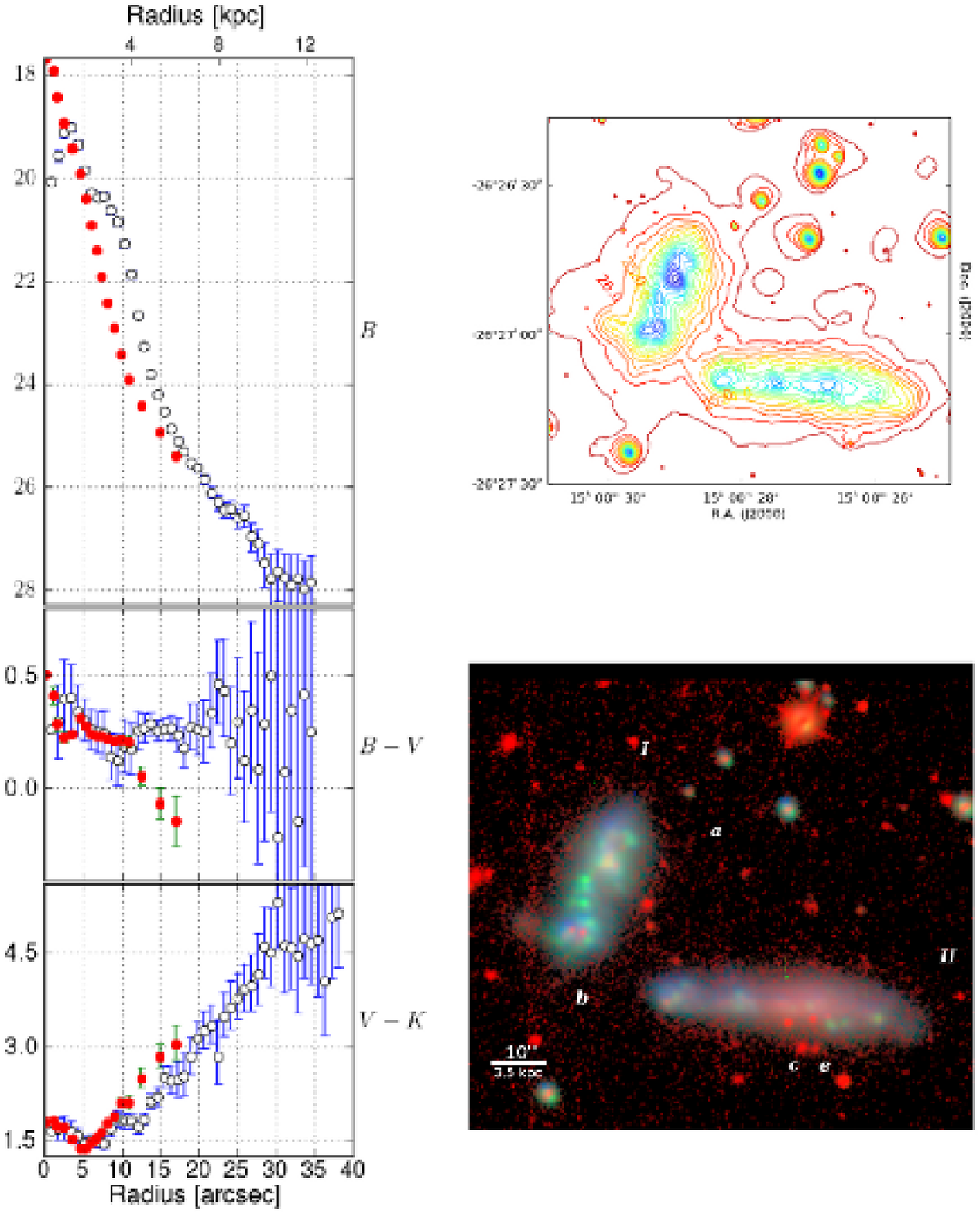}
\contcaption{\textbf{TOL1457--262 \emph{I}}. \textit{Left panel}: Surface brightness and color radial profiles for elliptical (open circles) and isophotal (red circles) integration. \textit{Upper right panel}: contour plot based on the $B$ band. Isophotes fainter than $24.0$, $25.5$ are iteratively smoothed with a boxcar median filter of size $5$, $15$ pixels respectively. \textit{Lower right panel}: A true color RGB composite image using the $B,V,K$ filters. Each channel has been corrected for Galactic extinction following \citet{1998ApJ...500..525S} and converted to the AB photometric system. The RGB composite was created by implementing the \citet{2004PASP..116..133L} algorithm.}
\end{minipage}
\end{figure*}

\clearpage

\begin{figure*}
\begin{minipage}{150mm}
\includegraphics[width=15cm,height=18cm]{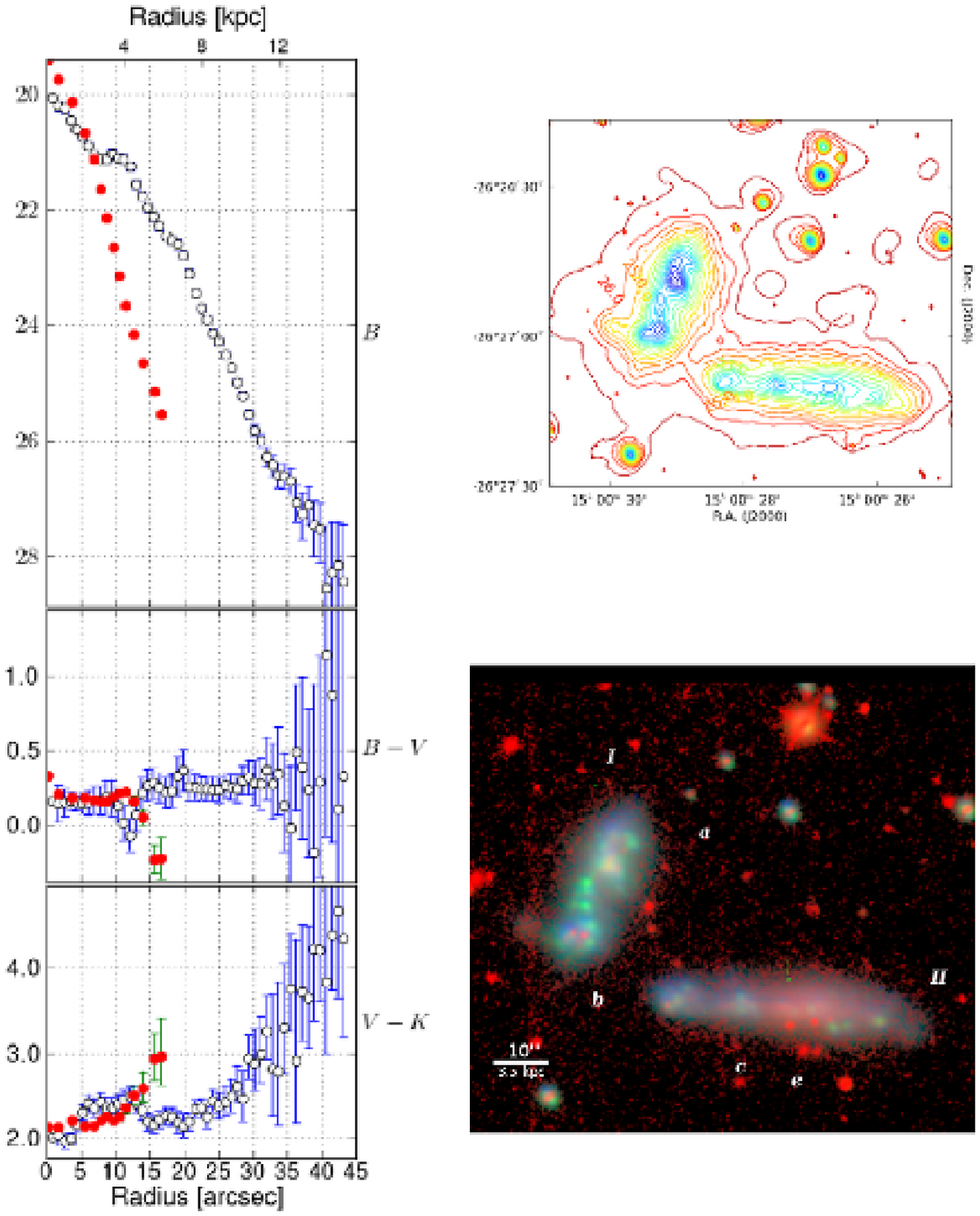}
\contcaption{\textbf{TOL1457--262 \emph{II}}. \textit{Left panel}: Surface brightness and color radial profiles for elliptical (open circles) and isophotal (red circles) integration. \textit{Upper right panel}: contour plot based on the $B$ band. Isophotes fainter than $24.0$, $25.5$ are iteratively smoothed with a boxcar median filter of size $5$, $15$ pixels respectively. \textit{Lower right panel}: A true color RGB composite image using the $B,V,K$ filters. Each channel has been corrected for Galactic extinction following \citet{1998ApJ...500..525S} and converted to the AB photometric system. The RGB composite was created by implementing the \citet{2004PASP..116..133L} algorithm.}
\end{minipage}
\end{figure*}

\clearpage

%

\begin{figure*}
\begin{minipage}{150mm}
\includegraphics[width=15cm,height=18cm]{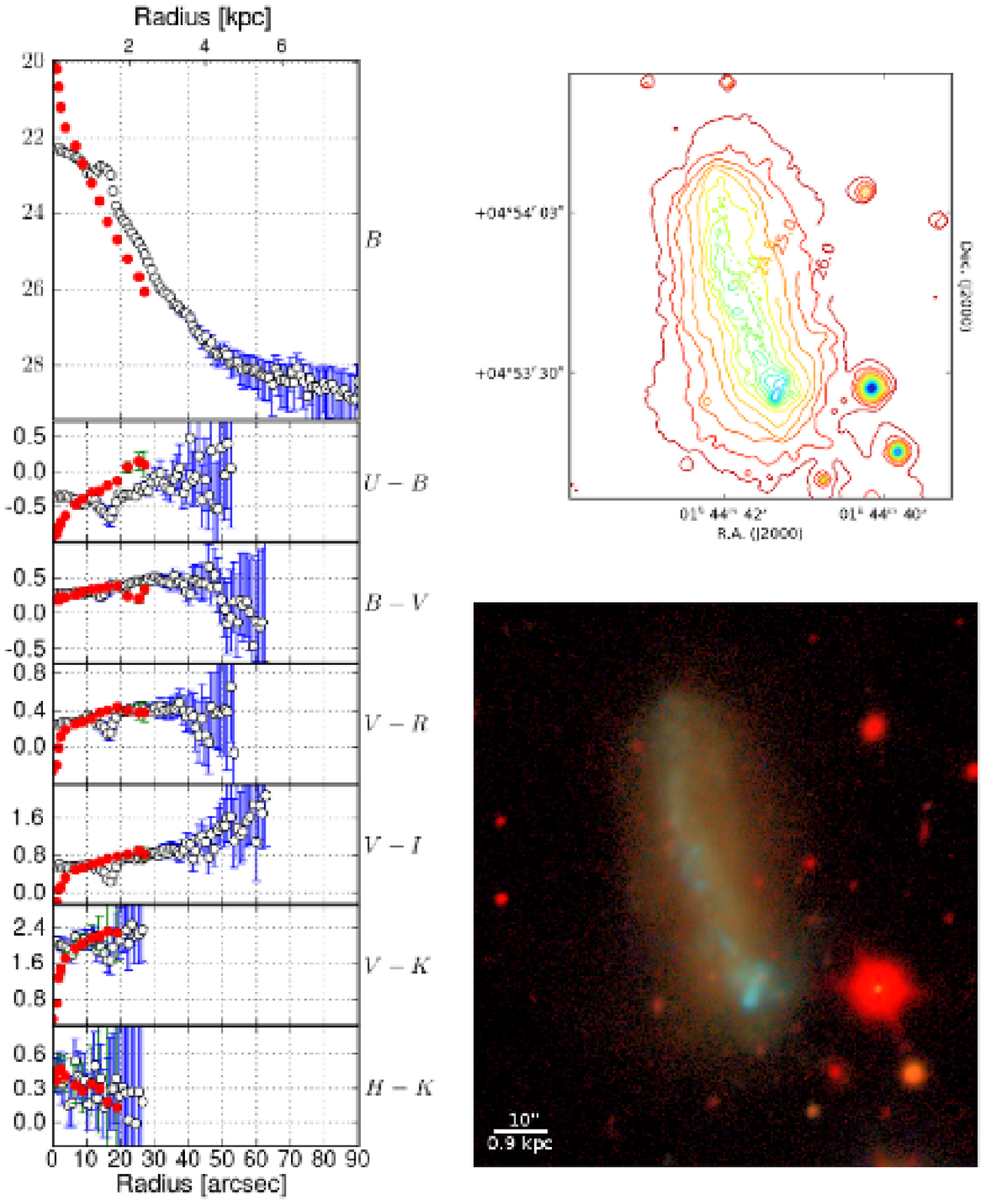}
\contcaption{\textbf{UM133}. \textit{Left panel}: Surface brightness and color radial profiles for elliptical (open circles) and isophotal (red circles) integration. \textit{Upper right panel}: contour plot based on the $B$ band. Isophotes fainter than $23.0$, $25.5$ are iteratively smoothed with a boxcar median filter of size $5$, $15$ pixels respectively. \textit{Lower right panel}: A true color RGB composite image using the $U,B,I$ filters. Each channel has been corrected for Galactic extinction following \citet{1998ApJ...500..525S} and converted to the AB photometric system. The RGB composite was created by implementing the \citet{2004PASP..116..133L} algorithm.}
\end{minipage}
\end{figure*}

\clearpage

\begin{figure*}
\begin{minipage}{150mm}
\includegraphics[width=15cm,height=18cm]{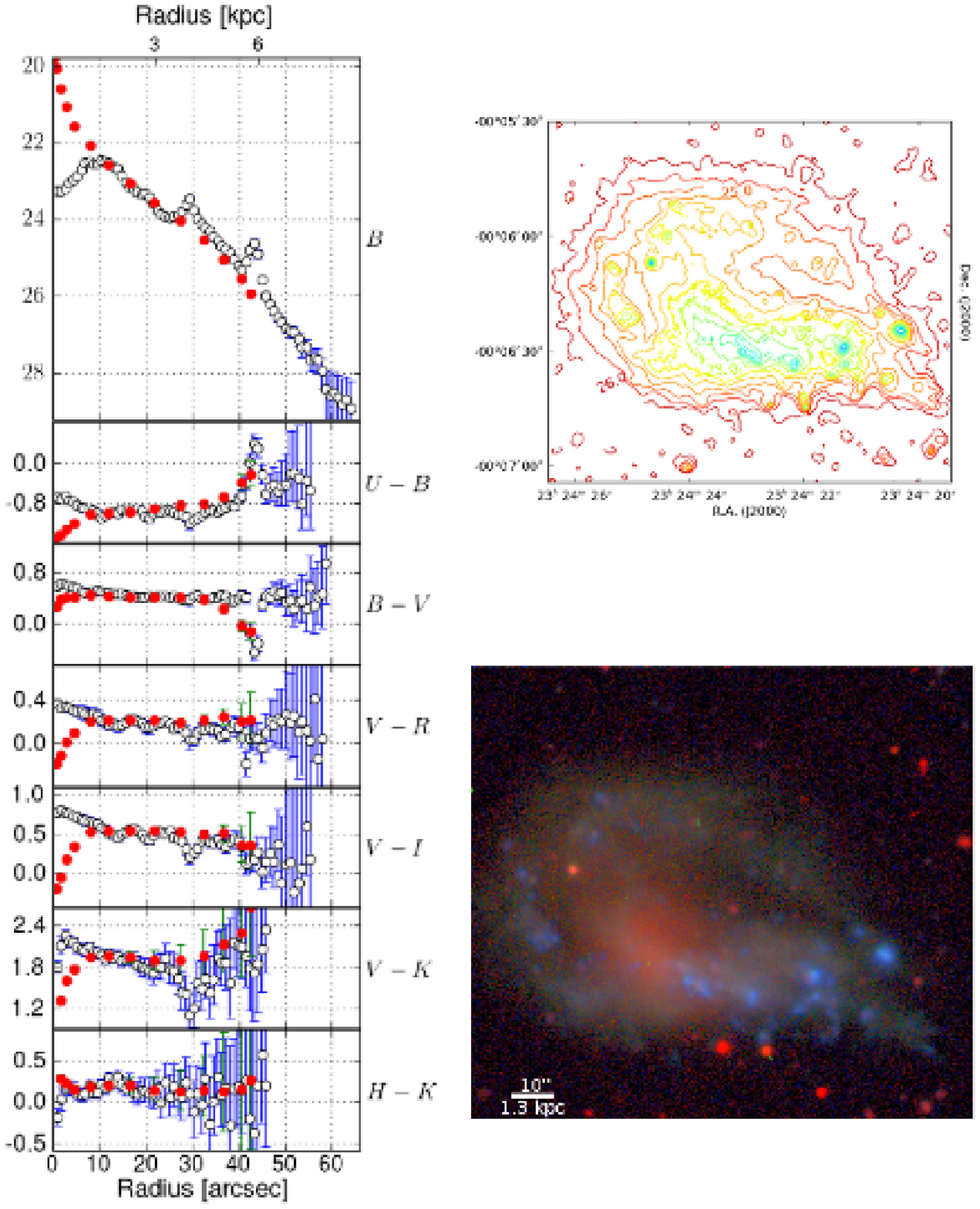}
\contcaption{\textbf{UM160}. \textit{Left panel}: Surface brightness and color radial profiles for elliptical (open circles) and isophotal (red circles) integration. \textit{Upper right panel}: contour plot based on the $B$ band. Isophotes fainter than $23.5$ are smoothed with a boxcar median filter of size $5$ pixels. \textit{Lower right panel}: A true color RGB composite image using the $U,B,I$ filters. Each channel has been corrected for Galactic extinction following \citet{1998ApJ...500..525S} and converted to the AB photometric system. The RGB composite was created by implementing the \citet{2004PASP..116..133L} algorithm.}
\end{minipage}
\end{figure*}

\clearpage

\begin{figure*}
\begin{minipage}{150mm}
\includegraphics[width=15cm,height=18cm]{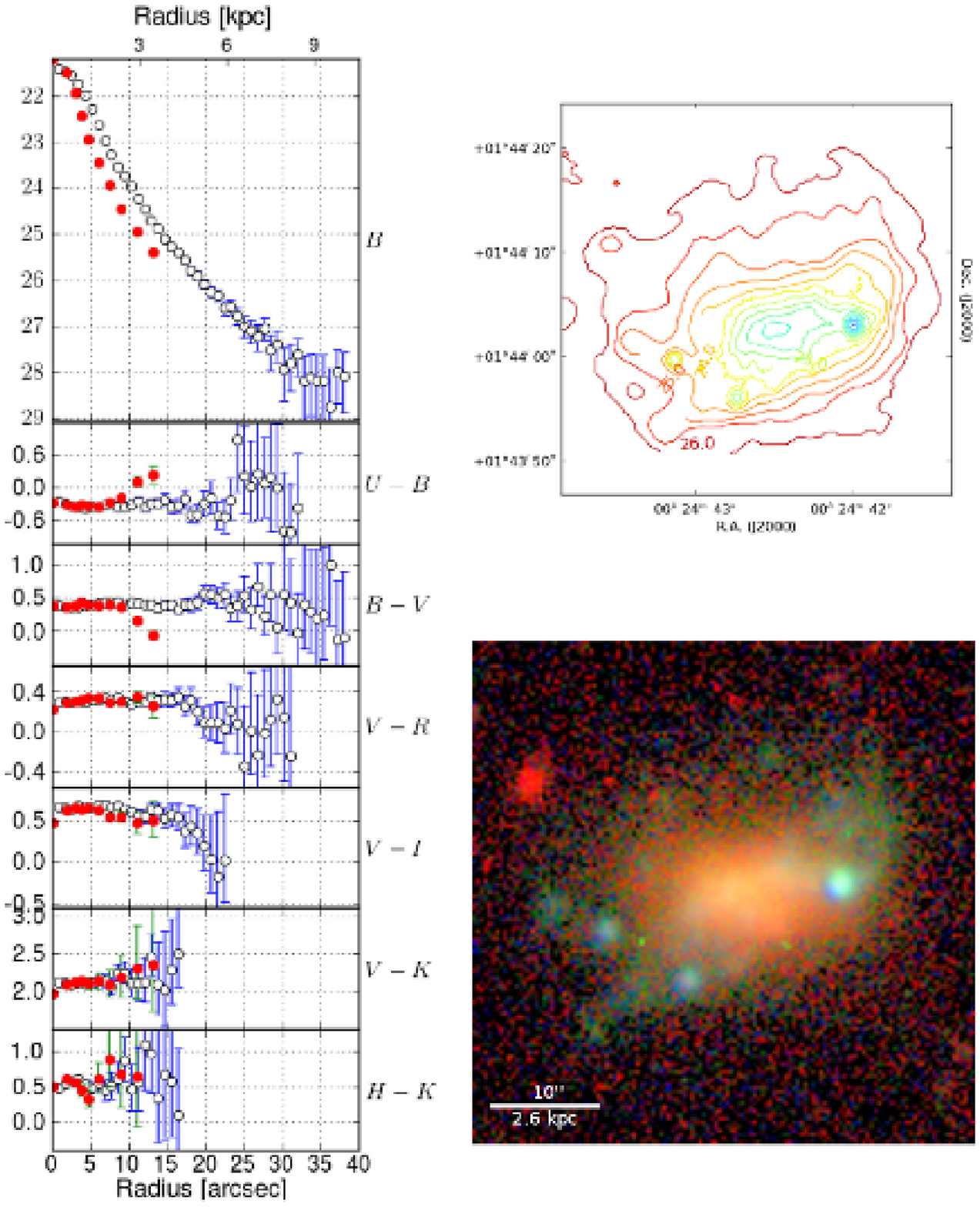}
\contcaption{\textbf{UM238}. \textit{Left panel}: Surface brightness and color radial profiles for elliptical (open circles) and isophotal (red circles) integration. \textit{Upper right panel}: contour plot based on the $B$ band. Isophotes fainter than $23.5$, $25.5$ are iteratively smoothed with a boxcar median filter of size $5$, $15$ pixels respectively. \textit{Lower right panel}: A true color RGB composite image using the $U,B,I$ filters. Each channel has been corrected for Galactic extinction following \citet{1998ApJ...500..525S} and converted to the AB photometric system. The RGB composite was created by implementing the \citet{2004PASP..116..133L} algorithm.}
\end{minipage}
\end{figure*}

\clearpage

\begin{figure*}
\begin{minipage}{150mm}
\includegraphics[width=15cm,height=18cm]{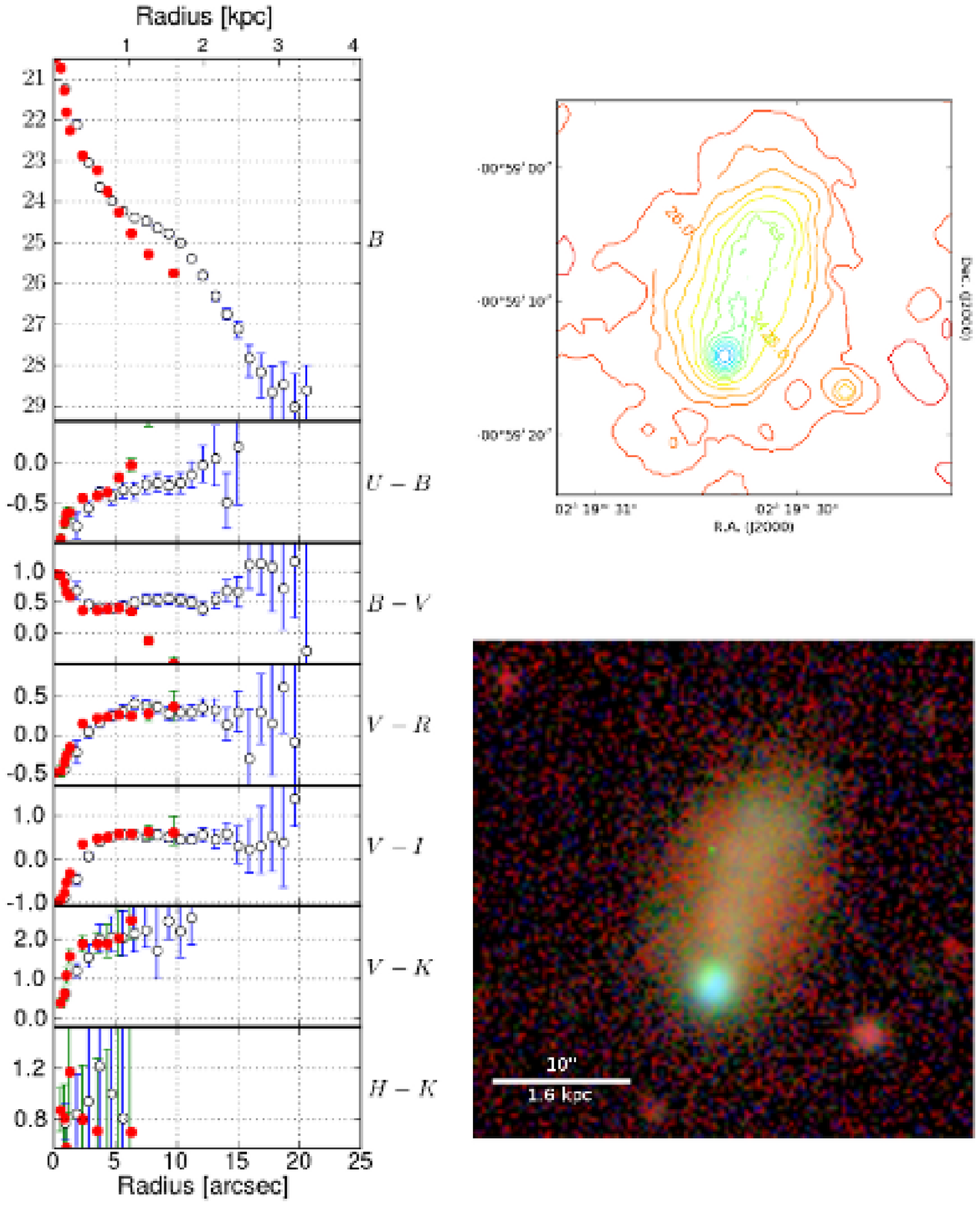}
\contcaption{\textbf{UM417}. \textit{Left panel}: Surface brightness and color radial profiles for elliptical (open circles) and isophotal (red circles) integration. \textit{Upper right panel}: contour plot based on the $B$ band. Isophotes fainter than $23.5$, $26.0$ are iteratively smoothed with a boxcar median filter of size $5$, $15$ pixels respectively. \textit{Lower right panel}: A true color RGB composite image using the $U,B,I$ filters. Each channel has been corrected for Galactic extinction following \citet{1998ApJ...500..525S} and converted to the AB photometric system. The RGB composite was created by implementing the \citet{2004PASP..116..133L} algorithm.}
\end{minipage}
\end{figure*}

\clearpage
\begin{figure*}
\begin{minipage}{150mm}
\includegraphics[width=15cm,height=18cm]{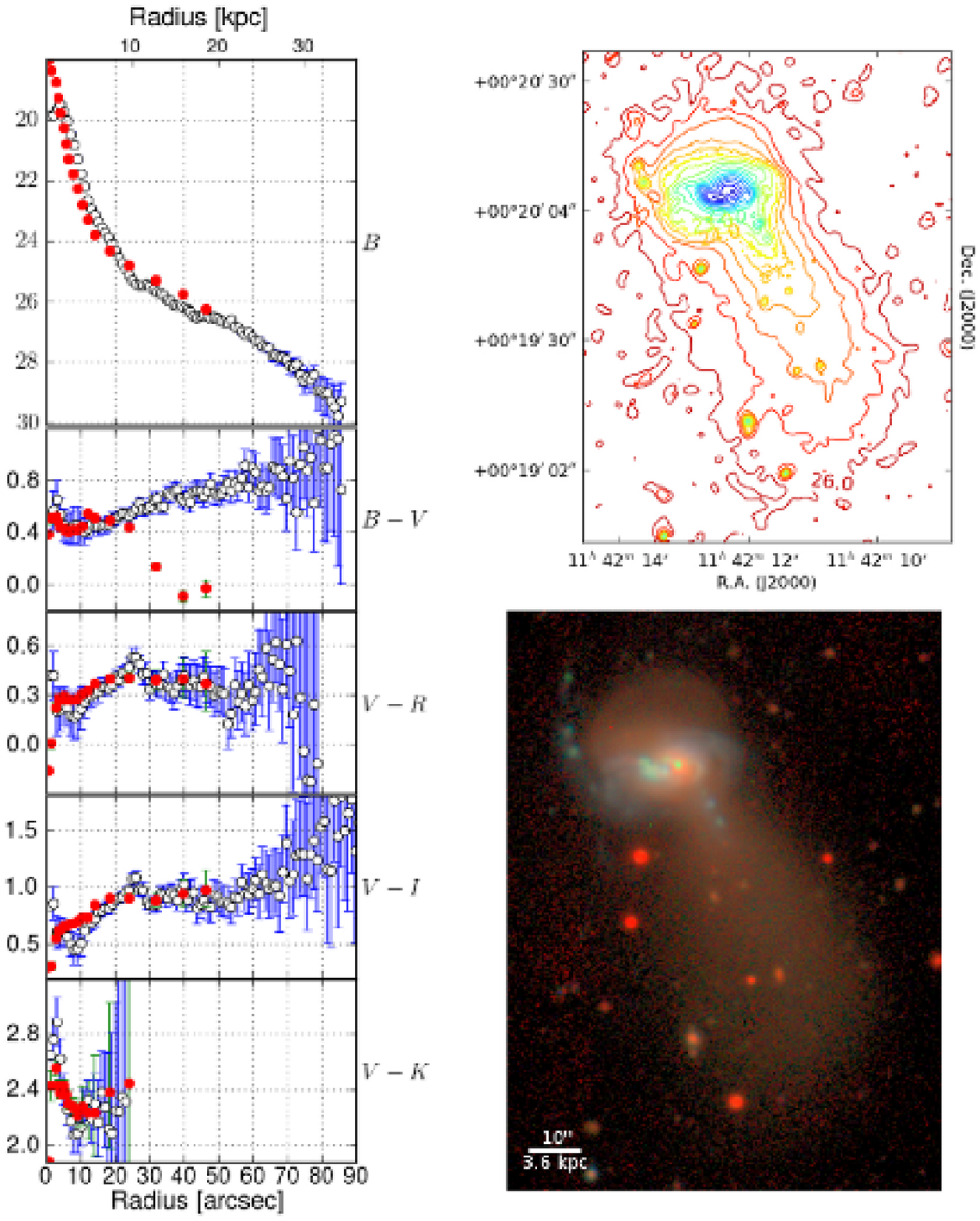}
\contcaption{\textbf{UM448}. \textit{Left panel}: Surface brightness and color radial profiles for elliptical (open circles) and isophotal (red circles) integration. \textit{Upper right panel}: contour plot based on the $B$ band. Isophotes fainter than $23.9$, $25.5$ are iteratively smoothed with a boxcar median filter of size $5$, $15$ pixels respectively. \textit{Lower right panel}: A true color RGB composite image using the $B,V,I$ filters. Each channel has been corrected for Galactic extinction following \citet{1998ApJ...500..525S} and converted to the AB photometric system. The RGB composite was created by implementing the \citet{2004PASP..116..133L} algorithm.}
\end{minipage}
\end{figure*}

\clearpage
\begin{figure*}
\begin{minipage}{150mm}
\includegraphics[width=15cm,height=18cm]{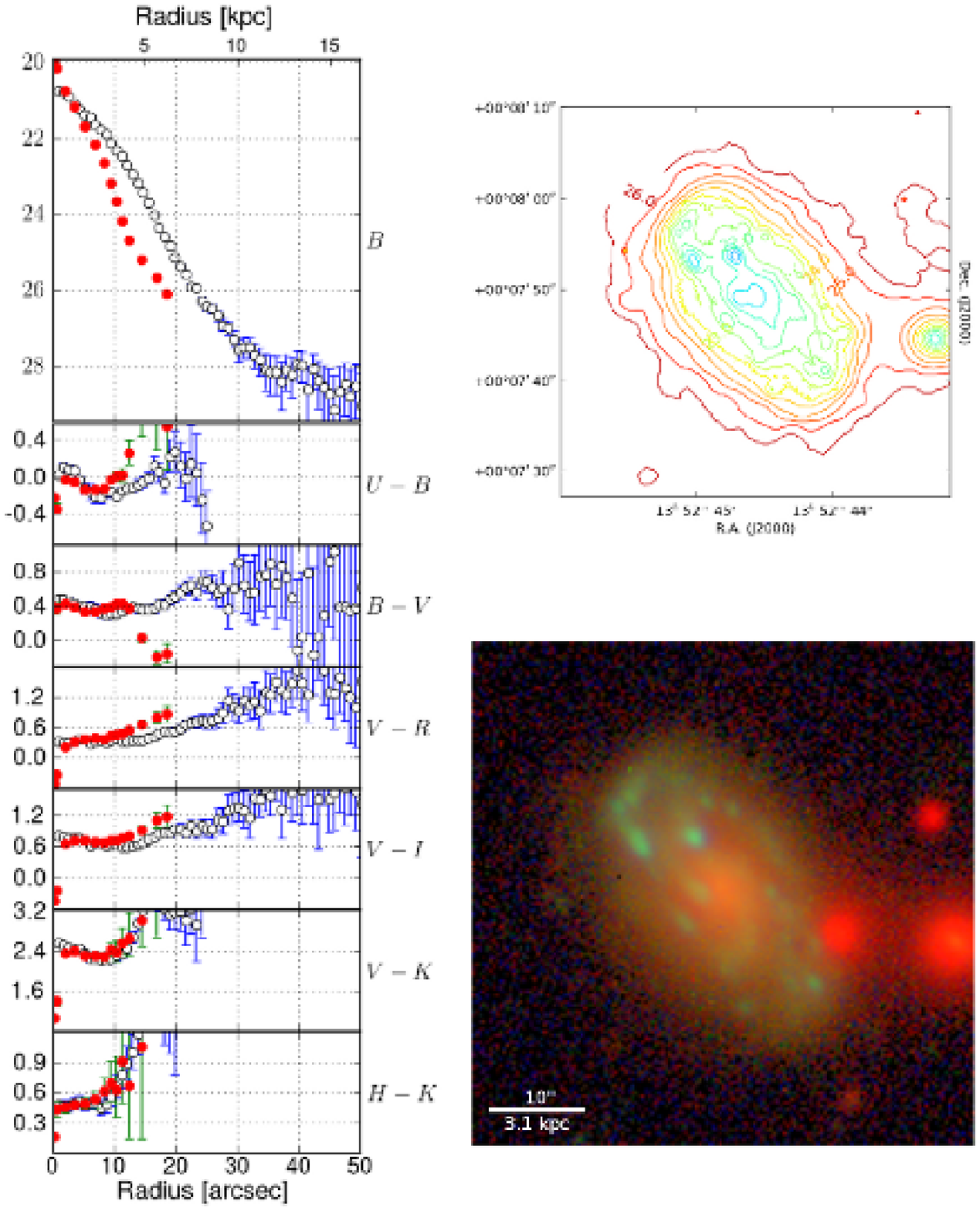}
\contcaption{\textbf{UM619}. \textit{Left panel}: Surface brightness and color radial profiles for elliptical (open circles) and isophotal (red circles) integration. \textit{Upper right panel}: contour plot based on the $B$ band. Isophotes fainter than $23.5$, $25.5$ are iteratively smoothed with a boxcar median filter of size $5$, $15$ pixels respectively. \textit{Lower right panel}: A true color RGB composite image using the $U,B,I$ filters. Each channel has been corrected for Galactic extinction following \citet{1998ApJ...500..525S} and converted to the AB photometric system. The RGB composite was created by implementing the \citet{2004PASP..116..133L} algorithm.}
\end{minipage}
\end{figure*}
\end{center}

\clearpage

\subsection{Data reductions}
\noindent In order to ensure the data consistently undergo the same processing, we have reduced all raw frames, except BVR data for \textit{ESO338--04}, with two pipelines specifically designed for this project - one for the optical and one for the NIR. This consistency ensures us that any difference in the results can not be attributed to the reduction process itself and instead reflects a true variation in properties between different galaxies. Both pipelines were already described in detail in~\citet{2010MNRAS.405.1203M}, so here we will only present additional points of interest relevant to the reduction. \\

\noindent The reductions were carried out on subsamples grouped by night and instrument. For every night and all instruments an individual sky background was fit to each reduced (i.e. bias--subtracted and flatfielded) frame \textit{before} combining them to a final stack, and at no point in the reduction process was a ``master sky'' ever used. After stacking, a final sky fitting and subtraction was performed. Calibration in both wavelength regimes is automatic, using Landolt standard stars in the optical and Two Micron All Sky Survey (2MASS) stars in the NIR.\\

\noindent \textbf{ALFOSC}~~~Some nights on this instrument were lacking usable standard stars. However, for all such observations short $3$ minute exposures were taken during photometric nights, or the same target was observed on several occasions. It was therefore always possible to calibrate problematic nights against photometric observations of the same object.\\

\noindent For $2004$ data there was a severe charge bleed in the frames of UM439, which the observers attempted to fix by varying the rotator field angle between consecutive frames of that target. This varying orientation of the images proved to be too much for the IRAF XYXYMATCH procedure, which the pipeline uses on SExtractor sources in order to align the frames. We chose to rotate all frames to the same orientation before reducing them, so there is one extra interpolation involved in the processing of these frames, as the angles were such that a simple flipping of the images in $x$, $y$ or both was not sufficient.\\

\noindent \emph{I} band frames taken with ALFOSC have pronounced fringes -- interference patterns due to the \textit{OH} emissions coming from the upper ionosphere. If the fringes are stable throughout the night then it would be sufficient to obtain a fringe map by stacking non-aligned dithered \emph{I} band frames and then subtract this from each individual image during reductions. This technique did not always remove the fringe pattern to our satisfaction, and residual ``hills'' and ``valleys'' could be clearly seen in the de-fringed images upon examination. Therefore we executed the pipeline in ``optical pair-subtraction mode'', using either dither or chopping mode pair-subtraction, as the data allowed. Since the fringes usually remain stable on such short time scales ($\sim10$ minutes from the start of one exposure to the next), this always produces clearly superior results compared to the average fringe map technique, and no fringe residuals can be seen on the de-fringed image even after smoothing it with a boxcar average (the smoothing is done solely for the purpose of inspection). \\

\noindent \textbf{EMMI}~~~There was a noticeable amplifier glow at varying positions along the edges of both CCDs in most \emph{I} band frames of the $2005$ data, which also seemed to change in time in position and intensity. If left uncorrected, this produces a significant non-flatness of the background in the stacked frame, which is impossible to remove with sky subtraction of a fitted plane. This issue is further enhanced by the fact that all four amplifiers were used during readout, each amplifier giving rise to its own varying amplifier glow. In such cases the pipeline was executed in ``optical pair-subtraction mode'', using either dither or chopping pair-subtraction, which successfully removed all amplifier glow.\\

\noindent \textbf{NOTCAM}~~~Approximately $8000$ of the raw NOTCAM images (years $2002$, $2004$ and $2005$) were lacking a world coordinate system (WCS) in the header. As the NIR pipeline relies on the WCS not only to align but also to identify 2MASS stars for calibration, we had to find a way to introduce a WCS for every individual raw frame. Obtaining a full plate solution in any kind of semi-interactive fashion for this many images is not practical. Instead we used the publicly available \textit{astrometry.net} software~\citep{2010AJ....139.1782L} to add the appropriate WCS to the headers. For most images the solver executed successfully using the standard indices built on the USNO-B catalog, but for a number of stubborn frames we successfully used 2MASS indices instead. Since \textit{astrometry.net} allows the user to search for solutions using already solved nearby fields, we were able to solve any remaining frames using local indices (e.g. an entire dither sequence of images of the same object can usually be solved as long as just one of those frames solves from USNO-B or 2MASS indices).\\

\noindent The Wide--Field camera of NOTCAM suffers from an optical distortion which becomes severe towards the edges of the array, and can shift the projected location of sources by several arcseconds. We have applied a correction for this using the available distortion map on the NOTCAM website. This map is from the year $2005$ and it worked well for our $2002$, $2004$, $2005$ and $2006$ NOTCAM data. We were unable to successfully remove all distortion for $2007$ data with the same distortion map. For that data we chose to instead align the images using only the central $160\times160$ arcsec ($\sim700\times700$ pixels) around the galaxy. Since all targets obtained in $2007$ have a fairly small apparent size, the lingering distortion along the edges is of no consequence for our analysis.\\

\noindent \textbf{SOFI}~~~There were no significant problems during the reductions of these data. All necessary calibration frames were of good quality, and a WCS was present in all frames.\\

\noindent \textbf{FORS}~~~The FORS exposures were taken for the purpose of re-calibrating existing observations, or to fill in for a number of filters that were missing in our table of observations. There were no significant problems during the reductions.\\
\begin{figure}
  \includegraphics[width=8cm,height=16cm]{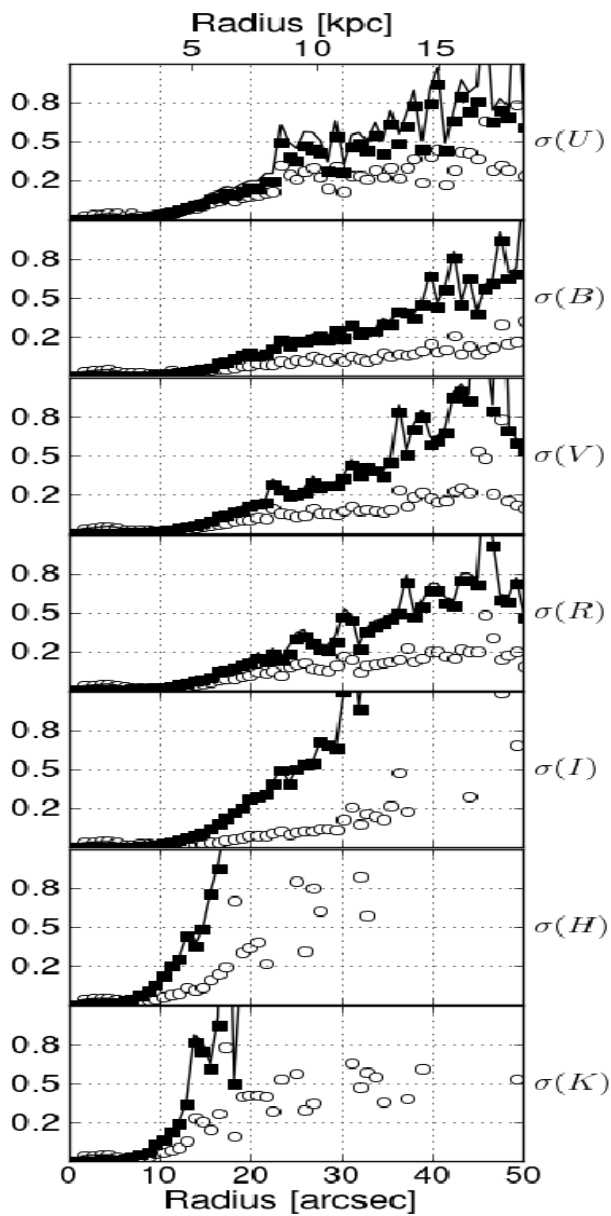}
\caption{A typical radial error composition (here for ESO185--13). The standard deviation of the mean inside a ring ($\sigma_{\textrm{sdom}}$, open circles) dominates the error budget in the central regions of the galaxy. As the signal to noise ratio decreases with increasing radius the uncertainty in the sky ($\sigma_{\textrm{sky}}$, filled squares) becomes the dominant error source. The composite error (solid line) is included in the errorbars of all surface brightness and color profiles.}\protect\label{errorplot}
\end{figure}
\subsection{Photometric calibration}\protect\label{photometry}
\noindent All data were calibrated in the Vega photometric system. \citet[][Northern]{1998AJ....115.2594H} and~\citet[][Southern]{1998AJ....116.2475P} photometric standards were obtained for the NIR regime. However, we decided early on that the pipelines should calibrate the NIR data using secondary 2MASS standards on each raw frame. This way the photometric accuracy of the stacked NIR frames is nearly independent of the actual photometric conditions during the observations. For some targets the field of view contained too few visible stars, and calibration of each individual frame before stacking was then impossible. In these cases a single zero point per galaxy per night was obtained and applied to the reduced stacked images since those always have more sources than the single exposure frames.\\

\noindent In the optical the data were calibrated against images of Landolt standard stars obtained during each observing night. Occasionally we ran into problems when calibrating. Some nights were clear but non-photometric, others allowed for observations only in the beginning and the end of the night, resulting in an unacceptably long time interval between galaxy and standard star observations, and yet others had too few or no standard stars available because of saturation or bad pixels. Data from nights with bad atmospheric conditions were re-calibrated using short $1$ or $3$ minute exposures, taken during photometric nights with FORS1 (VLT) or ALFOSC (NOT) in all cases where the zero point reliability was questionable.\\

\noindent As we aim to reach very faint surface brightness levels, we felt it necessary to perform a sanity check on our calibration. We therefore compared the photometry of the stars in our frames to that of values taken from the Pickles stellar library in various color-color plots, both in the optical and NIR. This comparison showed no significant offset between the two,  and we are therefore confident that our photometry is reliable\protect\footnote{\footnotesize These plots are available on demand. Email GM.}.\\

\noindent We estimated the zero point uncertainty for the optical data using the scatter of measured zero point values for each night. For the NIR we instead measured the mean offset in the magnitudes of stars on our frames when compared to 2MASS photometry. This uncertainty is represented by $\sigma_{zp}$ (\S~\ref{errorest}).\\
\begin{figure}
\begin{center}
  \includegraphics[width=8cm,height=6cm]{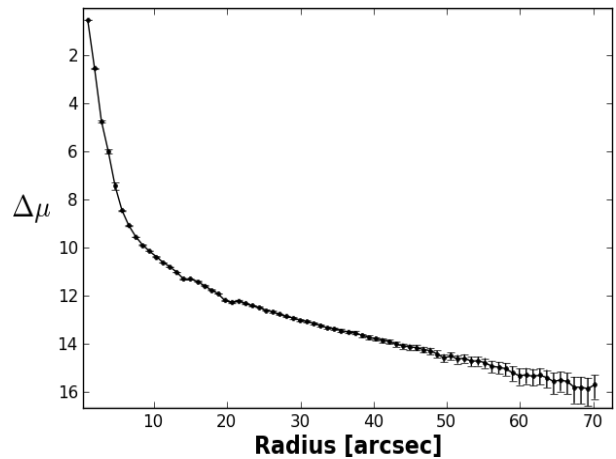}
\caption{ALFOSC PSF obtained from stars in our frames. The PSF is the average of three stars, where one saturated star was used to obtain the far wings of the profile beyond $\sim12$ arcsec. The individual star profiles were normalized to the total flux inside a $5$ arcsec radius before averaging.}.\protect\label{psffig}
\end{center}
\end{figure}
\section{Surface brightness profiles}\protect\label{sbsection}
\noindent BCGs often have irregular morphologies and it is therefore not always straightforward to decide on how best to obtain a radial distribution of the surface brightness. If the galaxy looks regular then ellipses will trace the isophotes quite well. If that is not the case then there are several options, developed specifically through the consideration of highly irregular starbursting galaxies, for how to approach that~\citep[e.g.][]{1996A&AS..120..207P,2001ApJS..133..321C,2002A&A...393..461P}. The usual assumption is that even if the starburst component is quite irregular, the underlying old stellar population is well-approximated by ellipses. We found, however, that if one goes deep enough, even that assumption is not always valid. We therefore decided to obtain the surface brightness profiles in two distinct, non-overlapping ways.\\

\noindent First, we looked at surface brightness as a function of radius, i.e. we performed the integration in concentric elliptical rings, where each new data point is obtained by stepping in radius along the semi-major axis. The geometric parameters for the ellipse were obtained by adopting the shape of the ellipse best fitting the $25{}^{th}$ isophote as given by IRAF ELLIPSE. The parameters (ellipticity, position angle, and (\textit{x},\textit{y}) ellipse center) were kept constant at each integration step. The position angle is measured counterclockwise from North to East. The best fitting ellipse was obtained for each target in a reference bandpass and then applied to the rest of the filters of that target. This way the physical area we integrate over at each radial step is identical in each filter. The reference bandpass is always taken to be the bluest available filter after the $U$ band, which usually means the $B$ band, or $V$ if $B$ is missing. We do not use the $U$ band as reference on purpose since the atmosphere has poor transmission at those wavelengths and as a rule the U band is never deep enough. The $B$ band on the other hand has long exposure times and a background which seems to be more stable than any other band at our disposal, which makes it a natural choice as a reference. Throughout this paper a step size of $1$ arcsec was adopted for every bandpass and every target. This stepsize was chosen because $1$ arcsec is on the order of the average seeing for all observations, which is $\sim1.1$ arcsec. The $B$ band ESO338--04 and HE2--10 observations are an exception with $\sim1.7$ arcsec seeing. Note that due to the different pixel scales between galaxies sometimes a $1$ arcsec step would result in a non-integer number of pixels, which is rounded down to an integer and hence the actual step size will be slightly less than an arcsecond. More precisely, data with a pixel scale of $0.288$ arcsec have an actual step size of $0.86$ arcsec. For each galaxy the images in all filters were resampled to a common (NIR) pixel scale.\\

\noindent We further obtained isophotal radial profiles of the galaxies. These are interesting to compare to the standard elliptical profiles because they are essentially independent of the location of the star--forming knots and the morphology of the galaxy at any surface brightness level. There can be no ``bumps'' or ``plateau'' structures in an isophotal profile since at each step it samples the target in a manner which always produces a monotonously declining brightness distribution. In contrast to the elliptical integration profiles, an isophotal profile shows radius as a function of surface brightness. One always starts at the brightest pixel, and then steps to fainter isophote levels in magnitude bins. At each bin the number of pixels with surface brightness inside the current magnitude range is identified. This area may have any shape and does not necessarily need to be circular, elliptical or even connected. One then sets the area equal to an equivalent circular area and thus obtains an equivalent radius for each magnitude bin. At fainter isophotes this procedure breaks down as one starts to run out of available pixels and the pixels actually included in a magnitude bin start to become dominated by the positive outliers of the background noise. At this point it is common practice to instead approximate the fainter isophotes with ellipses and present a composite isophotal/elliptical profile. We have chosen not to do this here at any point along the isophotal profile, firstly because we already present standard profiles with elliptical integration and so an isophotal/elliptical composite integration profile will not give us any new information at large radii, and secondly because even at the faintest isophotal levels we have targets that are very poorly approximated by ellipses, e.g. UM160 and IIZw40 in this sample and a number of galaxies in~\citet{Paper2}. Instead, we chose to truncate the profiles at brighter isophotes where we are still confident that they are reliable. A constant magnitude bin of $0.5$ mag was chosen for all galaxies and all filters. As for the elliptical profiles each isophotal area within the bins was defined using the reference filter (usually the $B$ band) and then applied to the rest of the filters. This means that the physical area included in each bin is identical across all filters. \\

\noindent Radial color profiles were obtained from both the elliptical integration and isophotal surface brightness profiles. The profiles are truncated when the composite error becomes too large, however that limit varies from color to color. Figure~\ref{datafig} presents the elliptical and isophotal surface brightness and color profiles for each galaxy. The profiles are reliable down to $\mu_K\lesssim23$ mag arcsec${}^{-2}$ in the NIR, and $\mu_B\lesssim28$ mag arcsec${}^{-2}$ in the optical. Judging by the behavior of the corresponding profile and its errors, these limits seem to fit most galaxies except ESO400--43 ($\mu_V\lesssim27$ mag arcsec${}^{-2}$), ESO462--IG 020 ($\mu_B\lesssim26.5$ mag arcsec${}^{-2}$), IIZw40 ($\mu_B\lesssim25.5$ mag arcsec${}^{-2}$), UM133 ($\mu_K\lesssim22.5$ mag arcsec${}^{-2}$), and UM417 ($\mu_K\lesssim22$ mag arcsec${}^{-2}$). On the other hand for several galaxies the profiles seem to reach much deeper than our imposed limit, with $\mu_K\lesssim24$ mag arcsec${}^{-2}$ and $\mu_B\lesssim29$ mag arcsec${}^{-2}$. We will further analyze the reliability of the data in \S~\ref{reliability}. Galactic extinction correction following~\citet{1998ApJ...500..525S} has been applied to all filters, however, internal extinction has not been taken into account. Further, none of the profiles have been inclination--corrected because we found it impossible to define an unambiguous disk for too many of the galaxies in the sample. \\

\noindent For every galaxy and every filter we examined the flatness of the residual background by smoothing the images with a $3\times3$ boxcar average. This was done solely for the purpose of looking for residual large scale structures, while unsmoothed images were used for the derivation of the surface brightness profiles.\\

\noindent Before obtaining elliptical and isophotal profiles all contaminating foreground and background objects were masked out. The masking process uses SExtractor output as an indication of each object's shape and size, and then increases each reported size by a factor of $2.5$. While this works in most cases in fully automatic mode, we visually examined the sizes and shapes of masks for each individual frame, and adjusted them where necessary. 
\subsection{Point spread function}\protect\label{psf}
\noindent To test whether the point spread function (PSF) of the instruments can affect the light distribution of a galaxy at faint levels, we obtained the PSF for ALFOSC, presented in Figure~\ref{psffig}. The radial light distribution of three stars, two faint and one very bright and saturated, were obtained from a random frame. Each profile was normalized by the flux sum inside a five arcsec radius, which scales the profiles down to the same magnitude levels. The average of the three profiles was taken where they overlap, which was out to a radius of $\sim12$ arcsec. The saturated star was naturally not used for the central region of the PSF. Beyond $\sim12$ arcsec the wings are based solely on the one saturated star. We then scaled the PSF up to the corresponding luminosity for a few random galaxies (observed with ALFOSC) to see if the instrument PSF has any influence on the surface brightness profiles at large radii. Our test showed that it does not. The PSF drops in brightness much faster than our galaxies, therefore any light we detect at faint isophotal levels is not a PSF effect.
\subsection{Error estimation}\protect\label{errorest}
\noindent A very detailed description of our error estimation is presented in~\citet{2010MNRAS.405.1203M}, however, here we will briefly outline how the errors are calculated. At each data point in the elliptical integration profiles we have accounted for 3 sources of uncertainty: $\sigma_{zp}$ -- the (constant) zero-point uncertainty, $\sigma_{sky}$ -- the uncertainty in the flatness of the sky, and $\sigma_{sdom}$ -- the uncertainty in the mean intensity level of each elliptical ring, where ``sdom'' stands for standard deviation of the mean. As mentioned in \S~\ref{photometry}, the zero-point uncertainty is the mean deviation in the magnitude of all stars in our frames from their corresponding tabulated values in the Sloan Digital Sky Survey (SDSS) and 2MASS catalogs. If the galaxy is not found in the SDSS, then this uncertainty is instead the standard deviation of the photometric zero points obtained during the corresponding observing night. The sky uncertainty is obtained by calculating the mean intensity inside of square apertures, placed on ``empty'' sky regions in the frame. The standard deviation of these mean intensities then estimates the departure from the flat zero sky and is assumed to be constant in each pixel. The size of the apertures does matter and we refer the interested reader to~\citet{2010MNRAS.405.1203M}. Since we cannot measure where the sky level should truly lie, presenting a $\sigma_{sky}$ which estimates the precision of the \emph{assumed} sky level is meaningless and misleading. Therefore, our $\sigma_{sky}$ is designed to only be a measure of the flatness of the sky and also directly carries information on when the assumed sky would start influencing the signal from a target. However, this should not be interpreted as an estimate of the \emph{accuracy} of the sky level, as it carries no information on whether or not the sky level in our images is true or not. The final source of uncertainty is in the accuracy with which we can measure the mean intensity inside each elliptical ring. This in itself is a composite error of the Poisson noise and the statistical fluctuations in flux across a ring and is given by the standard deviation of the mean (sdom) of each ring. It is a measure of the precision with which the mean intensity is known. This is not to be confused with the standard deviation inside each elliptical ring which is a measure of the scatter in flux -- a physical property, not an error source. An example radial error composition for elliptical integration profiles is shown in Figure~\ref{errorplot}. The sky uncertainty clearly dominates the error budget at faint intensity levels.\\

\noindent In an isophotal profile only the zero point and the sky background uncertainties are accounted for. This is because by design here the standard deviation of the mean at each radius will always be small and pretty much constant -- pixels that fall inside the same magnitude bin all have very similar intensities so any small scatter they might have would not be a measure of anything useful.

\subsection{Systematic errors}\protect\label{reliability}
\noindent In~\citet{2010MNRAS.405.1203M} we performed a test on the reliability of the reduction pipeline using synthetic disk galaxies. Specifically, we investigated at what surface brightness levels the systematic errors in the sky subtraction start dominating the observed flux. Since the systematic errors are generally unknown, this test allowed us to get a handle on their effects. We obtained a K band limit of $\mu_K=23$ mag arcsec${}^{-2}$ and $\mu_V=28.5$ mag arcsec${}^{-2}$, beyond which the surface brightness profile would show a systematic deviation from the synthetic exponential profile. We use the same reduction pipeline as ~\citet{2010MNRAS.405.1203M} but these limits cannot be simply carried over to other data, as they are dependent on the depth of each observation, the individual accuracy of the flatfielding, the typical behavior of the sky, etc. Regardless of the individual errorbars of each profile point if the data values appear to decrease/increase in an ordered fashion, then there is likely still useful signal measured at the corresponding radius. If one is instead sampling the sky, then the data values should reflect random statistical fluctuations. Therefore one could naively expect that any ordered behavior would indicate the continued presence of a signal from the target galaxy. We choose to test this explicitly for the $B$ and $K$ bands from the ALFOCS and NOTCAM instruments. Such a test already exists for the SOFI instrument~\citep{2010MNRAS.405.1203M}, while the dither pattern on EMMI data left no space around the primary galaxy for a synthetic disk to be inserted and visible in each frame. \\
\begin{figure}
\centering
\includegraphics[width=8cm,height=4.6cm]{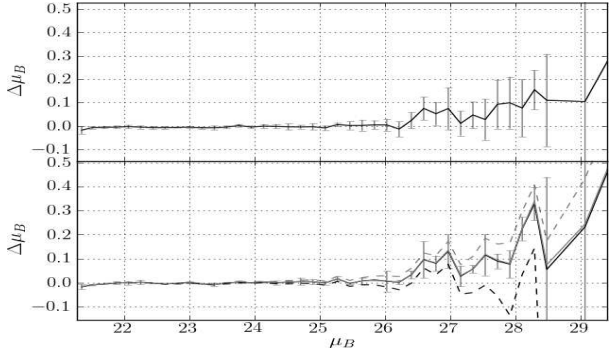}
\caption{\textbf{$B$ band estimation of systematic errors.} \emph{Upper panel:} The surface brightness difference between a noiseless analytic exponential disk and the profile of the latter after a run through the reduction pipeline. The profiles of six exponential disks, placed at different positions in the raw frames, were obtained and averaged, and then subtracted from the analytic disk profile to obtain $\Delta\mu_B$. \emph{Lower panel:} Same as above for three disk positions, each with four different simulated flatfielding errors of up to $2\%$ in amplitude. The errorbars are the standard deviation between the different profiles. }
\protect\label{fakediff}
\end{figure}
\begin{figure}
\centering
\includegraphics[width=8cm,height=4.6cm]{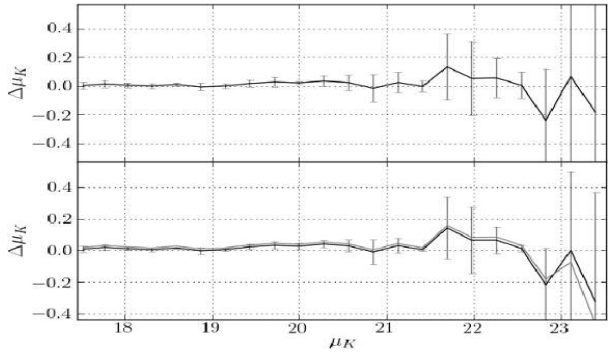}
\caption{\textbf{$K$ band estimation of systematic errors.} \emph{Upper panel:} Same idea as in Figure~\ref{fakediff} but for three exponential disks in the $K$ band. \emph{Lower panel:} Same as above for the three disk positions, but each with two different simulated flatfielding errors of up to $5\%$ in amplitude. The errorbar budget is the same as in Figure~\ref{fakediff}. }
\protect\label{fakediffNIR}
\end{figure}

\noindent In raw $B$ band frames we inserted synthetic disks at six different positions (not simultaneously), after having ``de-flattened'' the disks by multiplying with the corresponding flatfield before insertion. The reduction and calibration of these frames then proceeded as usual. We obtained a surface brightness profile for each disk position and compared it to the profile of the analytical exponential disk that was used. In all cases we recover the original properties of the disk -- scale length, central surface brightness, Holmberg radius, total luminosity down to the disk truncation -- to within $0.05-0.1$ units. In the top panel of Figure~\ref{fakediff} we show the magnitude difference between the original analytic exponential disk and the recovered surface brightness profile, where the latter is the average of the six different cases. The plot shows that there is no difference between the two down to $\mu_B\sim26$ mag arcsec${}^{-2}$, and a systematic brightening of the averaged recovered profile beyond this isophote, which is on average smaller than $\Delta\mu_B\sim0.1$ mag down to $\mu_B\sim28.5$ mag arcsec${}^{-2}$. Judging from the plot, the profile is probably still reliable even down to $\mu_B\sim29$ mag arcsec${}^{-2}$. To test the effect of enhanced flatfielding errors, for three of the positions we also simulated four cases of flatfielding errors of varying amplitude but with a maximum deviation of $2\%$, and then proceeded as described above to obtain the magnitude difference as a function of surface brightness level. This is presented in the lower panel of Figure~\ref{fakediff}. We reach $\mu_B\sim28.0$ mag arcsec${}^{-2}$ with the same excess of $0.1$ on average, and $\mu_B\sim29.0$ mag arcsec${}^{-2}$ with an average of $0.2$ mag offset. In light of these results, we are confident that we can trust the errorbars in our surface brightness profiles for the sample data, and we set our reliability limit in the $B$ band to a conservative $\mu_B\sim28.0$ mag arcsec${}^{-2}$.\\

\noindent For the $K$ band in a similar fashion we find negligible $\Delta\mu_K$ down to $\mu_K\sim22.5$ mag arcsec${}^{-2}$ (Figure~\ref{fakediffNIR}), and an offset of max $0.2$ mag down to $\mu_K\sim23.3$ mag arcsec${}^{-2}$.  Therefore we set the reliability limit in the $K$ band to $\mu_K\sim23.0$ mag arcsec${}^{-2}$.\\ 
\begin{center}
  \begin{table}
      \caption{Surface photometry parameters. The position angle ($PA^\circ$) in degrees and the ellipticity ($e$) were obtained from the $B$ band image and applied to the remaining filters. $PA^\circ$ is measured counterclockwise from North to East. The radius where the mean surface brightness is $\mu_B\approx26.5$ mag arcsec${}^{-2}$ is the Holmberg radius, $R_{H}$, and is given in arcseconds and kpc. The absolute $B$ magnitude, $M_B$, is calculated from the area inside $R_{H}$ and has been corrected for Galactic extinction~\citep{1998ApJ...500..525S}. The $M_B$ errors are identical to the errors of the apparent $B$ magnitude in Table~\ref{totlumtbl}.}
      \protect\label{pa_holm_tbl}
      \begin{tabular}{@{}|l|r|r|r|r|r|@{}}
        \hline
        Galaxy&$PA^\circ$&$e$&$R^{''}_{H}$&$R^{kpc}_{H}$&$M_B$\\\hline
        ESO185--13&$46$&$0.18$&$22.5$&$8.3$&$-19.48$\\
        ESO249--31&$45$&$0.39$&$129.6$&$6.8$&$-17.99$\\
        ESO338--04&$76$&$0.45$&$43.8$&$7.9$&$-19.31$\\
        ESO400--43$^\dagger$&$-15$&$0.22$&$23.9$&$9.2$&$-20.15$\\
        ESO421--02&$-89$&$0.26$&$56.2$&$3.4$&$-16.53$\\
        ESO462--20&$-86$&$0.40$&$27.6$&$10.6$&$-20.03$\\
        HE2--10&$28$&$0.15$&$96.8$&$7.4$&$-19.07$\\
        HL293B&$-2$&$0.12$&$11.2$&$0.9$&$-13.71$\\
        IIZW40&$-71$&$0.19$&$109.7$&$6.3$&$-18.68$\\
        MK600&$-51$&$0.48$&$44.9$&$2.4$&$-15.12$\\
        MK900&$41$&$0.27$&$49.6$&$2.7$&$-15.99$\\
        MK930&$41$&$0.19$&$19.7$&$6.8$&$-19.23$\\
        MK996&$20$&$0.29$&$27.6$&$2.4$&$-16.25$\\
        SBS0335--052E&$-61$&$0.11$&$11.2$&$2.9$&$-16.89$\\
        SBS0335--052W&$-50$&$0.22$&$8.6$&$2.3$&$-14.53$\\
        TOL0341--407&$-86$&$0.40$&$19.0$&$5.6$&$-17.50$\\
        TOL1457--262~\emph{I}&$-21$&$0.54$&$32.8$&$11.6$&$-20.02$\\
        TOL1457--262~\emph{II}&$85$&$0.74$&$56.2$&$19.8$&$-19.54$\\
        UM133&$16$&$0.40$&$37.4$&$3.3$&$-15.81$\\
        UM160&$80$&$0.14$&$48.4$&$6.6$&$-17.76$\\
        UM238&$-78$&$0.51$&$22.5$&$5.9$&$-16.76$\\
        UM417&$-17$&$0.40$&$14.0$&$2.3$&$-14.50$\\
        UM448&$11$&$0.24$&$43.8$&$17.5$&$-20.42$\\
        UM619&$45$&$0.46$&$26.8$&$8.9$&$-18.40$\\
        \hline
      \end{tabular}
      \medskip
      ~\\
      $\dagger$ -- $V$ instead of $B$ filter
  \end{table}
\end{center}
\begin{center}
  \begin{table*}
    \begin{minipage}{170mm}
      \caption{Integrated surface photometry for the sample. The integration is carried out down to the Holmberg radius $R^{''}_{H}$, which is defined from $\mu_B$ for each target and then applied to the remaining filters. All values have been corrected for Galactic extinction~\citep{1998ApJ...500..525S}.}
      \protect\label{totlumtbl}
      \begin{tabular}{@{}|l|l|r|r|r|r|r|r|r|@{}}
        \hline
        Galaxy&$B$&$U-B$&$B-V$&$V-R$&$V-I$&$V-K$&$H-K$\\\hline
        ESO185--13&$14.94 \pm 0.05$&$-0.17 \pm 0.16$&$0.26 \pm 0.05$&$0.19 \pm 0.03$&$0.49 \pm 0.03$&$1.89 \pm 0.06$&$0.19 \pm 0.06$\\
        ESO249--31&$12.18 \pm 0.04$&&$0.50 \pm 0.05$&$0.31 \pm 0.03$&$0.66 \pm 0.03$&$2.12 \pm 0.04$&$0.15 \pm 0.04$\\
        ESO338--04&$13.55 \pm 0.04$&&$0.34 \pm 0.05$&$0.07 \pm 0.05$&$-0.22 \pm 0.06$&$1.04 \pm 0.06$&$0.11 \pm 0.05$\\
        ESO400--43&$14.34\pm 0.19^\dagger$&&&&$0.38 \pm 0.20$&$1.83 \pm 0.20$&$0.25 \pm 0.05$\\
        ESO421--02&$13.94 \pm 0.04$&$-0.52 \pm 0.15$&$-0.15 \pm 0.05$&$0.63 \pm 0.03$&$0.97 \pm 0.03$&$2.66 \pm 0.07$&$0.19 \pm 0.07$\\
        ESO462--20&$14.47 \pm 0.04$&$-0.16 \pm 0.16$&$0.29 \pm 0.05$&&&$2.03 \pm 0.19$&$0.44 \pm 0.20$\\
        HE2--10&$11.93 \pm 0.04$&$0.11 \pm 0.14$&$0.48 \pm 0.05$&&&$2.78 \pm 0.06$&$0.45 \pm 0.08$\\
        HL293B&$17.35 \pm 0.05$&$-0.50 \pm 0.16$&$0.48 \pm 0.07$&$0.11 \pm 0.07$&$0.22 \pm 0.09$&$1.60 \pm 0.08$&$0.19 \pm 0.15$\\
        IIZW40&$11.68 \pm 0.04$&$-0.94 \pm 0.18$&$0.22 \pm 0.06$&&$-0.36 \pm 0.09$&$0.97 \pm 0.07$&$1.25 \pm 0.09$\\
        MK600&$15.07 \pm 0.06$&$-0.31 \pm 0.16$&$0.37 \pm 0.07$&$0.20 \pm 0.07$&$0.41 \pm 0.10$&$1.38 \pm 0.05$&$-0.10 \pm 0.11$\\
        MK900&$14.25 \pm 0.04$&$-0.20 \pm 0.15$&$0.58 \pm 0.06$&$0.44 \pm 0.07$&$0.81 \pm 0.09$&$2.34 \pm 0.10$&$0.20 \pm 0.10$\\
        MK930&$15.04 \pm 0.05$&$-0.77 \pm 0.16$&$0.41 \pm 0.06$&$0.07 \pm 0.07$&$0.17 \pm 0.09$&$1.86 \pm 0.09$&$0.42 \pm 0.09$\\
        MK996&$15.05 \pm 0.02$&$-0.30 \pm 0.15$&$0.57 \pm 0.05$&$0.42 \pm 0.06$&$0.78 \pm 0.09$&$2.47 \pm 0.05$&$0.28 \pm 0.09$\\
        SBS0335--052~E&$16.77 \pm 0.05$&&&&$-0.02\pm 0.10^\ddagger$&$1.19\pm 0.05^\ddagger$&$0.72 \pm 0.07$\\
        SBS0335--052~W&$19.14 \pm 0.05$&&&&$0.29\pm 0.09^\ddagger$&$1.62\pm 0.44^\ddagger$&$0.22 \pm 0.84$\\
        TOL0341--407&$16.41 \pm 0.04$&$-0.51 \pm 0.18$&$0.66 \pm 0.05$&&$0.20 \pm 0.50$&$1.63 \pm 0.06$&$0.17 \pm 0.07$\\
        TOL1457--262~\emph{I}&$14.29 \pm 0.05$&&$0.28 \pm 0.06$&&&$1.61 \pm 0.07$&\\
        TOL1457--262~\emph{II}&$14.78 \pm 0.20$&&$0.19 \pm 0.29$&&&$2.25 \pm 0.21$&\\
        UM133&$15.50 \pm 0.05$&$-0.43 \pm 0.16$&$0.34 \pm 0.06$&$0.30 \pm 0.07$&$0.58 \pm 0.10$&$2.14 \pm 0.09$&$0.34 \pm 0.10$\\
        UM160&$14.47 \pm 0.02$&$-0.95 \pm 0.06$&$0.42 \pm 0.04$&$0.16 \pm 0.05$&$0.47 \pm 0.05$&$1.83 \pm 0.16$&$0.13 \pm 0.16$\\
        UM238&$16.91 \pm 0.05$&$-0.33 \pm 0.16$&$0.40 \pm 0.07$&$0.30 \pm 0.07$&$0.62 \pm 0.10$&$2.12 \pm 0.16$&$0.57 \pm 0.19$\\
        UM417&$18.15 \pm 0.02$&$-0.44 \pm 0.04$&$0.58 \pm 0.04$&$0.13 \pm 0.04$&$0.22 \pm 0.06$&$1.82 \pm 0.45$&$0.98 \pm 0.48$\\
        UM448&$14.16 \pm 0.02$&&$0.49 \pm 0.04$&$0.26 \pm 0.06$&$0.64 \pm 0.05$&$2.37 \pm 0.12$&\\
        UM619&$15.78 \pm 0.03$&$-0.10 \pm 0.09$&$0.38 \pm 0.05$&$0.36 \pm 0.06$&$0.69 \pm 0.06$&$2.59 \pm 0.20$&$0.81 \pm 0.24$\\
        \hline
      \end{tabular}
      ~\\
      $\dagger$ -- $V$ instead of $B$ filter\\
      $\ddagger$ -- $B$ instead of $V$ filter
    \end{minipage}
  \end{table*}
\end{center}
\section[]{Contour plots}\protect\label{contourplots}
\noindent The isophotal contour plots for each galaxy were obtained from the deepest best image (the B band for the majority of the cases). The python \emph{astLib} package was used to resample the images to tangential projection so that North is up and East is to the left, and to obtain the axis labels in terms of RA and Dec, but the plotting of the contours was done with the built--in \emph{pylab} function \emph{'contour'} in bins of $0.5$ magnitudes. Beyond a certain surface brightness the isophotes become noisy as the magnitude bins start sampling the sky background. Therefore beyond this level, which is individually decided upon for each galaxy and ranges from $\mu_B\sim 22.0$ to $\mu_B\sim 27.5$ mag arcsec${}^{-2}$, the images were smoothed with a boxcar median filter of increasing size (e.g. $5\times5,~15\times15,~21\times21$, etc.) for each consecutive magnitude bin (e.g. $\mu_B=24$ -- $24.5$, $24.5$ -- $25$, $25$ -- $25.5$ mag arcsec${}^{-2}$, etc.). The upper right panel of Figure~\ref{datafig} shows B band contour plots for each galaxy. 
\section[]{RGB images}
\begin{figure}
\centering
\includegraphics[width=8cm,height=4.6cm]{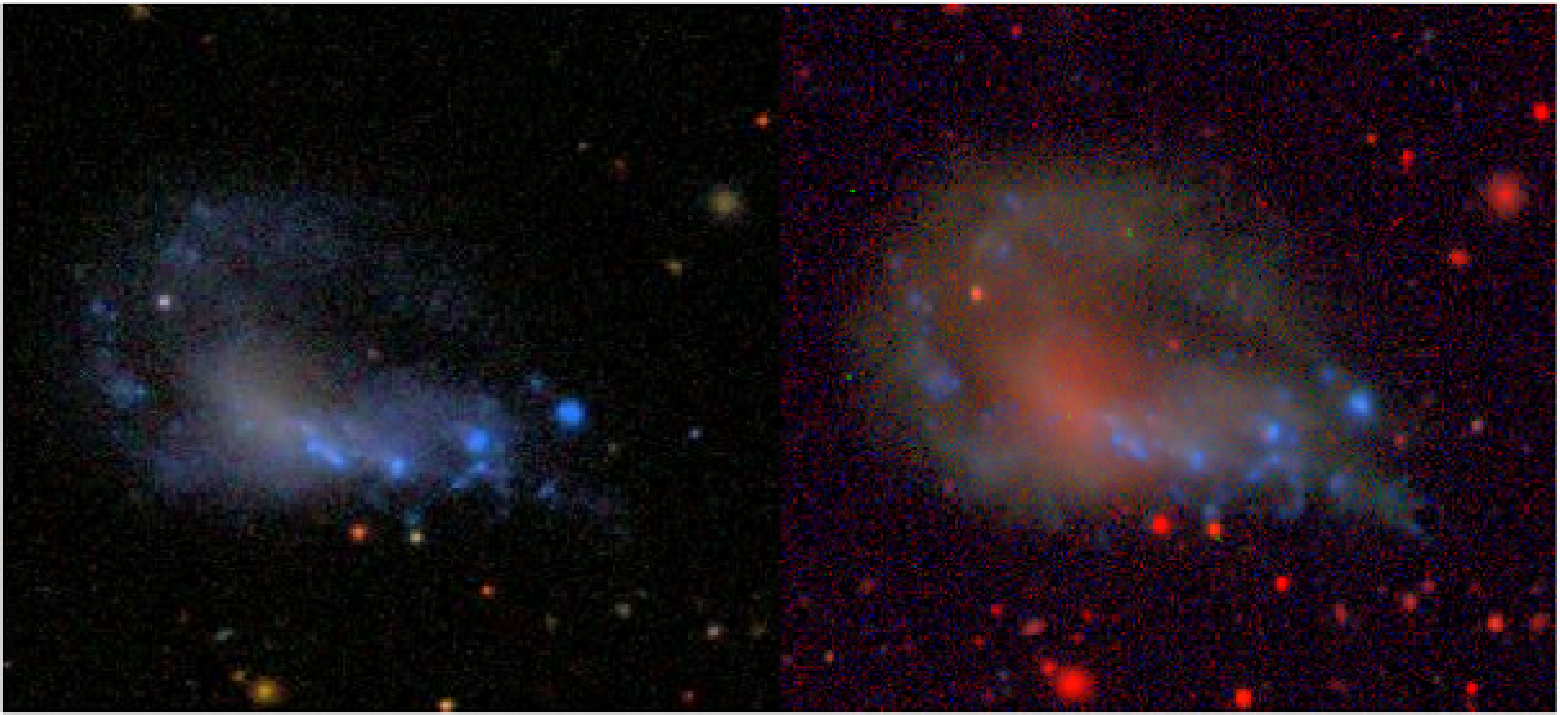}
\includegraphics[width=8cm,height=4.6cm]{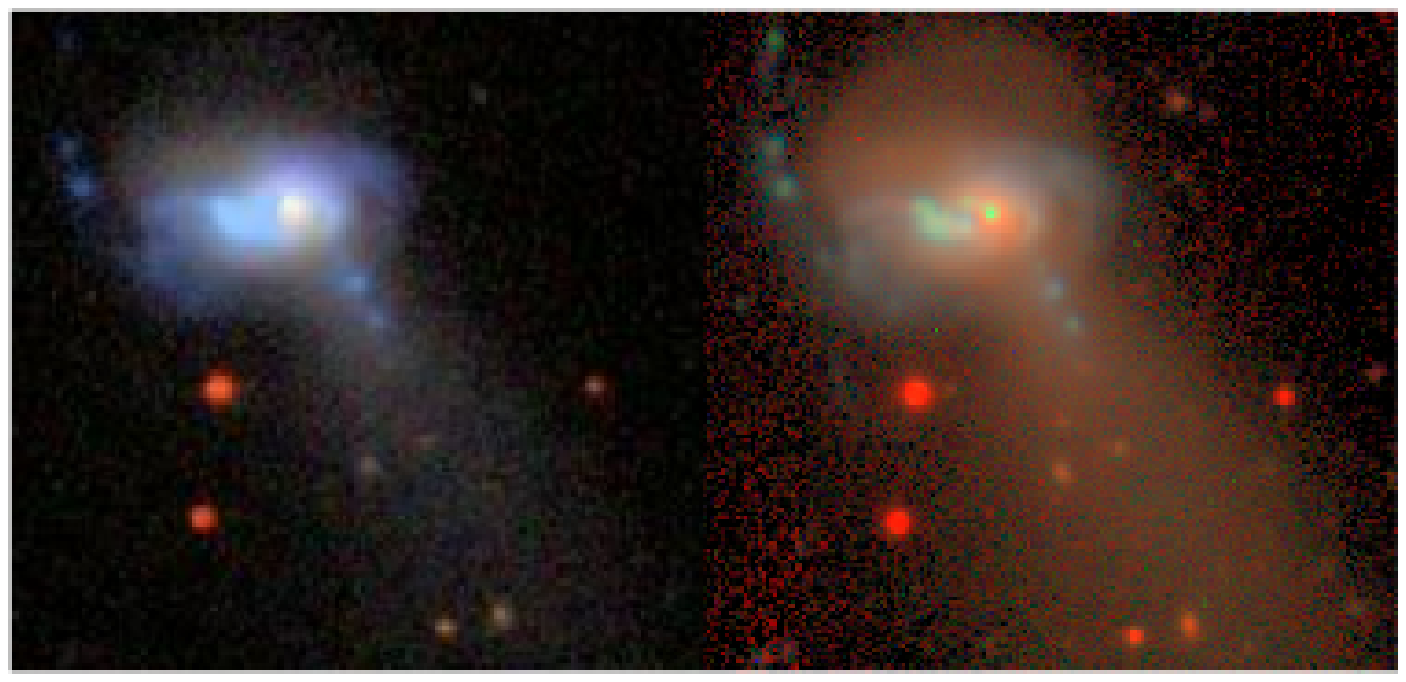}
\caption{\textbf{UM160} (top) and \textbf{UM448} (bottom) in SDSS (left) and our (right) RGB color schemes. }
\protect\label{sdssRGB}
\end{figure}

\protect\label{rgbimages}
\noindent RGB images give an invaluable overview of the morphology of a galaxy, and we provide such images for each galaxy with the three channels $(Blue, Green, Red)$ assigned to $(U,B,I)$ in the majority of the cases. We tested various available tools for scaling and combining the RGB channels, including \emph{GIMP}, \emph{FITS Liberator} with \emph{GIMP}, \emph{DS9}, \emph{Aladin}, and the python astronomical plotting package \emph{APLpy}. In most of these cases we find that it is very hard to select channel scaling factors that would universally give good results for all galaxies. In addition some of these algorithms regularly saturate the color of a channel in high--intensity areas of the image, thus obscuring all features. The best results were obtained with our own implementation of the~\citet{2004PASP..116..133L} algorithm. Only \emph{APLpy} gave results of comparable quality but required channel adjustments for individual galaxies. In order to facilitate direct comparison one must avoid having different channel scaling factors between galaxies, which the~\citet{2004PASP..116..133L} algorithm allows. The benefit of this algorithm is that the intensity of a channel may saturate in certain parts of the galaxy, but the color never does, thus preserving all interesting morphological features that some of the more irregular galaxies in our sample have. The RGB channels were corrected for Galactic extinction and converted to the AB magnitude system before scaling and combining. The lower right panel of Figure~\ref{datafig} shows the RGB image of each galaxy. To adjust the readers eyes to our color-scheme, in Figure~\ref{sdssRGB} we show an example of SDSS multicolor images and our own. 

\section[]{Integrated surface photometry}\protect\label{integrsurfphot}
\noindent Table~\ref{totlumtbl} summarizes the results of the integrated surface photometry for all targets. The total luminosity of a galaxy was obtained by summing the flux inside of circular/elliptical apertures up to the Holmberg radius $R_H$. For all targets except for ESO400--43 the B band was used to define the Holmberg radius and the shape and placement of the aperture, which was then applied to all remaining filters. The absolute magnitude was obtained with the cosmology--corrected ($H_0=73~km^{-1}s^{-1}Mpc^{-1}$, $\Omega_{\textrm{matter}}=0.27$) Heliocentric distances available at the NASA/IPAC Extragalactic Database (NED) for each target. We are missing the B filter for ESO400--43, so the Holmberg radius and the total luminosity were instead obtained from the V band. Similarly, for SBS0335--052E\&W we present $B-I$ and $B-K$ photometry due to lack of V band. The presented magnitudes have not been inclination--corrected because the perturbed morphology of some targets does not allow for an unambiguous identification of a disk structure. 
\subsection{Error estimation}
\noindent The errors for the total luminosities were estimated by varying the geometric parameters (ellipticity and position angle) of the aperture around the best--fitting values used in the surface brightness profile integration, listed in Table~\ref{totlumtbl}. The position angle was varied by $\pm5^\circ$, the ellipticity by $\pm0.1$. The standard deviation of the luminosity values obtained in such a fashion was added in quadrature to the photometric uncertainty for each filter and target. No other errors were included since in this case the sky uncertainty cannot significantly influence the flux sum of the entire galaxy, and the flux fluctuations inside the aperture reflect a physical property of the galaxy (radially changing morphology), rather than being a measure of uncertainty. The errors in the total apparent and total absolute magnitudes are the same since no distance errors were included in the error budget.
\section[]{Total colors}\protect\label{colors}
\noindent In order to interpret the observed total colors of the galaxies, we must first separate the stellar populations. In a BCG where the old population displays regular isophotes one can easily model it with a disk and subtract it in order to obtain the flux from the young population, regardless of how concentrated or extended the latter may be. Most of the galaxies in our sample, however, do not have regular isophotes even in the outskirts, and hence it would be difficult to estimate the reliability of such a subtraction. We further wish to subject all galaxies to the same treatment, so that any differences in results will unambiguously indicate differences in galaxy properties and morphology, and will not be simply the result of having individual definitions for each galaxy. However, upon examining the individual surface brightness profiles for all filters and galaxies, it became apparent that using one single definition for what comprises the outskirt regions of a galaxy, or how extended the starburst is, leads to less than informative results for some particularly troublesome targets. Furthermore, the sky brightness in the NIR images is higher than in the optical, and this limits the radial range to which we can trace the profiles in all filters simultaneously. For example, a change in population is clearly visible for ESO185--13 at $\mu_B\sim26.5$ mag arcsec${}^{-2}$ and radius $r\sim20$ arcsec, but at that radius either of the NIR profiles is completely dominated by sky fluctuations. For other targets, e.g. Tol1457--262\emph{I}\&\emph{II}, the NIR profiles are still reliably traceable at radii corresponding to $\mu_B\sim26$ -- $28$ mag arcsec${}^{-2}$. Despite the deep NIR images, which have systematically longer exposure times than in the optical by a factor of three on average, it is obviously impossible to both study the old extended population beyond $\mu_B\sim26$ mag arcsec${}^{-2}$ and simultaneously only use radial ranges with reliable NIR data. This is due to the fact that ground-based observations in the NIR are not photon--limited and increasing the exposure time cannot push down the detected isophotes to arbitrarily faint levels. Therefore we divide our investigation in separate radial ranges: the center region is defined as the ellipse whose semi--major axis is the radius corresponding to $\mu_B\sim24$ mag arcsec${}^{-2}$; the intermediate (young and old population mix) region as the elliptical ring between $24\lesssim\mu_B\lesssim26$ mag arcsec${}^{-2}$; and the galaxy outskirts region as between $26\lesssim \mu_B\lesssim28$ mag arcsec${}^{-2}$. The ellipse parameters used during this integration are the same as in Table~\ref{totlumtbl}. For color--color plots involving purely optical colors ($U-B,~B-V,~V-R,~V-I$) we can safely use data from the outermost region. For plots involving optical--NIR color combinations ($V-K,~B-K$) we restrict ourselves to using the intermediate region, even though it is often unclear whether the burst contribution in that region is negligible.\\
%
%
\begin{center}
  \begin{table*}
    \begin{minipage}{140mm}
      \caption{Total colors for radial ranges corresponding to $\mu_B\lesssim24$, $24\lesssim\mu_B<26$, and $26\lesssim\mu_B\lesssim28$ mag arcsec${}^{-2}$. All values have been corrected for Galactic extinction~\citep{1998ApJ...500..525S}. The errors include $\sigma_{sky}$, $\sigma_{sdom}$ and $\sigma_{zp}$ added in quadrature.}
      \protect\label{totclrtbl}
      \tiny
      \begin{tabular}{@{}|l|r|r|r|r|r|r|r|r|@{}}
        \hline
        Galaxy&$\mu_B$&$U-B$&$B-V$&$V-R$&$V-I$&$V-K$&$H-K$\\\hline
        ESO185--13&$\star$\textrm{--$24$}&$-0.16 \pm 0.16$&$0.26 \pm 0.06$&$0.15 \pm 0.05$&$0.45 \pm 0.05$&$1.87 \pm 0.07$&$0.21 \pm 0.07$\\
        &$24$--$26$&$-0.39 \pm 0.19$&$0.31 \pm 0.10$&$0.52 \pm 0.10$&$0.81 \pm 0.12$&$1.93 \pm 0.67$&$-0.13 \pm 0.94$\\&$26$--$28$&$-0.76 \pm 0.51$&$0.29 \pm 0.47$&$0.72 \pm 0.50$&$0.40 \pm 0.88$&&\\\hline
        ESO249--31&$\star$\textrm{--$24$}&&$0.47 \pm 0.05$&$0.34 \pm 0.03$&$0.65 \pm 0.03$&$2.22 \pm 0.05$&$0.20 \pm 0.04$\\
        &$24$--$26$&&$0.59 \pm 0.05$&$0.30 \pm 0.03$&$0.74 \pm 0.04$&$1.64 \pm 0.31$&$-0.25 \pm 0.35$\\
        &$26$--$28$&&$0.61 \pm 0.13$&$0.08 \pm 0.16$&$0.70 \pm 0.19$&&\\\hline
        ESO338--04&$\star$\textrm{--$24$}&&$0.30 \pm 0.09$&$0.04 \pm 0.06$&$-0.27 \pm 0.07$&$0.99 \pm 0.06$&$0.09 \pm 0.07$\\
        &$24$--$26$&&$0.27 \pm 0.12$&$0.36 \pm 0.09$&$0.66 \pm 0.11$&$1.62 \pm 0.41$&$-0.12 \pm 0.78$\\
        &$26$--$28$&&$0.10 \pm 0.59$&$0.22 \pm 0.66$&$1.06 \pm 0.58$&&\\\hline
        ESO400--43&$\star$\textrm{--}$24^{\dagger}$&&&&$0.37 \pm 0.20$&$1.85 \pm 0.20$&$0.28 \pm 0.07$\\
        &$24$--$26^{\dagger}$&&&&$0.58 \pm 0.22$&$1.51 \pm 0.50$&$-0.31 \pm 0.59$\\
        &$26$--$28^{\dagger}$&&&&$0.75 \pm 0.65$&&\\\hline
        ESO421--02&$\star$\textrm{--$24$}&$-0.54 \pm 0.15$&$-0.20 \pm 0.06$&$0.63 \pm 0.04$&$0.94 \pm 0.05$&$2.67 \pm 0.08$&$0.18 \pm 0.08$\\
        &$24$--$26$&$-0.35 \pm 0.17$&$0.08 \pm 0.11$&$0.64 \pm 0.11$&$1.08 \pm 0.11$&$2.72 \pm 0.16$&$0.25 \pm 0.20$\\
        &$26$--$28$&$-0.79 \pm 0.36$&$0.41 \pm 0.43$&$0.36 \pm 0.47$&$1.07 \pm 0.45$&$2.69 \pm 0.66$&$0.63 \pm 1.04$\\\hline
        ESO462--20&$\star$\textrm{--$24$}&$-0.11 \pm 0.16$&$0.26 \pm 0.07$&&&$2.06 \pm 0.20$&$0.48 \pm 0.20$\\
        &$24$--$26$&$-0.52 \pm 0.26$&$0.63 \pm 0.21$&&&$1.41 \pm 0.48$&$-0.08 \pm 1.01$\\
        &$26$--$28$&$-1.22 \pm 1.75$&$1.58 \pm 1.72$&&&&\\\hline
        HE2--10&$\star$\textrm{--$24$}&$0.05 \pm 0.14$&$0.41 \pm 0.06$&&&$2.85 \pm 0.07$&$0.46 \pm 0.09$\\&$24$--$26$&$0.26 \pm 0.24$&$0.68 \pm 0.11$&&&$2.65 \pm 0.30$&$0.50 \pm 0.41$\\&$26$--$28$&$0.27 \pm 0.88$&$0.62 \pm 0.45$&&&$2.38 \pm 1.67$&$0.45 \pm 2.30$\\\hline
        HL293B&$\star$\textrm{--$24$}&$-0.52 \pm 0.17$&$0.47 \pm 0.09$&$0.07 \pm 0.10$&$0.14 \pm 0.11$&$1.43 \pm 0.15$&$0.24 \pm 0.19$\\&$24$--$26$&$-0.46 \pm 0.17$&$0.51 \pm 0.08$&$0.33 \pm 0.09$&$0.57 \pm 0.12$&$2.25 \pm 0.90$&$0.09 \pm 1.16$\\&$26$--$28$&$-0.39 \pm 0.37$&$0.69 \pm 0.20$&$0.28 \pm 0.24$&$0.60 \pm 0.28$&&\\\hline
        IIZW40&$\star$\textrm{--$24$}&$-0.82 \pm 0.23$&$0.18 \pm 0.07$&&$-0.27 \pm 0.11$&$0.89 \pm 0.09$&$0.34 \pm 0.31$\\
        &$24$--$26$&&$0.34 \pm 0.74$&&&$1.23 \pm 1.26$&\\
        &$26$--$28$&&$0.07 \pm 2.61$&&&&\\\hline
        MK600&$\star$\textrm{--$24$}&$-0.36 \pm 0.16$&$0.37 \pm 0.07$&$0.16 \pm 0.07$&$0.37 \pm 0.09$&$1.60 \pm 0.09$&$0.20 \pm 0.12$\\
        &$24$--$26$&$-0.12 \pm 0.21$&$0.39 \pm 0.09$&$0.29 \pm 0.09$&$0.59 \pm 0.14$&$1.14 \pm 1.99$&$-0.56 \pm 2.15$\\
        &$26$--$28$&$-0.14 \pm 0.83$&$0.34 \pm 0.42$&$-0.26 \pm 0.50$&$-0.84 \pm 2.16$&&\\\hline
        MK900&$\star$\textrm{--$24$}&$-0.24 \pm 0.16$&$0.55 \pm 0.07$&$0.43 \pm 0.07$&$0.76 \pm 0.09$&$2.26 \pm 0.11$&$0.18 \pm 0.10$\\
        &$24$--$26$&$0.02 \pm 0.17$&$0.70 \pm 0.07$&$0.51 \pm 0.08$&$0.92 \pm 0.10$&$2.54 \pm 0.27$&$0.22 \pm 0.34$\\
        &$26$--$28$&$-0.02 \pm 0.48$&$0.66 \pm 0.20$&$0.49 \pm 0.23$&$1.01 \pm 0.26$&$2.67 \pm 1.45$&$0.43 \pm 2.05$\\\hline
        MK930&$\star$\textrm{--$24$}&$-0.79 \pm 0.16$&$0.43 \pm 0.08$&$0.05 \pm 0.08$&$0.14 \pm 0.11$&$1.82 \pm 0.10$&$0.48 \pm 0.10$\\
        &$24$--$26$&$-0.68 \pm 0.18$&$0.28 \pm 0.12$&$0.39 \pm 0.13$&$0.58 \pm 0.16$&$2.44 \pm 0.72$&$-0.00 \pm 0.85$\\
        &$26$--$28$&$-0.31 \pm 0.55$&$-0.03 \pm 0.67$&$0.15 \pm 0.92$&$-0.03 \pm 1.41$&&\\\hline
        MK996&$\star$\textrm{--$24$}&$-0.32 \pm 0.16$&$0.56 \pm 0.06$&$0.42 \pm 0.07$&$0.77 \pm 0.10$&$2.48 \pm 0.07$&$0.32 \pm 0.11$\\
        &$24$--$26$&$-0.09 \pm 0.16$&$0.65 \pm 0.05$&$0.47 \pm 0.07$&$0.91 \pm 0.10$&$2.44 \pm 0.15$&$0.15 \pm 0.52$\\
        &$26$--$28$&$-0.28 \pm 0.32$&$0.59 \pm 0.16$&$0.47 \pm 0.16$&$0.88 \pm 0.23$&$2.13 \pm 1.33$&$-1.02 \pm 2.01$\\\hline
        SBS0335--052E&$\star$\textrm{--$24$}&&&&$-0.07 \pm 0.15{}^\ddagger$&$1.10 \pm 0.14{}^\ddagger$&$0.54 \pm 0.17$\\
        &$24$--$26$&&&&$0.30 \pm 0.18{}^\ddagger$&$1.99 \pm 0.86{}^\ddagger$&\\
        &$26$--$28$&&&&$0.65 \pm 0.74{}^\ddagger$&&\\\hline
        SBS0335--052W&$\star$\textrm{--$24$}&&&&$0.18 \pm 0.13{}^\ddagger$&$1.32 \pm 0.22{}^\ddagger$&$0.53 \pm 0.42$\\
        &$24$--$26$&&&&$0.30 \pm 0.20{}^\ddagger$&$1.97 \pm 0.77{}^\ddagger$&$0.15 \pm 1.25$\\
        &$26$--$28$&&&&$0.60 \pm 0.64{}^\ddagger$&$2.86 \pm 1.81{}^\ddagger$&$-0.03 \pm 2.63$\\\hline
        TOL0341--407&$\star$\textrm{--$24$}&$-0.51 \pm 0.16$&$0.67 \pm 0.06$&&$0.14 \pm 0.50$&$1.63 \pm 0.09$&$0.23 \pm 0.09$\\
        &$24$--$26$&$-0.57 \pm 0.17$&$0.67 \pm 0.07$&&$0.55 \pm 0.50$&$2.01 \pm 0.42$&$0.25 \pm 0.46$\\
        &$26$--$28$&$-0.53 \pm 0.55$&$0.72 \pm 0.30$&&$0.89 \pm 0.60$&&\\\hline
        TOL1457--262\emph{I}&$\star$\textrm{--$24$}&&$0.28 \pm 0.07$&&&$1.61 \pm 0.07$&\\
        &$24$--$26$&&$0.27 \pm 0.10$&&&$2.77 \pm 0.22$&\\
        &$26$--$28$&&$0.24 \pm 0.48$&&&&\\\hline
        TOL1457--262\emph{II}&$\star$\textrm{--$24$}&&$0.19 \pm 0.06$&&&$2.22 \pm 0.06$&\\
        &$24$--$26$&&$0.23 \pm 0.09$&&&$2.56 \pm 0.19$&\\
        &$26$--$28$&&$0.35 \pm 0.37$&&&&\\\hline
        UM133&$\star$\textrm{--$24$}&$-0.48 \pm 0.16$&$0.30 \pm 0.06$&$0.24 \pm 0.06$&$0.48 \pm 0.09$&$1.95 \pm 0.23$&$0.30 \pm 0.28$\\
        &$24$--$26$&$-0.25 \pm 0.16$&$0.40 \pm 0.07$&$0.36 \pm 0.07$&$0.75 \pm 0.10$&$2.33 \pm 0.95$&$0.36 \pm 1.16$\\
        &$26$--$28$&$-0.05 \pm 0.39$&$0.39 \pm 0.18$&$0.37 \pm 0.23$&$0.97 \pm 0.27$&&\\\hline
        UM160&$\star$\textrm{--$24$}&$-0.98 \pm 0.06$&$0.45 \pm 0.04$&$0.21 \pm 0.05$&$0.55 \pm 0.05$&$1.92 \pm 0.13$&$0.15 \pm 0.13$\\
        &$24$--$26$&$-0.89 \pm 0.09$&$0.40 \pm 0.06$&$0.19 \pm 0.09$&$0.44 \pm 0.11$&$1.99 \pm 0.56$&$0.08 \pm 0.61$\\
        &$26$--$28$&$-0.76 \pm 0.59$&$0.47 \pm 0.34$&$0.18 \pm 0.54$&$-0.01 \pm 1.06$&$2.78 \pm 1.93$&$0.08 \pm 2.11$\\\hline
        UM238&$\star$\textrm{--$24$}&$-0.32 \pm 0.16$&$0.40 \pm 0.07$&$0.32 \pm 0.07$&$0.64 \pm 0.10$&$2.13 \pm 0.10$&$0.55 \pm 0.16$\\&$24$--$26$&$-0.30 \pm 0.17$&$0.40 \pm 0.07$&$0.30 \pm 0.08$&$0.54 \pm 0.14$&$1.98 \pm 0.57$&$0.46 \pm 0.99$\\&$26$--$28$&$-0.16 \pm 0.44$&$0.40 \pm 0.22$&$0.04 \pm 0.34$&$-0.52 \pm 1.40$&$2.82 \pm 1.38$&\\\hline
        UM417&$\star$\textrm{--$24$}&$-0.58 \pm 0.09$&$0.62 \pm 0.09$&$-0.08 \pm 0.09$&$-0.06 \pm 0.09$&$1.48 \pm 0.31$&$0.92 \pm 0.66$\\&$24$--$26$&$-0.28 \pm 0.08$&$0.42 \pm 0.05$&$0.29 \pm 0.06$&$0.59 \pm 0.08$&$2.20 \pm 0.75$&$0.91 \pm 1.67$\\&$26$--$28$&$0.05 \pm 0.54$&$0.66 \pm 0.21$&$0.23 \pm 0.21$&$0.47 \pm 0.34$&&\\\hline
        UM448&$\star$\textrm{--$24$}&&$0.47 \pm 0.06$&$0.24 \pm 0.07$&$0.60 \pm 0.07$&$2.43 \pm 0.14$&\\&$24$--$26$&&$0.58 \pm 0.05$&$0.41 \pm 0.07$&$0.93 \pm 0.07$&$2.09 \pm 1.57$&\\&$26$--$28$&&$0.74 \pm 0.09$&$0.33 \pm 0.16$&$0.96 \pm 0.15$&&\\\hline
        UM619&$\star$\textrm{--$24$}&$-0.10 \pm 0.10$&$0.36 \pm 0.06$&$0.34 \pm 0.05$&$0.68 \pm 0.06$&$2.31 \pm 0.10$&$0.52 \pm 0.18$\\&$24$--$26$&$0.05 \pm 0.18$&$0.44 \pm 0.07$&$0.59 \pm 0.07$&$0.92 \pm 0.08$&$2.62 \pm 0.64$&$0.89 \pm 1.13$\\&$26$--$28$&$-0.13 \pm 0.64$&$0.59 \pm 0.18$&$0.82 \pm 0.19$&$1.17 \pm 0.22$&$3.05 \pm 1.89$&\\\hline
        \hline
      \end{tabular}
      \medskip
      ~\\
      $\dagger$ -- $V$ instead of $B$ filter\\
      $\ddagger$ -- $B$ instead of $V$ filter
    \end{minipage}
  \end{table*}
\end{center}
\noindent The total colors for these radial ranges are presented in Table~\ref{totclrtbl}. Each value was calculated by summing up all pixels inside the region, after foreground stars and background galaxies have been masked out. As already mentioned above, each region was defined from a reference filter, usually the B band, and then applied to the rest of the filters. Additionally, a union of the source masks in all filters was created and applied to the region in each filter before summation. Note that this means that the summation is carried over the same physical region of the galaxy in each filter and involves the exact same number of pixels. While it pains us to throw away perfectly good pixels this way, possibly with high signal-to-noise, not doing so could bias the results and unnecessarily complicate the interpretation.\\

\noindent At faint surface brightness levels the flux in some pixels will start randomly falling below a value of $0$, which is an effect of the sky subtraction. To avoid a systematic reddening of colors, \textit{all} pixels are summed, positive and negative, and no clipping of any kind is applied. Since the mean sky level is $\sim0$ in each frame, the negative sky pixels cancel out the positive ones and we recover total colors consistent with the trends observed in the radial color profiles for a given magnitude range. There is of course always a sky residual present in the frames, the total sum of the sky is never exactly $0$ regardless of what sky subtraction technique is used, but the influence of that sky residual is estimated and included in the total errors in the form of $\sigma_{sky}$.
\subsection{Error estimation}
\noindent The errors in the total colors were obtained in a fashion similar to \S~\ref{errorest} for each filter and then added in quadrature, $\sigma(X-Y)=\sqrt{\sigma_X^2+\sigma_Y^2}$, where $\sigma_X$ and $\sigma_Y$ each include all three sources of uncertainty -- $\sigma_{sky}$, $\sigma_{sdom}$ and $\sigma_{zp}$. Note that only $\sigma_{sdom}$ varies with each region (center, intermediate, and outskirts), while the other two are constant over the entire frame in each filter.
\section[]{Host structural parameters}\protect\label{strucparam}
\noindent The scale length $h_r^B$  and the central surface brightness $\mu_0^B$ are obtained for each galaxy by a weighted least squares fit of an exponential disk to the reference profile in the two radial ranges $\Delta R_1$ and $\Delta R_2$ defined by $\mu_{B}=24$--$26$ and $\mu_{B}=26$--$28$ mag arcsec${}^{-2}$ respectively. When applied to the NIR filters, the radial ranges $\Delta R_1$ and especially $\Delta R_2$ usually encompass a very noisy part of the profile. Even a weighted fit to such data points will give exponential disk parameters with arbitrarily large uncertainties. In other words, we often cannot measure how the light distribution at NIR wavelengths behaves at such large radial distances from the center. Therefore fitting individual scale lengths for each filter at the same $\Delta R_1$ and $\Delta R_2$ will give essentially random $h_r$ and $\mu_0$. Since we do not know the true profile shapes at these radii, we make the assumption that the $h_r^B$ scale length, which is known with a fairly high accuracy as the B band profile is usually the deepest, is valid for all wavelengths and only fit the central surface brightness $\mu_0$ in each filter. This assumption is not unreasonable, e.g.~\citet{2004A&A...428..837F} find that the disk scale length of minor mergers does not change in $BVI$. The same appears to be true for normal ellipticals, as seen from the radial color profiles of the red and dead elliptical galaxy $KIG~732$ in Appendix~\ref{kig}. In hybrid optical--NIR colors strong gradients can still exist as demonstrated by the $V-K$ radial color profile for $KIG~732$. This means that our burst estimate at NIR wavelengths is systematically overestimated. Further, in the NIR regime burst estimation often fails due to the complete breakdown of the NIR profile inside $\Delta R_1$ or $\Delta R_2$, but it gives reasonable estimates of $\mu_0$ for profiles which still exhibit a statistically significant ordered signal, only one with very large errors. We will refrain from overanalysing the NIR burst colors, and will predominantly use optical burst estimates. Host structural parameters for all galaxies are presented in Table~\ref{scalelentbl}.
\subsection{Error estimation}
\noindent The $h_r^B$ errors are the propagated errors of the fitted slope, scaled to units of the profile errors at those radii. The $\mu_0^B$ errors include the error of the fit and the zero point errors $\sigma_{zp}$. The $\mu_0$ errors for all other filters also account for the unscaled $h_r^B$ error since the uncertainty in the assumed slope directly influences the extrapolated $\mu_0$ value. 
\section[]{Estimation of the burst component}\protect\label{burstestim}
\noindent The central region of the profiles is usually dominated by the strong signal from the starburst population but it also contains a non-negligible contribution from the underlying old population. Notable exceptions to this are irregular morphology targets (e.g. Tol0341--407, UM160, UM238, etc.) for which the center of light ($r=0$ arcsec) is not at all located at the burst as there are several knots of star formation spread around the galaxy. If one can disentangle the contributions from the two populations one can obtain the colors of the burst. This is a rather complicated problem without a universal solution and lies beyond the scope of this paper. However, a simplified estimate of the light of the burst in each filter can be obtained by a 1D subtraction of the fitted exponential disk from the profile data. For each filter, we use the B band $h_r$ and the fitted $\mu_0$ for the individual filters to interpolate along the exponential disk line, thus obtaining an estimate of the contribution of the old population to the total flux at that radius. The excess in luminosity above the exponential disk is an estimate of the burst luminosity. Relative burst contribution to the $B$ band and apparent burst luminosity obtained in such a fashion for all filters and galaxies are presented in Table~\ref{burstclrtbl}. We note that the morphology of some galaxies clearly contradicts the assumption of an underlying regular disk, even though an exponential profile can still be fitted to the light distribution at large radii. One should keep this in mind and refrain from overinterpreting the data.
\subsection{Error estimation}
\noindent The errors of the burst component parameters are estimated from the fitting errors of the exponential disk, scaled to units of the profile errors, and combined with the zero point uncertainty. Thus they may be underestimated since there is no measure of the uncertainty in the exact flux level. There is also no sky uncertainty contribution to the total error budget of the burst, but as Figure~\ref{errorplot} shows, at such bright isophotal levels is it always the uncertainty in the flux level that would dominate. All errors are assumed to be independent and added in quadrature. 
\section[]{Asymmetry and concentration}\protect\label{asymparam}
\noindent Asymmetry, clumpiness and concentration have been argued to contain information on the past and present evolution history of a galaxy even when measured in only one filter~\citep{2003ApJS..147....1C}, which makes them uniquely economical parameters to obtain. The asymmetry is thought to correlate with galaxy colors in a way which can separate recent major mergers from non--mergers~\citep{2000ApJ...529..886C}. To investigate this, in Table~\ref{petrasymtbl} we compute the asymmetries for all targets using the method
\begin{equation} 
\label{eq:asym}
A_{abs}=\frac{\sum\lvert I_0-I_\phi\rvert}{2\sum\lvert I_0\rvert}
\end{equation}
because it correlates well with the colors, and hence is a good indicator of star formation~\citep{2000ApJ...529..886C}. $I_0$ is the original (processed) image, and $I_\phi$ is rotated $180$ degrees. Decreasing signal--to--noise (S/N) in the fainter isophotal levels can greatly affect the minimum asymmetry, hence normally one would apply a noise correction. All galaxies in the sample are either starbursting or star--forming at moderately high SFRs, however, and the composite asymmetry due to both star formation (flocculent asymmetry) and merger (dynamical asymmetry) events is likely to be characterized by the presence of the bright star forming regions and not dominated by any possible faint extended tidal structures. Hence, instead of applying a noise correction to reach fainter isophotal levels, we choose to compute the asymmetry inside the Petrosian radius $r[\eta=0.2]$, which typically gives asymmetry errors lower than $0.02$ rms for the galaxies in the sample~\citep{2000ApJ...529..886C}. $\eta$ denotes the inverted $\eta$, defined as the ratio between the local surface brightness at some radius and the average surface brightness inside that radius. The location of the starburst regions in the individual targets is varied, and they can be highly dispersed around the galaxy. In our procedure for calculating the (minimum) asymmetry of each target we therefore shift the initial center of the asymmetry box by $\pm30$ pixels in each direction, and compute individual asymmetries for rotation around each of the points inside the central $60\times60$ box. This way we obtain asymmetry maps of size $60\times60$ pixels for each target and filter. If the apparent size of the galaxy was too large to safely encompass the asymmetry minimum, we instead obtained a $120\times120$ asymmetry map. For all galaxies and all filters there is always a well--defined global minimum in the asymmetry maps, though multiple local minima exist for targets with highly irregular burst morphologies. In Table~\ref{petrasymtbl} we show the Petrosian minimum asymmetry in each filter and the corresponding Petrosian radius used to calculate that minimum. \\

\noindent As Table~\ref{petrasymtbl} indicates, the Petrosian $r[\eta(0.2)]$ radius is usually very small. It includes only the brightest isophotes, while faint extended irregular features like tidal tails and arms are thus excluded from the asymmetry measurement. It seems therefore prudent to have a second measure of asymmetry, with a different (larger) limiting radius, in order to capture any effect on the asymmetry that faint irregular structures might have. We chose the Holmberg radius $r(\mu_B=26.5)$ for the $B$ band, and in the rest of the optical filters the radius at which the corresponding surface brightness profiles reach $\mu\sim26.5$ mag arcsec${}^{-2}$. In the NIR regime the $H$ and $K$ profiles are much shallower, so a limiting radius of $r(\mu_{NIR}=23)$ was chosen instead. The area enclosed in these radii is much larger than inside the $r[\eta(0.2)]$ radius, however, the sample galaxies are at significantly differing redshifts. The difference in distance between the closest and the farthest galaxy is a factor of $7$. This will lower the asymmetry measured from galaxies further away and thus skew the asymmetry distribution of the sample. To even the play field and allow direct comparison between the galaxies, all images were first smoothed by a boxcar average of $1\times1$ kpc. The asymmetry minimum was then obtained in the same way as described above for the case of Petrosian asymmetries. These values are shown in Table~\ref{holmasymtbl}.\\

\noindent The concentration index was calculated following~\citet{2003ApJS..147....1C} as
\begin{equation} 
\label{eq:conc}
C=5\times\log{\frac{r_{80\%}}{r_{20\%}}}
\end{equation}
\noindent where $r_{80}$ is the radius which encompasses $80\%$ of the light inside the Petrosian $1.5\times r[\eta(0.2)]$ radius, and $r_{20}$ is $20\%$ of $r_{80}$. Measurements for each galaxy and filter are given in Table~\ref{concentr}.
\begin{center}
  \begin{table}
      \caption{\textbf{Host structural parameters.} Absolute $B$ magnitude of the host ($M_B^{host}$) obtained by subtracting the burst luminosity. The scale length $h_r$ in arcseconds and kpc, and the central surface brightness $\mu_0$ are based on a weighted least squares fit to the deepest image ($B$ band) for two radial ranges derived from $24\lesssim\mu_B<26$ and $26\lesssim\mu_B\lesssim28$ mag arcsec${}^{-2}$. Correction for Galactic extinction~\citep{1998ApJ...500..525S} has been applied.}
      \protect\label{scalelentbl}
      \tiny
      \begin{tabular}{@{}|l|r|r|r|r|r|r|@{}}
        \hline
        Galaxy&$\mu_B$&$M_B^{host}$&$h_r^{''}$&$h_r^{kpc}$&$\mu_0^\textrm{B}$\\\hline
        ESO185--13&$24$--$26$&$-17.7$&$4.18\pm0.26$&$1.55\pm0.10$&$20.87\pm0.27$\\
        &$26$--$28$&$-15.9$&$15.97\pm0.94$&$5.91\pm0.35$&$24.78\pm0.12$\\\hline
        ESO249--31&$24$--$26$&$-17.0$&$31.48\pm0.87$&$1.65\pm0.05$&$21.40\pm0.11$\\
        &$26$--$28$&$-15.7$&$74.85\pm1.92$&$3.92\pm0.10$&$24.23\pm0.08$\\\hline
        ESO338--04&$24$--$26$&$-17.7$&$7.74\pm0.22$&$1.40\pm0.04$&$20.68\pm0.14$\\
        &$26$--$28$&$-16.2$&$18.93\pm1.36$&$3.43\pm0.25$&$23.84\pm0.22$\\\hline
        ESO400--43${}^\dagger$&$24$--$26$&$-18.4$&$4.31\pm0.10$&$1.65\pm0.04$&$20.64\pm0.22$\\
        &$26$--$28$&$-16.9$&$9.39\pm0.75$&$3.61\pm0.29$&$23.60\pm0.32$\\\hline
        ESO421--02&$24$--$26$&&$11.97\pm0.09$&$0.72\pm0.01$&$20.57\pm0.05$\\
        &$26$--$28$&&$11.11\pm0.28$&$0.67\pm0.02$&$20.20\pm0.18$\\\hline
        ESO462--20&$24$--$26$&$-17.9$&$6.33\pm0.30$&$2.44\pm0.12$&$21.27\pm0.17$\\
        &$26$--$28$&$-16.7$&$16.49\pm2.33$&$6.36\pm0.90$&$24.29\pm0.33$\\\hline
        HE2--10&$24$--$26$&$-18.0$&$17.94\pm0.22$&$1.37\pm0.02$&$20.49\pm0.07$\\
        &$26$--$28$&$-16.6$&$41.13\pm1.59$&$3.15\pm0.12$&$23.53\pm0.13$\\\hline
        HL293B&$24$--$26$&$-12.7$&$2.25\pm0.01$&$0.18\pm0.00$&$20.66\pm0.05$\\
        &$26$--$28$&$-12.9$&$2.28\pm0.08$&$0.18\pm0.01$&$20.73\pm0.20$\\\hline
        IIZW40&$24$--$26$&$-17.1$&$29.26\pm1.83$&$1.67\pm0.10$&$21.60\pm0.20$\\
        &$26$--$28$&&&&\\\hline
        MK600&$24$--$26$&$-13.9$&$11.56\pm0.41$&$0.61\pm0.02$&$21.58\pm0.12$\\
        &$26$--$28$&$-13.4$&$25.94\pm0.70$&$1.37\pm0.04$&$23.79\pm0.09$\\\hline
        MK900&$24$--$26$&$-14.8$&$14.69\pm0.15$&$0.80\pm0.01$&$21.61\pm0.05$\\
        &$26$--$28$&$-14.5$&$22.07\pm0.43$&$1.20\pm0.02$&$23.03\pm0.09$\\\hline
        MK930&$24$--$26$&$-18.3$&$3.84\pm0.07$&$1.33\pm0.02$&$19.59\pm0.11$\\
        &$26$--$28$&$-16.0$&$11.30\pm0.61$&$3.91\pm0.21$&$23.85\pm0.16$\\\hline
        MK996&$24$--$26$&$-15.4$&$5.91\pm0.03$&$0.52\pm0.00$&$20.69\pm0.03$\\
        &$26$--$28$&$-14.9$&$7.35\pm0.20$&$0.65\pm0.02$&$21.78\pm0.14$\\\hline
        SBS0335--052E&$24$--$26$&$-14.9$&$2.59\pm0.04$&$0.68\pm0.01$&$21.66\pm0.07$\\
        &$26$--$28$&$-14.7$&$3.93\pm0.30$&$1.03\pm0.08$&$23.07\pm0.30$\\\hline
        SBS0335--052W&$24$--$26$&$-13.2$&$2.62\pm0.01$&$0.69\pm0.00$&$22.60\pm0.05$\\
        &$26$--$28$&$-13.8$&$2.81\pm0.20$&$0.74\pm0.05$&$22.95\pm0.27$\\\hline
        TOL0341--407&$24$--$26$&&$2.54\pm0.13$&$0.75\pm0.04$&$18.71\pm0.32$\\
        &$26$--$28$&$-15.4$&$7.79\pm0.26$&$2.29\pm0.08$&$23.56\pm0.11$\\\hline
        TOL1457--262~\emph{I}&$24$--$26$&&$5.90\pm0.25$&$3.63\pm0.07$&$20.84\pm0.18$\\
        &$26$--$28$&$-18.3$&$12.55\pm0.87$&$3.39\pm0.22$&$23.62\pm0.22$\\\hline
        TOL1457--262\emph{II}&$24$--$26$&$-18.5$&$10.29\pm0.20$&$2.08\pm0.09$&$20.46\pm0.10$\\
        &$26$--$28$&$-16.9$&$9.61\pm0.63$&$4.43\pm0.31$&$20.08\pm0.45$\\\hline
        UM133&$24$--$26$&&$9.84\pm0.25$&$0.87\pm0.02$&$20.89\pm0.11$\\
        &$26$--$28$&$-14.8$&$15.53\pm0.33$&$1.38\pm0.03$&$22.80\pm0.10$\\\hline
        UM160&$24$--$26$&&$10.36\pm0.62$&$1.41\pm0.08$&$20.27\pm0.28$\\
        &$26$--$28$&&$7.80\pm0.28$&$1.06\pm0.04$&$18.70\pm0.30$\\\hline
        UM238&$24$--$26$&$-16.0$&$5.20\pm0.14$&$1.37\pm0.04$&$21.51\pm0.10$\\
        &$26$--$28$&$-15.7$&$8.30\pm0.40$&$2.18\pm0.11$&$23.11\pm0.18$\\\hline
        UM417&$24$--$26$&$-12.7$&$7.28\pm0.54$&$1.19\pm0.09$&$22.96\pm0.14$\\
        &$26$--$28$&&$3.10\pm0.12$&$0.51\pm0.02$&$19.76\pm0.26$\\\hline
        UM448&$24$--$26$&$-18.1$&$10.73\pm0.80$&$4.29\pm0.32$&$22.53\pm0.22$\\
        &$26$--$28$&$-17.9$&$23.18\pm0.89$&$9.27\pm0.36$&$24.33\pm0.09$\\\hline
        UM619&$24$--$26$&&$4.22\pm0.11$&$1.40\pm0.04$&$19.18\pm0.16$\\
        &$26$--$28$&$-17.4$&$6.45\pm0.29$&$2.14\pm0.10$&$21.48\pm0.25$\\\hline
        \hline
      \end{tabular}
      \medskip
      ~\\
      $\dagger$ -- $V$ instead of $B$ filter
  \end{table}
\end{center}
\begin{center}
  \begin{table*}
    \begin{minipage}{150mm}
      \caption{Estimated luminosity in excess of the exponential disk defined by $h_r^{''}$ and $\mu_0$. The upper and lower numbers for each galaxy are for disk properties derived from the radial ranges corresponding to $24\lesssim\mu_B<26$ and $26\lesssim\mu_B\lesssim28$ mag arcsec${}^{-2}$, respectively. Fields are left blank where the estimation method failed. All values have been corrected for Galactic extinction~\citep{1998ApJ...500..525S}. The $\%$ column gives the relative burst contribution to the total galaxy luminosity in the $B$ band.}
      \protect\label{burstclrtbl}
      \tiny
      \begin{tabular}{@{}lllrrrrrrr@{}}
        \hline
        Galaxy&$\%$&$B_\star$&$(U-B)_\star$&$(B-V)_\star$&$(V-R)_\star$&$(V-I)_\star$&$(V-K)_\star$&$(H-K)_\star$\\\hline
        ESO185--13&$74$&$15.3 \pm 0.3$&$-0.12 \pm 0.31$&$0.24 \pm 0.27$&$0.04 \pm 0.05$&$0.32 \pm 0.05$&$1.77 \pm 0.08$&$0.29 \pm 0.09$\\
        &$95$&$15.0 \pm 0.1$&$-0.13 \pm 0.19$&$0.26 \pm 0.13$&$0.16 \pm 0.05$&$0.48 \pm 0.05$&$1.78 \pm 0.06$&$0.08 \pm 0.17$\\\hline
        ESO249--31&$45$&$13.1 \pm 0.1$&&$0.43 \pm 0.11$&$0.37 \pm 0.03$&$0.58 \pm 0.04$&$2.43 \pm 0.05$&$0.36 \pm 0.05$\\
        &$86$&$12.4 \pm 0.1$&&$0.48 \pm 0.08$&$0.33 \pm 0.03$&$0.62 \pm 0.04$&$2.14 \pm 0.07$&$0.21 \pm 0.07$\\\hline
        ESO338--04&$73$&$13.9 \pm 0.1$&&$0.37 \pm 0.14$&$-0.04 \pm 0.01$&$-0.74 \pm 0.05$&$0.74 \pm 0.03$&$0.31 \pm 0.04$\\
        &$94$&$13.6 \pm 0.2$&&$0.35 \pm 0.22$&$0.05 \pm 0.04$&$-0.34 \pm 0.06$&$0.93 \pm 0.05$&$0.07 \pm 0.13$\\\hline
        ESO400--43&$76^\dagger$&$14.6 \pm 0.2^\dagger$&&&&$0.31 \pm 0.22$&$1.83 \pm 0.23$&$0.35 \pm 0.08$\\
        &$94^\dagger$&$14.6 \pm 0.3^\dagger$&&&&$0.35 \pm 0.33$&$1.88 \pm 0.33$&$0.27 \pm 0.28$\\\hline
        ESO421--02&&&&&&&&\\
        &&&&&&&&\\\hline
        ESO462--20&$80$&$14.7 \pm 0.2$&$-0.07 \pm 0.23$&$0.22 \pm 0.17$&&&$2.09 \pm 0.20$&$0.56 \pm 0.21$\\
        &$95$&$14.5 \pm 0.3$&$-0.09 \pm 0.36$&$0.23 \pm 0.33$&&&$2.00 \pm 0.22$&$0.44 \pm 0.27$\\\hline
        HE2--10&$53$&$12.6 \pm 0.1$&$0.04 \pm 0.15$&$0.22 \pm 0.07$&&&$2.98 \pm 0.06$&$0.51 \pm 0.09$\\
        &$89$&$12.1 \pm 0.1$&$0.13 \pm 0.18$&$0.44 \pm 0.13$&&&$2.78 \pm 0.06$&$0.45 \pm 0.09$\\\hline
        HL293B&$14$&$19.5 \pm 0.1$&$-0.76 \pm 0.16$&$0.42 \pm 0.07$&$-1.32 \pm 0.07$&&&\\
        &$16$&$19.4 \pm 0.2$&$-0.60 \pm 0.27$&$0.05 \pm 0.21$&&&&\\\hline
        IIZW40&$68$&$12.1 \pm 0.2$&&$0.09 \pm 0.20$&&$-0.10 \pm 0.09$&$0.78 \pm 0.07$&$-0.16 \pm 0.19$\\
        &$87$&$11.8 \pm 0.9$&$-0.56 \pm 0.97$&$0.23 \pm 0.95$&&$-0.16 \pm 0.10$&$1.08 \pm 0.30$&$0.43 \pm 0.30$\\\hline
        MK600&$49$&$15.9 \pm 0.1$&$-0.46 \pm 0.20$&$0.35 \pm 0.14$&$0.01 \pm 0.07$&$0.18 \pm 0.10$&$1.45 \pm 0.06$&$0.52 \pm 0.09$\\
        &$76$&$15.4 \pm 0.1$&$-0.40 \pm 0.18$&$0.41 \pm 0.10$&$0.25 \pm 0.07$&$0.44 \pm 0.10$&$1.77 \pm 0.11$&$0.57 \pm 0.16$\\\hline
        MK900&$49$&$15.0 \pm 0.1$&$-0.37 \pm 0.16$&$0.42 \pm 0.07$&$0.35 \pm 0.07$&$0.60 \pm 0.09$&$1.95 \pm 0.11$&$0.06 \pm 0.10$\\
        &$73$&$14.6 \pm 0.1$&$-0.23 \pm 0.17$&$0.54 \pm 0.10$&$0.43 \pm 0.07$&$0.71 \pm 0.09$&$1.88 \pm 0.12$&$-0.08 \pm 0.12$\\\hline
        MK930&$52$&$15.7 \pm 0.1$&$-0.80 \pm 0.19$&$0.49 \pm 0.13$&$-0.25 \pm 0.07$&$-0.33 \pm 0.10$&$0.94 \pm 0.10$&$1.17 \pm 0.13$\\
        &$94$&$15.1 \pm 0.2$&$-0.78 \pm 0.22$&$0.42 \pm 0.17$&$0.05 \pm 0.08$&$0.14 \pm 0.11$&$1.81 \pm 0.31$&$0.75 \pm 0.35$\\\hline
        MK996&$42$&$16.0 \pm 0.1$&$-0.54 \pm 0.15$&$0.45 \pm 0.05$&$0.33 \pm 0.06$&$0.49 \pm 0.09$&$2.54 \pm 0.05$&$0.85 \pm 0.09$\\
        &$68$&$15.5 \pm 0.1$&$-0.22 \pm 0.21$&$0.54 \pm 0.15$&$0.40 \pm 0.06$&$0.70 \pm 0.09$&$2.44 \pm 0.09$&\\\hline
        SBS0335--052E&$75$&$17.1 \pm 0.1$&&&&$-0.14 \pm 0.11^\ddagger$&$0.75 \pm 0.24^\ddagger$&$0.50 \pm 0.24$\\
        &$86$&$16.9 \pm 0.3$&&&&$-0.19 \pm 0.32^\ddagger$&$0.69 \pm 0.37^\ddagger$&$0.44 \pm 0.26$\\\hline
        SBS0335--052W&$26$&$20.6 \pm 0.1$&&&&$-0.11 \pm 0.10^\ddagger$&$0.57 \pm 0.22^\ddagger$&\\
        &$39$&$20.2 \pm 0.3$&&&&$-2.33 \pm 0.29^\ddagger$&$0.52 \pm 0.56^\ddagger$&\\\hline
        TOL0341--407&&&&&&&&\\
        &$83$&$16.6 \pm 0.1$&$-0.50 \pm 0.19$&$0.65 \pm 0.12$&&$0.01 \pm 0.50$&$1.40 \pm 0.08$&$0.25 \pm 0.10$\\\hline
        TOL1457--262~\emph{I}&$73$&$14.5\pm 0.2$&&$0.29 \pm 0.18$&&&$1.26 \pm 0.11$&\\
        &$94$&$14.3\pm 0.2$&&$0.27 \pm 0.22$&&&$1.22 \pm 0.10$&\\\hline
        TOL1457--262~\emph{II}&$47$&$15.6\pm 0.1$&&$-0.06\pm 0.10$&&&$1.91\pm 0.10$&\\
        &$57$&$15.9 \pm 0.5$&&$-0.65 \pm 0.45$&&&&\\\hline
        UM133&&&&&&&&\\
        &$56$&$16.1 \pm 0.1$&$-0.61 \pm 0.18$&$0.26 \pm 0.10$&$0.18 \pm 0.06$&$0.09 \pm 0.09$&&\\\hline
        UM160&&&&&&&&\\
        &&&&&&&&\\\hline
        UM238&$21$&$18.6 \pm 0.1$&$-0.40 \pm 0.18$&$0.38 \pm 0.11$&$0.31 \pm 0.07$&$0.83 \pm 0.10$&$1.97 \pm 0.08$&$0.77 \pm 0.15$\\
        &$60$&$17.5 \pm 0.2$&$-0.32 \pm 0.24$&$0.34 \pm 0.19$&$0.44 \pm 0.07$&$0.88 \pm 0.10$&&\\\hline
        UM417&$23$&$19.8 \pm 0.1$&$-0.85 \pm 0.15$&$0.80 \pm 0.16$&$-0.70 \pm 0.09$&$-1.55 \pm 0.09$&$-0.29 \pm 0.17$&$0.71 \pm 0.25$\\
        &&&&&&&&\\\hline
        UM448&$78$&$14.4 \pm 0.2$&&$0.47 \pm 0.23$&$0.20 \pm 0.08$&$0.53 \pm 0.07$&$2.43 \pm 0.13$&\\
        &$88$&$14.3 \pm 0.1$&&$0.46 \pm 0.10$&$0.24 \pm 0.06$&$0.59 \pm 0.06$&$2.50 \pm 0.20$&\\\hline
        UM619&&&&&&&&\\
        &$60$&$16.3 \pm 0.3$&$0.08 \pm 0.28$&$0.20 \pm 0.25$&$-0.36 \pm 0.07$&$0.21 \pm 0.07$&$0.77 \pm 0.10$&$-0.22 \pm 0.17$\\\hline
        \hline
      \end{tabular}
      \medskip
      ~\\
      $\dagger$ -- $V$ instead of $B$ filter\\
      $\ddagger$ -- $B$ instead of $V$ filter
    \end{minipage}
  \end{table*}
\end{center}
\section[]{Characteristics of individual galaxies}\protect\label{individ}
\noindent We will adhere to the morphological classification of~\citet[][LT86]{1986sfdg.conf...73L} and the~\citet{1989ApJS...70..479S} classification based on spectral features, and either reiterate the BCG class where available from LT86,~\citet{2001ApJS..133..321C} or~\citet{2003ApJS..147...29G} or, for the galaxies in our sample without published classification, we will obtain a morphological classification by analyzing the contour and RGB plots for our data. The LT86 classes we apply are \emph{iE} (irregular inner and elliptical outer isophotes), \emph{nE} (central nucleus in an elliptical host), \emph{iI,C} (off--center nucleus in a cometary host), and \emph{iI,M} (off--center nucleus in an apparent merger). The~\citet{1989ApJS...70..479S} classes, available only for UM galaxies, are \emph{DANS} (dwarf amorphous nuclear starburst galaxies), \emph{Starburst nucleus galaxies}, \emph{H\texttt{II}H} (\HII hotspot galaxies), \emph{DH\texttt{II}H} (dwarf \HII hotspot galaxies), and \emph{SS} (Sargent--Searle objects). \\

\noindent Unless otherwise specified, the isophotal and elliptical surface brightness profiles agree very well. For some targets the isophotal profile is slightly fainter at large radii than the elliptical one. This is a natural consequence of the galaxy having a significant nebular component, but which is not dominating the luminosity at every radius. For the rest of the galaxies there is no discernible difference between the two profiles. This happens in the two extreme cases where either the dominance of the nebular emission is complete also at larger radii, or there is very little nebular emission in the galaxy. Depending on individual morphologies the centers of integration may be identical or vary in physical location, which we will mention in the cases where it has any bearing on the analysis. We interpret the color radial profiles using the~\citet{2008A&A...482..883M} stellar evolutionary tracks for a Salpeter IMF stellar population with metallicities $Z=0.001,~0.004,~0.02,~0.2$. These tracks are predominantly used when comparing with the old host population. Here we will briefly mention the main guidelines such evolutionary tracks can offer. Some combinations of colors, e.g. $B-V$ vs. $V-I$, suffer from a severe age--metallicity degeneracy, getting simultaneously redder with increased age and with increased metallicity in a fashion which often makes it impossible to distinguish between the two. For old populations ($\gtrsim1$ Gyr) combinations involving $V-K$ are useful in breaking the metallicity degeneracy because $V-K$ is very weakly dependent on age but has a strong $Z$ dependence, getting redder with increased $Z$. For all color combinations there are limiting cases which can sometimes be useful to indicate $Z$ and age with a fair certainty, e.g. $V-I>1.1$ always indicates a high $Z$ (close to solar or above) and an age older than $5$ Gyr. Similar limits exist for the rest. $H-K$ is the only color that  beyond $1$ Gyr gets bluer with increased age for any $Z$ but the tendency is weaker for solar and super--solar $Z$.\\

\noindent In all cases nebular emission and the presence of dust modify the observed colors, e.g.~\citet{2003APJ...588..281H} find that $H-K$ can redden to $\sim0.5$, while $V-I$ can get~$\sim0.8$ mag bluer if the region is dominated by nebular emission. Since BCGs are anticipated to have a significant nebular component, we also look at stellar evolutionary tracks with added nebular emission. These models are based on the \emph{Yggdrasil} spectral synthesis code~\citep{2011ApJ...740...13Z} with \emph{Starburst99} Padova-AGB stellar population spectra~\citep{1999ApJS..123....3L,2005ApJ...621..695V} at metallicities $Z=0.0004$, $Z=0.004$, $Z=0.008$ and $z=0.020$. The nebular contribution to the overall spectral energy distribution is computed using the photoionization code \emph{Cloudy}~\citep{1998PASP..110..761F}, assuming a spherical geometry for the photoionized gas and a nebular metallicity identical to that of the stars. With the help of these two models we carry out a separate analysis for each galaxy and their respective total, central (down to $\mu_B\sim24$ mag arcsec${}^{-2}$), and burst colors, as well as the colors of the $\mu_B=24-26$ and $\mu_B=26-28$ mag arcsec${}^{-2}$ host regions. The total number of separate color--color diagrams we have analyzed is $\sim450$, and we do not present those due to space considerations, but the plots are available on demand. From the individual analysis of the color--color diagrams we will attempt to give indications of the age and metallicity of both the underlying host and the young populations for each galaxy. An example color--color diagram can be seen in Figure~\ref{SEMs} for the whole sample. \\

\noindent We have briefly looked at $H\alpha$ equivalent width radial profiles for eight of the galaxies in the sample, however, detailed analysis of $H\alpha$ data is beyond the scope of this paper. Here we only note that if nebular emission is detected at large radii, then there is a tendency for the equivalent width to be significantly large ($\gtrsim400$--$\sim2000$\AA) at all radii, meaning the colors in every radial bin we study will have nebular emission contribution. For galaxies with central star forming regions the effect of the nebular component is negligible at large radii ($\mu_B=26-28$ mag arcsec${}^{-2}$). If nebular emission is indicated by the stellar evolutionary models even at these radii, the $H\alpha$ data show that it is in-situ for the galaxies we have looked at. \\

\noindent In our data the color profiles for most targets show a radial reddening which is likely to be mainly due to the increase of stellar age at large radii, unless otherwise specifically mentioned. In some cases we observe a flattening of the color profiles beyond some radius. This is expected in the case of a sustained low star--formation activity randomly percolating across the disk when there are no favored locations in terms of gas density~\citep{2001AJ....121.2003V}. In other words, such profiles indicate that we are sampling a population which is homogeneous in the sense that its different constituents have similar metallicities and ages. Note that this assumed homogeneity applies to the host, and not to the composite galaxy. The flattening of the color profiles is more easy to see in optical--NIR colors than in the purely optical, for which the host is usually sampled at much larger radii. In some cases the isophotal color profiles are better tracers of radial age or metallicity evolution than the elliptical integration ones because the irregular morphology of the targets will dilute any accumulative effect that the e.g. increasing age has on the profile shape. The effects of extinction occasionally show up in the region of the burst, typically reddening the innermost parts of the profile at and around the brightest isophotes. In the event that this is not observed, compact starburst knots would typically still give rise to strong fluctuations in the colors at small radii and those colors will be strongly affected by nebular emission. We will be more prolific in our analysis for galaxies which have been sparsely studied in the literature, or for which there is little or no analysis of the radial surface brightness and color distributions. 
\subsection*{ESO185--13}
\noindent This strongly starbursting galaxy~\citep[][$SFR=5.4~M_\odot/$yr]{2001A&A...374..800O} is likely the product of collision as evidenced by the detection of numerous bright compact star clusters~\citep{2000ASPC..211...63O} and by the galaxy morphology. Beyond $\mu_B\sim 26.0$ mag arcsec${}^{-2}$ the elliptical surface brightness profile reveals a prominent low surface brightness (LSB) component, which was previously undetected. In Appendix~\ref{eso185plume} we show how much detailed structure of this component we actually detect. Clearly visible in all optical filters is the flux overdensity to the North--East of the prominent tail. This is not seen in the NIR data, which is not deep enough to reach such large radii. In the NIR we only detect a single exponential disk (with no central excess) down to $\mu_{H,K}=23$ arcsec${}^{-2}$ at $r\sim15$ arcsec, which is where the $B$ band profile break occurs. We classify this galaxy as \emph{iI,M} as it is an apparent merger. The color--color diagrams indicate the presence of an old population ($\gtrsim2$ Gyr), as well as strong nebular emission contamination for the star--forming regions, which drops considerably beyond $\mu_B>24$ and becomes negligible for $\mu_B>26$ mag arcsec${}^{-2}$. While we were unable to constrain the age of the starburst in this case, it is quite young, as indicated by the $3.5$ Myr old peak of star cluster formation~\citep{2011MNRAS.414.1793A}.
\subsection*{ESO249--31}
\noindent This peculiar spiral with extended arms shows clear signs of an ongoing merger event~\citep[][]{2003A&A...405...31B}. We thus classify it as~\emph{iI,M} BCG. The profiles should be affected by a significant amount of extinction~\citep[][$H_\alpha/H_\beta\sim4.1$]{2003A&A...405...31B}. The star--forming regions are spread out at various radii resulting in many small scale fluctuations in the optical color profiles but no clear large--scale gradients. In hybrid and purely NIR colors, both the radial color profiles and the integrated colors in color--color diagrams show a negative radial metallicity gradient. The presence of an old stellar population $>2$ Gyr with low or intermediate metallicity ($Z<0.008$) is indicated, as well as a younger population with metallicity closer to solar ($0.008\lesssim Z\lesssim0.02$). 
\subsection*{ESO338--I04}	
\noindent This Wolf-Rayet galaxy~\citep{1999A&AS..136...35S} has been thoroughly studied in the optical broadband and narrowband~\citep[e.g.][]{1984Msngr..38....2B,1986A&AS...64..469B,1985A&A...146..269B,2002A&A...390..891B}. It has undergone previous and rather recent star--formation events, as evidenced by its rich population of intermediate--age globular clusters~\citep{1998A&A...335...85O}, which are unresolved in our images. Based on its morphology,~\citet{2003ApJS..147...29G} classify it as \emph{iE} BCG, but its chaotic velocity field is significant indication that this galaxy is in fact a recent merger~\citep{2001A&A...374..800O}. It has a spectroscopically confirmed companion at $70$ kpc projected distance~\citep{1999A&AS..137..419O}. Strong nebular emission contamination is visible for all optical colors, but the effect is negligible for the old population region ($\mu_B\gtrsim26$ mag arcsec${}^{-2}$). Optical--NIR and pure NIR colors detect the presence of a young $<100$ Myr stellar population. Note that there is a foreground star South--West of the central star-forming region, which we have masked out before any integrated colors or profiles were obtained.
\subsection*{ESO400--G043}
\noindent There are two prominent star--forming regions in this galaxy, both of which are off-center. We classify it as \emph{iE} BCG due to its regular elliptical outer isophotes. It has a \emph{dI} companion at a projected distance of $70$ kpc~\citep{1988Natur.331..589B} with a comparable \HI mass.~\citet{1984Msngr..38....2B,1986A&AS...64..469B} and~\citet{2002A&A...390..891B} demonstrate the presence of an old stellar population. We cannot completely isolate this population in any of the investigated regions, though the measured $V-I\gtrsim0.5$ at large radii would require stars $\gtrsim1$ Gyr of age. Additionally, the strong negative $H-K$ gradient we observe indicates a radial decrease in metallicity and/or an increase in age. The color--color diagrams indicate that the integrated colors for all regions contain young and old stars, mixed with gas. Specifically, contribution from gas is clearly visible in $V-K$ which is too blue to be coming solely from stars. The dust content in this galaxy is low~\citep{2002A&A...390..891B}.
\subsection*{ESO421--02}
\noindent Together with ESO462--IG 020, this galaxy was first studied by~\citet{1986A&AS...64..469B}. It shows a spatially extended starburst with multiple nuclei, embedded in a common elliptical host. We therefore classify it as \emph{iE} BCG. Neither of our population synthesis models fits the observed colors for the burst or the two outer regions. This is probably due to contribution from nebular emission which is not in--situ, which the models do not account for. The color profiles are predominantly flat. $U-B$ shows gradients, but this color is very sensitive to an aging young population and would reflect such aging by changing on shorter time scales than the redder colors. Flat color profiles, minor local color variations notwithstanding, are a signature of a percolating SF at a low but persistent rate~\citep{2001AJ....121.2003V}. However, the RGB image suggests that this is a merger since two very similar--looking dwarf galaxies are seen in physical contact with each other, with the starburst located between them along the region of contact. This is further supported by unpublished kinematic data, which give a clear indication that this is indeed an on--going merger between two dwarfs, one to the East, the other to the West. The Eastern part is falling into the Western part because the too regular (blue) velocity curve of the latter, implies that it could not have suffered a fly--through of the Eastern dwarf (which has a red velocity curve). 
\subsection*{ESO462--IG 020}
\noindent This galaxy has been classified as a distorted contact pair in the surface photometry catalog of ESO--Uppsala galaxies. We see indications of the ongoing merger in the optical and optical--NIR isophotal color profiles, which show sharp color changes visible as discontinuity jumps at $r\sim1, 2,$ and $4$ kpc. These jumps in color imply the presence of more than two distinct stellar populations with different SFH which have mixed spatially on large but not on small scales. We therefore classify this galaxy as \emph{iI,M} BCG. The color--color diagrams indicate that perhaps more than two populations are present and influencing the integrated colors. Nebular emission is implied, but the current SF is poorly modeled with an instantaneous burst. The only definite conclusion we can draw from the color--color diagrams for this galaxy is that a population older than $1$ Gyr is present, but we cannot narrow down its metallicity. There is a bright nearby star to the North--East of the galaxy which could be contaminating the galaxy light beyond $\mu_B\sim26$ mag arcsec${}^{-2}$.
\subsection*{HE2--10 (ESO495-G021)}
\noindent He2--10 is the first Wolf--Rayet galaxy ever discovered~\citep{1976MNRAS.177...91A}. Contrary to popular belief this is not a dwarf as it is too luminous ($M_B=-19.1,~M_K=-22.3$ mag) and has an extended regular envelope spanning $\sim13.8$ kpc. The observed $V-K$ color profile is centrally very red, reflecting both a higher $Z$, measured to be close to solar by~\citet{1999ApJ...514..544K}, and a larger internal extinction. There is dust clearly visible in the RGB image (dark lanes in the North and South directions). The negative $V-K$ gradient is thus due to a strong metallicity and extinction decrease with radius. In fact we can model the regular envelop surrounding the starburst with a metal--poor stellar population of $Z\sim0.004$ - a sharp decrease from $Z_\odot$. The $H-K$ profile on the other hand is essentially flat beyond $\sim15$ arcsec. It is also too red at any radius for any age and metallicity, as is often the case for Wolf-Rayet galaxies~\citep{2005APJS..157...30B}. Its mean value of $0.5$ requires a large contribution from nebular continuum. If we are sampling a population older than $\sim4.5$ Gyr beyond the central $1.5$ kpc then the expected radial decrease in $Z$, gas density, and dust content, together with the simultaneous increase in age will all enhance each other to produce a strong negative gradient. The flatness we are seeing must therefore be due to an old homogeneous population, or to a population of an intermediate or younger age, which is well--mixed. The regular contours of the outer isophotes support the former case. We classify this galaxy as \emph{iE} BCG. 
\subsection*{HL 293B (SDSS J223036.79-000636.9)}
\noindent This faint dwarf is a Haro-Luyten object, quasi-stellar in appearance. It was first classified as a BCG by~\citet{1981ApJ...247..823T}, which was later refined to \emph{nE} BCG by~\citet{2001ApJS..133..321C}. This galaxy is currently undergoing a starburst only a few Myr old, as evidenced by the detection of luminous blue variable stars~\citep{2009ApJ...690.1797I}. It seems to have a lot less NIR than optical flux, even though the exposure times are reasonably large ($42$ minutes in $K$). The NIR flux is coming from a very compact region spatially coextensive with the burst nucleus.~\citet{2001ApJS..133..321C} similarly do not find an extended underlying component in this galaxy. There is nevertheless an old stellar population present, although it does not appear to be spatially dominant. While all other colors show signs of strong contamination by nebular emission, $V-K\gtrsim2.0$ at $V-I\gtrsim0.5$ would require the presence of old stars. 
\subsection*{IIZw40}
\noindent This \emph{iI,M} BCG galaxy is very blue and strongly starbursting~\citep{1996ApJ...466..150V}. Its optical and NIR spectra have been extensively studied~\citep[e.g.][]{1982MNRAS.198..535B,1983ApJ...268..667T}, as well as its morphology and companions~\citep[e.g.][]{1991A&A...241..358C}, \HI gas content~\citep[e.g.][]{1992A&A...265...19S} and metallicity~\citep[e.g.][]{1983ApJ...273...81K,1991A&AS...91..285T,1994MNRAS.270...35M}. IIZw40 is the result of a merger of two dwarf galaxies, judging by its optical morphology~\citep[e.g.][]{1970ApJ...162L.155S,1997MNRAS.286..183T} and its extreme \HI tidal tail extending South--East approximately $8.5'$ in diameter~\citep{1998AJ....116.1186V} which does not seem to have an optical counterpart. The central colors are consistent with a very young stellar population of a few Myr, while at large radii all colors are strongly contaminated by nebular emission. The dust mass found in this galaxy is rather large~\citep[][$M_{dust}\approx10^{6}M_\odot$]{1998ApJ...496..145L}, which is comparable to the stellar mass~\citep[e.g.][$M_\star\approx2.5\times10^{6}M_\odot$]{1996ApJ...466..150V}, so the reddening of the isophotal $U-B$ in the starburst regions is likely the effect of internal extinction. \\

\noindent There is no radial range in our data at which we can sample a region dominated by an old population. Since IIZw40 has such low Galactic latitude it suffers great extinction ($3.5$ mag in B) and the otherwise quite long exposure times of e.g. $40$ minutes in the B band only allow us to reach $\mu_B\sim25.0$ mag arcsec${}^{-2}$. Judging from the rest of our sample, the surface brightness profile in the region around $\mu_B\sim 25$ mag arcsec${}^{-2}$ is still very much contaminated by the burst, especially if the galaxy has disturbed morphology. It is therefore entirely possible that we are only seeing the current young starburst and, at large radii, the well-mixed remnants of some intermediate--age population from a previous burst. There could be an old host dominant beyond $\mu_B=26$ mag arcsec${}^{-2}$, as is the case for e.g. ESO185--13 and UM448, and this would be entirely invisible to us.
\subsection*{MK600}
\noindent At first glance this \emph{iE} BCG has a double exponential disk profile, with a faint population which dominates the surface brightness profile beyond $\mu_B\sim26$ mag arcsec${}^{-2}$. Previous studies of this galaxy do not mention such a structure~\citep{2006ApJS..162...49H}. Our B band profile is the deepest to date, however, MK600 is situated between three very bright stars and the second exponential disk component could be due to contamination from these stars. This does indeed seem to be the case judging by the deep contour image in Appendix~\ref{genappednix}. We have applied heavy masking of these stars. The profile data beyond $\mu_B\sim26$ mag arcsec${}^{-2}$ are based on a narrow unmasked cone to the South--West, with minimum cone radius $\sim1'$ at $\mu_B\sim26$ mag arcsec${}^{-2}$. As this area is seemingly unaffected by the stars, this structure could still be real, but we remain suspicious of its existence and will not dwell on it further. In the $H$ and $K$ bands we see a single exponential disk profile down to $\mu_{H,K}=23$ mag arcsec${}^{-2}$ at radius $r\sim25$ arcsec, with no indication of a central excess.~\citet{1990A&A...233..348A} find strong evidence against a single starburst episode in MK600, and favor instead a mode of sustained star formation over $\sim100$ Myr. We find some supporting evidence for this since the color--color diagrams strongly indicate that the current star formation is poorly modeled with an instantaneous burst. The galaxy is a faint IRAS source~\citep{1995ApJS...98..369B}, with low extinction~\citep[][$E(B-V)=0.05$]{2004AJ....128.2170H}, and indeed there does not appear to be a lot of dust obscuring the young blue regions, as evidenced by the location of the burst colors in a $U-B$ or $B-V$ vs. $H-K$ diagrams, which are both consistent with negligible amounts of dust. Nebular emission contribution is visible when comparing $B-V$ with $U-B$, $V-R$, $V-I$, or $V-K$ at all radii, though the effect is considerably smaller beyond $20$ arcsec.  
\subsection*{MK900}
\noindent The star forming regions in this \emph{nE} BCG seem to follow the semi--minor axis of the underlying galaxy and are found symmetrically on both sides of the geometric center. All color--color diagrams ubiquitously indicate the presence of a homogeneous old ($\gtrsim4$ Gyr) metal--poor ($Z\lesssim0.004$) host. In this case the instantaneus burst assumption fits the observed colors very well. $U-B$ shows no extinction even in the SF regions, which is to be expected as this is another faint IRAS source~\citep{1995ApJS...98..369B}. Contamination from nebular emission is confined to the star formation region, with its effect on the colors being negligible beyond the central $10$ arcsec. This is consistent with the galaxy being moderately gas--rich~\citep[][$M_{\HI+He}/M_{\star}=0.34$]{2010ApJ...708..841W}. The nuclear star formation has an $H_\alpha$ luminosity comparable to that of 30 Doradus~\citep{2000AJ....119.2757V}.
\subsection*{MK930}
\noindent This is a Wolf-Rayet galaxy~\citep{1999A&AS..136...35S} whose multiple nuclei~\citep[e.g.][]{2004AJ....128...62G} indicate the advanced stages of a merger. We therefore classify it as an \emph{iI,M} BCG. The presence of an old ($\gtrsim4$ Gyr) stellar population can be deduced from all color profiles. Nebular emission is implied by all color--color diagrams, but at the same time neither of the models is well--suited for the burst, total, or central mixed colors, indicating either a prolonged burst, or severe dust effects. The latter seems more likely since non-negligible amounts of warm dust are present in this galaxy~\citep{2008ApJ...676..970W}. We were unable to impose useful constraints on the age of the young population, however,~\citet{2011MNRAS.415.2388A} place it at $\lesssim10$ Myr based on the ages of super star clusters. Note that the flattening of the elliptical B band profile beyond $\mu_B\sim27$ mag arcsec${}^{-2}$ is due to the sky, and is not an indication of an additional underlying component. 
\subsection*{MK996}
\noindent This \emph{nE} BCG is also a Wolf-Rayet galaxy with a compact off--center star formation region. It is old, as evidenced by~\citet{1996ApJ...463..120T} who find an old globular system ($\sim10$ Gyr) distributed around the galaxy. All of our data are consistent with an old ($>3$ Gyr) metal--poor ($Z\lesssim0.004$) stellar population with moderate contamination of nebular emission predominantly visible in $U-B$, $B-V$, and $H-K$ diagrams, which is negligible outside the compact star formation region. The presence of a dust lane has previously been confirmed by both~\citet{1996ApJ...463..120T} and~\citet{1999A&AS..138..213D}, and is also visible in our RGB image.
\subsection*{SBS 0335--052 E}\protect\label{sbsE}
\noindent The K band exposure time for this galaxy is the longest one in the entire sample ($\sim5$ hours with NOTCAM), yet we detect very little K band flux. The surface brightness quickly drops to $\mu_K\sim22.0$ mag arcsec${}^{-2}$ within the central $2$ kpc ($r\approx4$ arcsec), beyond which radius the errors become too large ($\sigma\gtrsim1$ mag). We performed a sanity check using the $\sim2$ hour SOFI K band image, but the result was the same and we could only reach $\mu_K\sim22.0$ mag arcsec${}^{-2}$. The K band flux is concentrated in the compact regular elliptical region visible in the contour plot, with no measurable flux beyond this region. Considering this it is difficult to speak of a host galaxy in the traditional sense of the term since the B band, which contains contributions from both stars and gas, seems to be much more extended than the K band, which is considered a good proxy for stellar mass. The radius $r\sim2.5$ arcsec marks the ``end'' of the compact K band flux region. The abrupt onset of this change testifies that the two regions contain populations with very different physical properties. A very strong contribution by nebular emission is necessary to reconcile the observed $H-K$ colors with $B-I$ or $B-K$, the latter two colors indicating a very young population of a few $10$ Myr. This is consistent with the $\sim10$ Myr age of super star clusters found by~\citet{2010ApJ...725.1620A}. Indeed, researchers have argued that this is a primordial galaxy~\citep[e.g.][]{1997ApJ...476..698I,1997ApJ...477..661T,1998A&A...338...43P,2000A&A...363..493V} with the estimated age of the extended component dramatically decreasing from $\sim100$ Myr~\citep[e.g.][]{1997ApJ...476..698I} to $5$-$6$ Myr in more recent research~\citep{2000A&A...363..493V} (see, however,~\citealt{2001A&A...371..429O} for a contrary opinion). It also has surprisingly large amounts of hot dust~\citep[][$3$-$5\times10^5M_\odot$]{1999ApJ...516..783T,2000A&A...363..493V}.
\subsection*{SBS 0335--052 W}
\noindent This galaxy was first discovered as a peak in the \HI maps of the companion galaxy.~\citet{1997AstL...23..308P} detected an optical counterpart, and~\citet{2005ApJ...632..210I} measured its record--low metallicity, $\sim10\%$ lower than the previous record--holder I Zw 18. This galaxy is in the same field of view as SBS 0335--052 E, so the same remarks on K band exposure time apply here. Similarly, we find very little K band flux, again located only in the small compact region visible in the contour plot. The K band profile reaches $\mu_K\sim22.5$ mag arcsec${}^{-2}$ but we can barely resolve the galaxy in this filter -- there are only 3 reliable data points in the profile. Due to this we shall not attempt to interpret the $B-K$ and $H-K$ color profiles. The $B-I$ profile behaves similarly to SBS 0335--052 E, with a pronounced blue minimum at $r\sim1.5$ arcsec, which, again, marks the ``end'' of the compact bright region seen in the contour plot.
\subsection*{Tol 0341--407}
\noindent The blue star--forming regions in the East and West are two separate galaxies with distinctly different photometric and kinematic properties~\citep{1999A&AS..137..419O}, hence we classify this merger as \emph{iI,M}. Due to the very irregular morphology of the galaxy the isophotal and elliptical centers of integration are very different and the two surface brightness profiles do not trace each other at any radius. The elliptical color profiles start at the center of mass (using the center of light as proxy), which happens to be close to the geometric center due to the axisymmetric placement of the two starburst regions. Judging by $B-V$ or $U-B$ vs $V-I$ diagrams, the central region is completely dominated by a population younger than $1$ Gyr, while beyond $\mu_B\sim26$ mag arcsec${}^{-2}$ we detect another population, $\gtrsim5$ Gyr. Thus the color--color diagrams are consistent with the existence of three separate stellar populations in this galaxy, as already suggested by~\citet{2004A&A...423..133W}. Strong nebular emission is contaminating the colors of the two starbursting regions in $U-B$ vs $B-V$ and $B-V$ vs $V-K$ diagrams. 
\subsection*{Tol 1457--262 \emph{I}}
\noindent This is a merger between two distinct galaxies, Tol1457--262a (North) and Tol1457--262b (South). We thus classify it as \emph{iI,M}. The lack of $U$, $I$ and $H$ filters prevents us from imposing any kind of limit on gas content, stellar age or metallicity. The burst region is spatially dominating and shows strong nebular emission contamination in a $B-V$ vs. $V-K$ diagram. The $\mu_B=24$--$26$ mag arcsec${}^{-2}$ region shows an older metal--rich ($\sim Z_\odot$) population, however we cannot confirm this with only three filters. Note that here the $K$ band is suspect in the faint regions, since upon heavily smoothing the $K$ band data to closer examine the sky we detected a residual background structure, which is not additive in nature, and clearly indicates an illumination/flatfielding problem. This residual structure completely dominates the $K$ band profile beyond $\sim18$ arcsec for Tol 1457--262 \emph{I} and beyond $\sim35$ arcsec for Tol 1457--262 \emph{II}.
\subsection*{Tol 1457--262 \emph{II}}
\noindent At first glance it looks deceptively easy to fit this object with elliptical isophotes, however, the isophotal and elliptical surface brightness profiles are very different from each other, though with similar centers of integration. The reason for this discrepancy becomes obvious once we compare the distances to the different components of this object. Tol1457--262c (roughly, the Eastern region) and Tol1457--262e (the Western region) have a tabulated \emph{NED} velocity difference of $\sim600$ km s${}^{-1}$. In other words, these are two different galaxies superposed along the line of sight with a velocity different large enough to make it unlikely that they are interacting with each other since this is significantly higher than the limit for Hickson compact groups of galaxies~\citep{1997ARA&A..35..357H}. The galaxies are superposed in a fashion which makes it impossible for us to separate them using just broadband imaging data. As a consequence, both surface brightness profiles are sampling a mixture of stellar populations at different redshifts. 
\subsection*{UM133}
\noindent This is an \emph{iI,C} BCG, with no evidence of interaction~\citep{1997MNRAS.288...78T}. The elliptical color profiles are influenced by the off--center location of the strongest starburst knot.~\citet{2000A&A...361...33N} suggest that cometary BCGs like UM133 and UM417 could be an evolutionary link between BCGs and other dwarf galaxies because their structural parameters ($\mu_0$ and $h_R$) are intermediate between those inferred for iE/nE BCGs and the dIrrs/dEs. Color--color diagrams indicate moderate contamination from nebular emission. Due to the morphology of the galaxy we cannot give separate metallicity and age limits for the different populations, but we do detect at least two, one $\gtrsim2$ Gyr, and the other younger. 
\subsection*{UM160}
\noindent Numerous very blue star forming regions are spread out around the red center of this Wolf-Rayet~\citep{1999A&AS..136...35S} galaxy. The starbursting knots are tracing what looks like distorted spiral arms. This is a gas--rich ($\sim9\times10^9M_\odot$) galaxy with a low--mass ($\sim3\times10^8M_\odot$) dwarf companion $\sim1$ arcmin to the North~\citep{1990ApJ...364...23S}. The color profiles are consistent with a mixed stellar population contaminated by nebular emission but with no detectable dust effects. Due to the inside--out distribution of the burst regions the color--color diagrams show increasing nebular emission contamination at larger radii. The irregular morphology further prevents us from separating the burst from the older population, and we cannot confirm the presence of three distinct populations in this galaxy, as suggested by~\citet{2004A&A...423..133W}. Our population separation method rests on simplified assumptions which naturally do not hold for galaxies with peculiar morphology. We do observe a population of $\lesssim50$ Myr in the burst region and $\sim300$ Myr in the central red region, however the latter certainly suffers from contamination from the burst, and we cannot sample a purely old population dominated region at any radius. We classify this galaxy as \emph{iI,M} BCG.
\subsection*{UM238}
\noindent Previous observations have compared this galaxy to UM 133 because its West starburst knot was the only one visible, giving the galaxy a cometary appearance~\citep{1997MNRAS.286..183T}. Our contour and RGB plots indicate at least six such regions, which must be super star cluster complexes. The brightest one is to the West, as already observed by the previous authors, one is to the North, and the remaining four to the South--East and East of the geometric center. Both~\citet{2004A&A...423..133W} and~\citet{2006A&A...455..825C} find three distinct populations in this galaxy, young ($\sim6.8$ Myr), intermediate ($\sim10$ Myr) and old ($\gtrsim1$ Gyr), with the old one clearly dominating the total mass of the galaxy. Due to the morphology of the galaxy we detect a population $\gtrsim1$ Gyr in our color--color diagrams but are unable to impose any limits on the age of the star forming regions. There is an excess of $\sim0.2$ mag in all $H-K$ diagrams, which could be due to nebular continuum emission, however, we see only very mild nebular emission contribution in $B-V$ vs. $V-I$ or $V-K$ diagrams. We classify this galaxy as \emph{iI} BCG.
\subsection*{UM417}
\noindent This is an isolated \emph{iI,C} BCG in the sense that there are no detected companions within $1$ Mpc projected distance~\citep{1991A&A...241..358C}.~\citet{1989ApJS...70..479S} classify it as a \emph{SS} object. The color--color diagrams show a large contribution from nebular emission, though the effect is negligible outside of the starburst region. These diagrams also suggest that the burst is of intermediate ($Z\sim0.008$) or low ($Z\sim0.004$) metallicity, and no older than a few Myr, which makes it the youngest burst in the sample. We see no clear indication of dust in the same diagrams, which is consistent with the very low extinction values found in this galaxy by~\citet{1991A&AS...91..285T}, who measure $H_\alpha/H_\beta=3.71$. Further, the instantaneous burst assumption does not work well for this galaxy since the integrated colors for some regions are too red in $B-V$ for a given $V-I$ or $V-K$, which must indicate prolonged star formation since it cannot be due to dust. 
\subsection*{UM448}
\noindent The star formation in this Wolf-Rayet galaxy is occurring only in the spiral--like structure visible in the RGB plot since those are the only regions with detectable $H_\alpha$ emission~\citep{1997MNRAS.286..183T}.~\citet{1989ApJS...70..479S} classify it as a \emph{H\texttt{II}H} galaxy due to its spectral properties.~\citet{2003A&A...410..481N} classify it as {\em iI} based on NIR images. In our contour plot it is evident that beyond $\mu_B\sim25.5$ mag arcsec${}^{-2}$ the isophotes become regular and elliptical, albeit somewhat boxy. This would make an~\emph{iE} BCG classification more suitable. However, further examination of this faint structure shows that we can trace it out to an additional $\sim35$ arcsec beyond the $\mu_B\sim25.5$ isophote. Figure~\ref{um448plume} in Appendix~\ref{um448} demonstrates this, clearly showing a broad tidal arm to the North--West. It is visible in all optical filters. In the NIR the extended component as a whole is much less prominent, however, non-negligible signal is still detected from several areas. $U-B$ or $B-V$ vs. $V-I$ colors reveal nebular emission contamination from the star formation region, and an old ($\gtrsim5$ Gyr) metal--poor ($Z\lesssim0.001$) stellar population in the regular elliptical region beyond $\mu_B\sim25.5$ mag arcsec${}^{-2}$. We are lacking $U$ and $H$ band observations so we cannot get a feeling for the dust content, however,~\citet{1998ApJ...496..145L} obtain $M_{dust}=1.4\times10^4M_\odot$ within the optical disk of the galaxy, which is a factor of $\sim1000$ lower than the \HI content inside the same region($M_{\HI}=1.7\times10^7M_\odot$). The $V-K$ reddening we observe at small radii must then be primarily due to gas, rather than extinction. We can deduce the existence of three distinct populations from the behavior of the $V-R$ and $V-I$ profiles. The first population dominates the range $r=0$--$10$ arcsec, the second $r=10$--$25$ arcsec, and the third beyond $r\gtrsim25$ arcsec. This is consistent with the results for this galaxy of~\citet{2004A&A...423..133W}, who obtain a best fit of the SED with three distinct populations. \\
\subsection*{UM619}
\noindent This is a \emph{DANS} galaxy~\citep{1989ApJS...70..479S} with no detected companions~\citep{2001A&A...371..806N}. It has an \HI mass of $6.9\times10^{8}M_\odot$~\citep{2001AJ....122.2341I}. The contribution from nebular emission is moderate to negligible in the measured colors, judging by how well they are modeled by a pure stellar population. There is, however, a $\sim0.2$ mag excess in all $H-K$ diagrams coming from areas brighter than $\mu_B\sim24$ mag arcsec${}^{-2}$, likely due to nebular emission. The color--color diagrams consistently reveal an old population of $\gtrsim8$ Gyr. In fact, we do not measure integrated colors indicative of a population younger than $1$ Gyr for any color combination in any of our radial bins. The SFR is reasonably high~\citep[][$SFR=0.12 M_\odot/$yr]{2001AJ....122.2318B} but the star forming regions tracing the spiral arms are not dominant over the underlying host when the radial bins of integration are significantly larger than the angular diameter of the star--forming knots in the RGB image. We classify it as \emph{iE} BCG due to its regular outer isophotes.
\begin{figure}
\begin{center}
  \includegraphics[width=7cm,height=6cm]{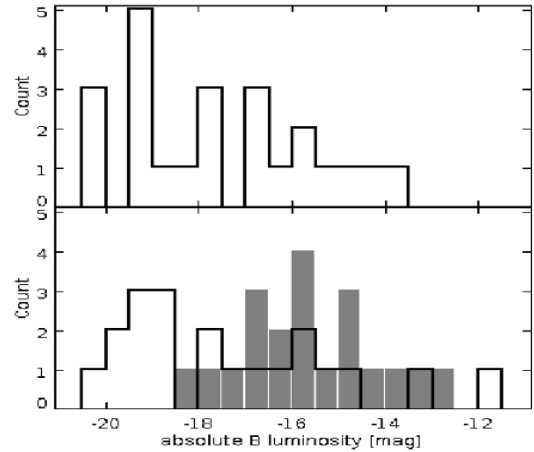}
\caption{Absolute $B$ magnitude distribution of the sample for the composite galaxy (upper panel), only the burst (black contour, lower panel), and the host galaxy (gray filled spikes, lower panel). }\protect\label{maghist}
\end{center}
\end{figure}
\begin{figure}
\begin{center}
  \includegraphics[width=8cm,height=6cm]{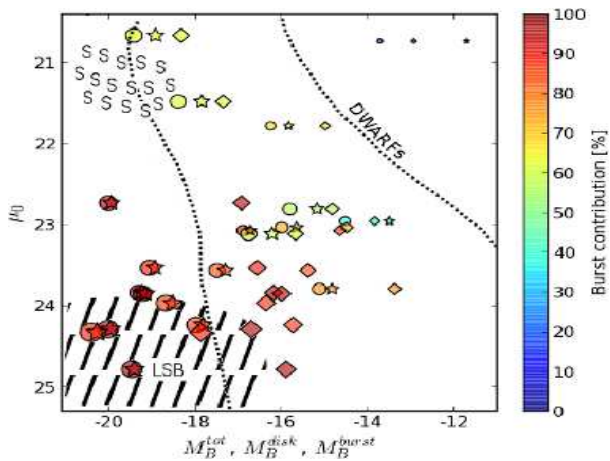}
\caption{Binggeli diagram: Central surface brightness of the host vs. the absolute $B$ magnitude of the whole galaxy (circle), the burst (star), and the host (diamond). The size of the markers reflects the scale length of the host. Markers are color--coded by the percentage of total light that is attributed to the burst. The approximate locations of dwarfs (the area between the two dotted lines), spirals (S--shaped marker), and LSB galaxies (hatched area) are marked for convenience.}\protect\label{binggeli}
\end{center}
\end{figure}
\begin{figure}
\begin{center}
  \includegraphics[width=8cm,height=10cm]{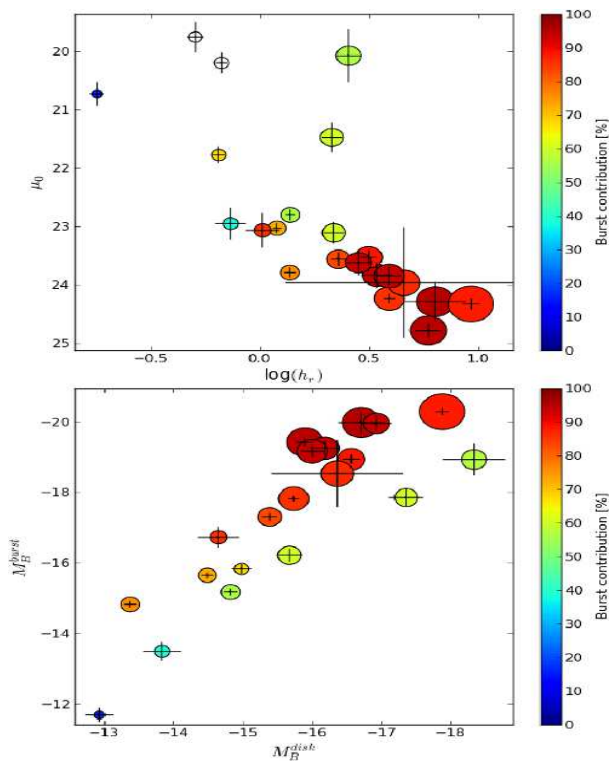}
\caption{Upper panel: $B$--band host $\mu_0$ in mag arcsec${}^{-2}$ vs. $\log_{10}{h_r}$ in kpc. For reasons explained in \S~\ref{clrclrdiags} we use the parameters obtained from the $\mu_B=26$--$28$ mag arcsec${}^{-2}$ region. Lower panel: Absolute magnitude of the burst vs. absolute magnitude of the (assumed exponential) host. Marker size in both panels reflects the scale length. The colors indicate the relative burst contribution to the total (burst+host) luminosity of each target. Errorbars are overplotted on each marker.}\protect\label{disk_burst_hr}
\end{center}
\end{figure}
\section[]{Results and discussion}\protect\label{discuss}
\noindent Before we delve into the discussion we should mention that the total apparent magnitudes in Table~\ref{totlumtbl} are consistent with tabulated photometry measurements on NED. The largest offsets between our measurements and other authors are $\sim0.2$ mag for IIZw40, MK600, and Tol0341--407. This difference is independent of the assumed distance to the targets, and is likely due to the small variations in Galactic extinction corrections applied by the individual authors, and the fact that the used $B$ filter is not identical, and can on average result in differences of $\sim0.15$ in integrated magnitude. In the case of MK600 the three bright nearby stars may have influenced our total luminosity measurement. 
\subsection{Notes on isophotal profiles}\protect\label{isoproblems}
\noindent The weakness of purely isophotal profiles is that they become unreliable at fainter isophotes much sooner than in the case of elliptical integration. The way we have chosen to obtain the isophotal profiles produces a systematic effect clearly visible in all radial surface brightness and color profiles involving the reference filter, usually B, or V if B is missing. The isophotal profile in the reference filter becomes systematically brighter as one reaches deeper into the sky, then becomes fainter again as the sky residual peak is passed, and eventually starts randomly fluctuating as the pixel to pixel sky statistics start dominating at the faintest levels. The profiles are truncated before the latter happens, as the errors become too large beyond $28~B$ magnitudes for all targets. There is no way, however, to prevent bright sky pixels in the range $\sim24.5$--$28~B$ magnitudes from entering the respective magnitude bins, as there is nothing, other than spatial position on the image, to identify them as sky instead of galaxy. Naturally, this magnitude range is approximate and varies with the depth of the observations, but for the vast majority of our images the bulk of the sky lies in the interval $\sim24.5$--$28~B$ magnitudes, with the peak at $\sim26~B$ magnitudes. To reduce this effect, the images used for isophotal profiles were masked out as close to the galaxy as possible, without risking removing too much of the fainter galaxy isophotes. This, however, has another, equally undesirable effect of systematically biasing the size of the equivalent radius -- beyond a certain surface brightness (usually $\mu_B\gtrsim26$ mag arcsec${}^{-2}$) each new radius artificially decreases as the consecutive isophotes run out of pixels, which causes the faint end of the profile to decrease steeply in an unphysical fashion. Both of these effects are much more pronounced in surface brightness or color profiles involving the reference filter because the areas defined by each magnitude bin were tailored to those frames, and hence the effect of the sky is systematic, accumulative and pronounced. The rest of the filters, which use isophotal areas defined by the reference, have their sky brightness peaks at different wavelengths from the reference filter and hence the effect is diluted, albeit still systematic, and is completely undetectable in color combinations which do not involve the reference filter. \\

\noindent For the majority of our targets the isophotal profiles trace the elliptical integration profiles fairly accurately. If the slopes (scale lengths) of these two profiles agree then that is a good indication that the star formation has not changed the galaxy morphology beyond recognition, i.e. that the host galaxy has not been completely transformed by the presence of the star forming regions, and it retains its structural properties ($\mu_0$ and $h_r$) even after the star formation has been switched on.
\subsection{General properties of the sample}\protect\label{generaldiss}
\noindent A dwarf galaxy has a total luminosity fainter than $-18~B$ magnitudes, however, this magnitude cut--off varies from author to author and has no physical basis. The general consensus is that galaxies fainter than this are guaranteed to be dwarfs. With this cut--off approximately half of our sample consists of luminous BCGs (Figure~\ref{maghist}, upper panel), while the other half are blue compact dwarfs (BCDs). We refer to both types as BCGs. Separating the ``burst'' and ``host'' populations in each target, as described in \S~\ref{strucparam}, reveals that the total luminosity of galaxies brighter than $\sim-18$ is completely burst dominated and the hosts of all targets, except the discordant redshift system Tol1457--262\emph{II}, have total luminosities consistent with being dwarfs, with $M_B^{disk}\gtrsim-18$ (Figure~\ref{maghist}, lower panel). This is also seen in the Binggeli diagram in Figure~\ref{binggeli}, where the hosts of the more massive galaxies are clearly located in the indicated dwarf--occupied area, with their corresponding burst luminosities constituting more than $80\%$ of the total galaxy luminosities. In fact, the burst makes up for the dominant percentage of the total light output for most targets. Only for HL293B and SBS0335--052West, does the host constitute the dominant luminosity contribution. These are, however, very compact and faint, so it is unclear from the surface brightness profiles in Figure~\ref{datafig} if we can sample the old population at all, even at large radii. Figure~\ref{binggeli} further shows that, with diagonally decreasing host scale length from left to right, it is the most extended LSB dwarfs that host the most dominant bursts in relative terms and the brightest bursts in absolute terms. This can also be seen in Figure~\ref{disk_burst_hr}. If all hosts with $\mu_0\gtrsim23$ mag arcsec${}^{-2}$ had dominating relative burst contribution regardless of $h_r$ then this would simply be due to the fact that the contrast between the bright burst and the fainter host is much more stark if the latter is a LSB galaxy. As such a relation seems to be simultaneously $h_r$-- and $\mu_0$--dependent, however, this is not a trivial scaling correlation. The more extended the host, the lower the central surface brightness (Figure~\ref{disk_burst_hr}, upper panel) and while this also usually means the host is brighter than the more compact objects, it is not the brightest host that has the brightest burst in absolute terms and it is not the most extended host that has the highest burst contribution in relative terms (Figure~\ref{disk_burst_hr}, lower panel). We explain this in terms of star formation efficiency. When they are not undergoing a starburst episode, the more extended dwarf hosts spend most of their lives forming stars at low efficiencies and hence they retain a larger \HI reservoir. Once the starburst is induced this larger available gas mass contributes to a more vigorous star formation. In contrast, compact higher surface brightness hosts have higher star formation efficiencies when outside of starburst mode and thus would end up with smaller \HI gas masses, i.e. they are more evolved systems. The lower panel of Figure~\ref{disk_burst_hr} further suggests that there are lines of constant relative burst contribution which seem to be predominantly defined by the similar scale length of the galaxies, and along which both the burst and host luminosities can vary in a linear fashion. The sample size is rather small, and we can only clearly detect the $\sim60\%$ (yellow--green) and $\sim85\%$ (red--orange) lines. There are very few targets with burst contribution $<40\%$ and hence no low burst contribution lines, however, this is a selection effect -- the galaxies of this sample were predominantly chosen to be bright starbursts. \\

\begin{figure}
  \begin{center}
    \includegraphics[width=7.5cm,height=6cm]{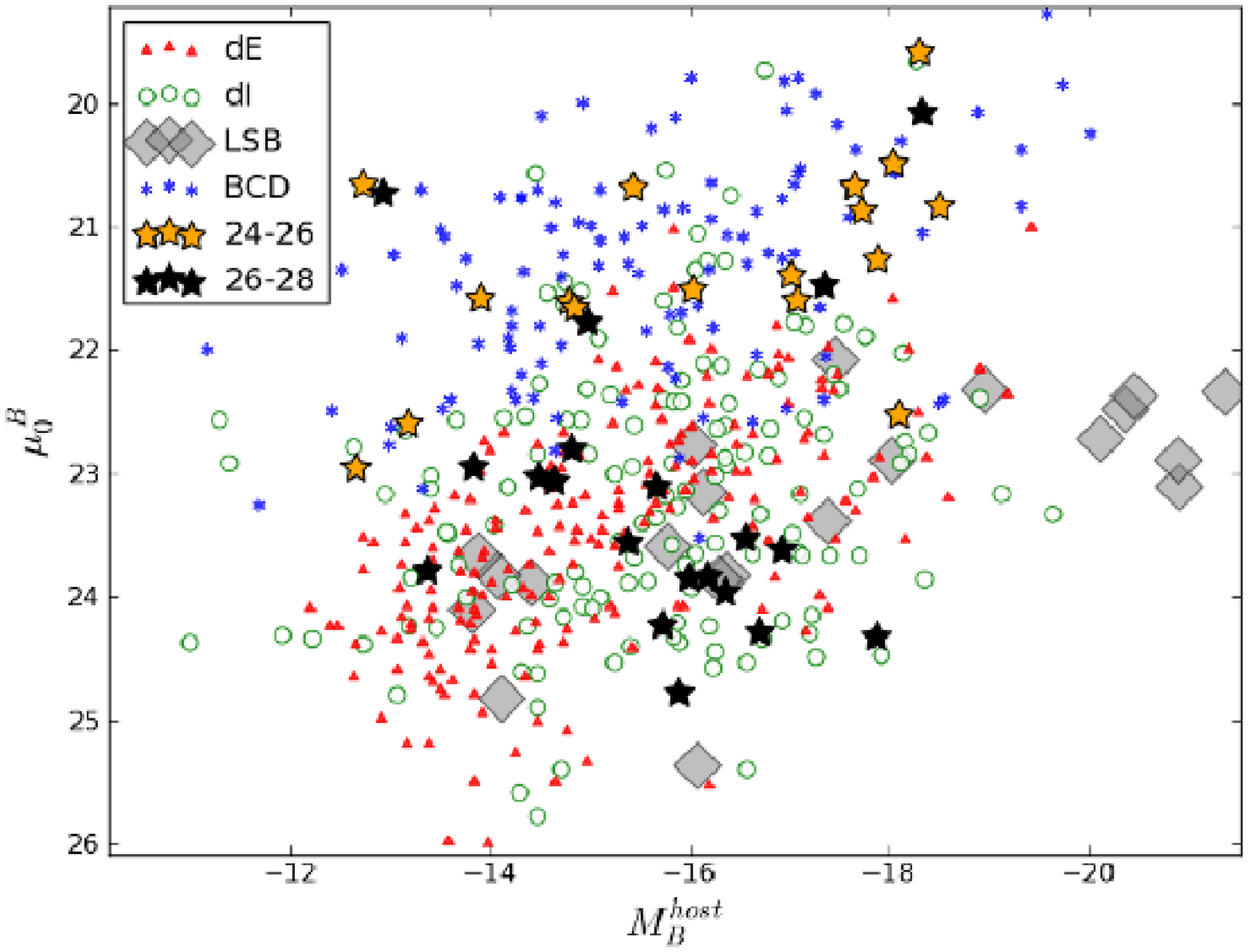}
    \includegraphics[width=7.5cm,height=6cm]{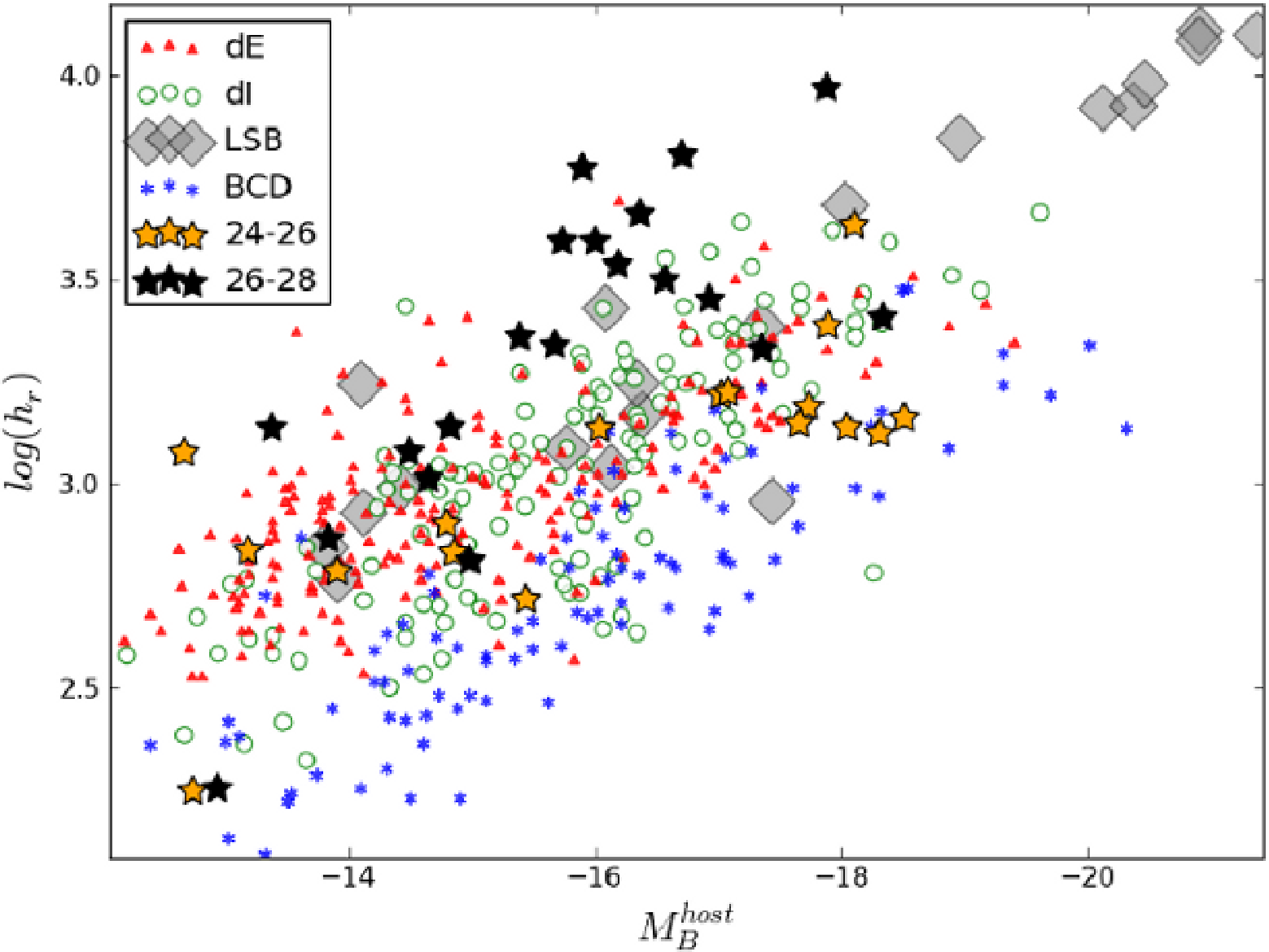}
    \caption{Comparing the structural properties of the host galaxies in our BCG sample (orange and black stars) to those of dEs, dIs, LSB and BCD galaxies from the literature. dE, dI, BCD and some LSB data were obtained from~\citet{2008A&A...491..113P} and references therein, while the giant LSB galaxies were taken from~\citet{1995AJ....109..558S} and references therein. The orange stars indicate structural parameters derived from the physical region corresponding to $\mu_B=24-26$ mag arcsec${}^{-2}$, while the black stars are from the fainter $\mu_B=26-28$ region. Note that $h_r$ is here in units of pc, not kpc.}\protect\label{papaderos}
  \end{center}
\end{figure}
\noindent In Figure~\ref{papaderos} we compare the structural parameters of our sample to other types of dwarfs, and also to observations of blue compact dwarfs (BCD) from other authors. The data in the figure were taken from~\citet{2008A&A...491..113P} for the dE, dI, BCD, and some LSB galaxies, and from~\citet{1995AJ....109..558S} and references therein for (giant) LSBs. We have obtained structural parameters from two separate outer regions, corresponding to $\mu_B=24-26$ and $\mu_B=26-28$ mag arcsec${}^{-2}$. For the $\mu_B=24-26$ mag arcsec${}^{-2}$ region the central surface brightness $\mu_0$ for our sample is consistent with other BCD measurements from the literature, as seen in the upper panel of Figure~\ref{papaderos}. If we use the fainter $\mu_B=26-28$ mag arcsec${}^{-2}$ region we instead obtain $\mu_0$ consistent with those measured for dE and dI galaxies. Our profiles are among the deepest BCG profiles available in the current literature. Previous BCG studies, which rarely present data significantly beyond the Holmberg radius, claim that for a given total host luminosity $M_{host}$, the hosts of BCGs tend to on average have a $\mu_0$ brighter than dEs and dIs~\citep[e.g.][]{2008A&A...491..113P}. Our data clearly demonstrate that if one samples fainter host regions there is no separation between dEs, dIs and BCGs in terms of $\mu_0$ vs. $M_{host}$. A similar trend is observed when we compare the scale length derived from $\mu_B=24-26$ and $\mu_B=26-28$ mag arcsec${}^{-2}$ to literature data. The brighter region gives $h_r$ values consistent with the BCD occupied region in the lower panel of Figure~\ref{papaderos}, although the average $h_r$ is slightly larger for our sample even in this region. The fainter $\mu_B=26-28$ mag arcsec${}^{-2}$ region moves the galaxies to scale lengths which are inconsistent with the BCD observations from the literature. While some data points are now clearly in the region occupied by dEs and dIs, quite a few have $h_r$ even larger than that, with $\log{h_r}\gtrsim3.5$. This is not a problem since the dE and dI observations themselves rarely probe surface brightness levels as faint as our $\mu_B=26-28$ mag arcsec${}^{-2}$ region. While the dE profiles are unlikely to change slope in deeper observations, this is entirely possible for many dIs.\\

\noindent Another possibility exists, however, when we consider the giant LSB galaxies from~\citet{1995AJ....109..558S}. Both our extreme scale lengths ($\log{h_r}\gtrsim3.5$) and their corresponding central surface brightnesses ($\mu_0\gtrsim24$ mag arcsec${}^{-2}$) are perfectly consistent with the structural parameters of these giant LSB galaxies (LSBG). There is no convention on the definition of LSBG, although usually galaxies with central surface brightness fainter than $\mu_B\sim23$ mag acrsec${}^{-2}$ and more luminous than $M^{tot}_B\sim-14$ mag are considered to fall in this category. The location of our galaxies in Figure~\ref{papaderos} suggests a tentative connection between LSBG and (at least some) BCG hosts. This result is consistent with~\citet{1999Ap&SS.269..625B}, who argue that luminous starbursts in BCGs are fed by LSBGs and not dIs due to luminosity, metallicity and \HI mass considerations. Additionally, this result further exemplifies the importance of sampling the host well away from the effects of the burst. While this may vary in individual cases, the surface brightness profile of BCGs seems significantly affected by the burst almost down to the Holmberg radius. However, while a difference in burst contribution would explain the varying $\mu_0$ and $h_r$ between the two regions, it cannot be excluded that the physical structure is indeed changing at the faintest levels, and that many luminous BCGs have double--disk profiles~\citep[e.g.][]{2006ApJS..162...49H}.\\

\begin{figure*}
\begin{center}
  \includegraphics[width=16cm,height=20cm]{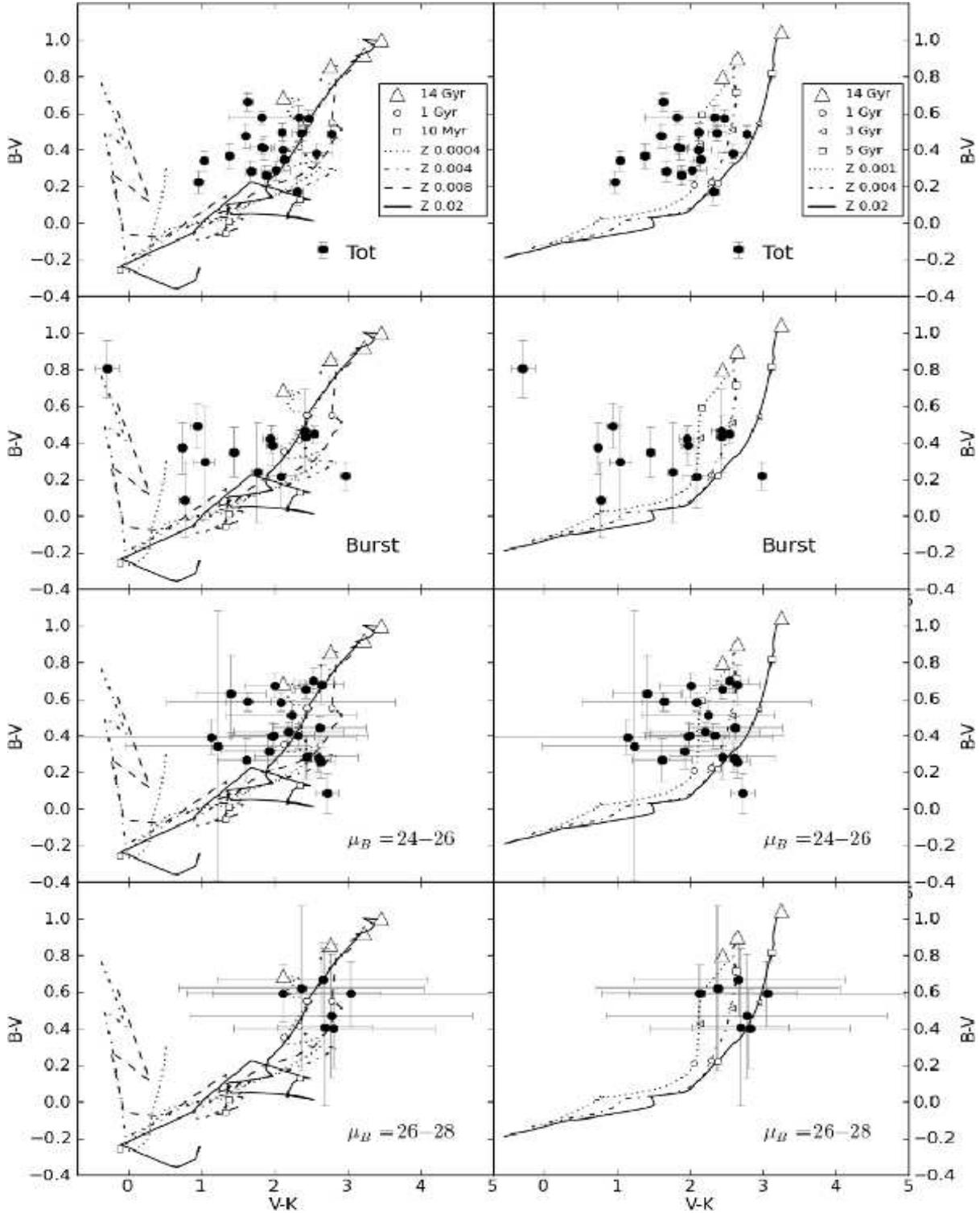}
\caption{Integrated optical/NIR colors with SEMs tracks with (left panel) and without (right panel) nebular emission. From top to bottom in both panels the plotted data points are the total colors over the entire area of the galaxy, the estimated burst colors, the colors of the $\mu_B\sim24-26$ mag arcsec${}^{-2}$ region, and the colors of the $\mu_B\sim26-28$ mag arcsec${}^{-2}$ region. The data are corrected for Galactic extinction. The errorbars include different uncertainties depending on the case (see \S~\ref{clrclrdiags} for details). The number of targets in the plots is not equal since the burst/host separation fails completely for a number of objects with irregular morphology. The tracks in the left panel include nebular emission and assume an instantaneous burst, with metallicities $Z=0.0004$ (dotted), $Z=0.004$ (dash--dotted), $Z=0.008$ (dashed), and $Z=0.02$ (line). The ages of $10$ Myr (open square), $1$ Gyr (open circle), and $14$ Gyrs (open triangle) are marked for convenience. In the right panel the tracks are for a pure stellar population with a Salpeter IMF, $M_{min}=0.08M_\odot$, $M_{max}=120M_\odot$, an $e$--folding time of $10^9$ yr, and with metallicity $Z=0.001$ (dotted), $Z=0.004$ (dash--dotted), and $Z=0.02$ (line). The ages of $1$ (open circle), $3$ (open left triangle), $5$ (open square), and $14$ Gyrs (open triangle) are marked for convenience. }\protect\label{SEMs}
\end{center}
\end{figure*}
\noindent The majority of the BCGs show color gradients in purely optical ($B-V$), purely NIR ($H-K$), and optical--NIR hybrid ($V-K$) colors, and flat profiles at large radii.  In terms of the luminosity the highest burst contributions are found in galaxies which light distribution we can trace down to $\mu_B\sim 28$ mag arcsec${}^{-2}$, and for which we observe a clear break in this distribution (i.e. a change in stellar population) beyond $\mu_B\gtrsim 26$ mag arcsec${}^{-2}$. These are ESO185--13 ($B$ band burst contribution $95\%$), ESO249--31 ($86\%$), ESO400--43 ($94\%$ in $V$), HE2--10 ($89\%$), Tol0341--407 ($83\%$), and UM448 ($88\%$). The burst dominance is not solely due to defining everything above this new very faint host as a burst. A profile with constant slope in the range $\mu_B\sim24.0$--$28.0$ would indicate that there is no change in population over that region and since the radial distances are large, presumably this single population is the host. Such targets also display varying burst--domination percentages. Some are rather high, e.g. SBS335--052East ($86\%$),  MK930 and ESO338--04 (both with $94\%$), etc., while others are much less, e.g. UM238 ($60\%$), or host dominated instead, e.g. HL293B ($16\%$) and SBS335--052West ($39\%$). Note that previous studies rarely present the surface brightness profile much further than the Holmberg radius with comparably small errors~\citep[e.g.][]{1996A&AS..120..207P,1999A&AS..138..213D,2001ApJS..133..321C}, so for targets in our sample which have a profile break around $\mu_B\sim26.0$ what we here define as host is a previously undetected population of different structural parameters and photometric properties than what is measured at the smaller radii between $\mu_B\sim24$--$26$ mag arcsec${}^{-2}$. For ESO185--13 and UM448 the now revealed host is a true LSB galaxy, with $\mu_0=24.7\pm0.12,~24.33\pm0.09$ mag arcsec${}^{-2}$, and $h_r=5.9\pm0.3,~9.3\pm0.4$ kpc, respectively.\\
\subsection{Color--color diagrams}\protect\label{clrclrdiags}
\noindent The parameters that can influence the colors of a galaxy in the optical and NIR regimes are the metallicity, the age of the stellar population, the internal extinction, the star formation history, and the nebular component, be it ionized in situ or by photons ``leaking'' from nearby star forming regions. These can conspire to either enhance or counteract each other's influence on the galaxy colors, which is why the best way to analyze the colors we have obtained for various physical areas and stellar populations is to use spectral evolutionary models (SEMs) based on stellar population synthesis. We have chosen to present only $B-V$ vs. $V-K$ because this diagram helps break the metallicity degeneracy more than the rest.\\

\noindent In Figure~\ref{SEMs} we compare our data to the two cases of a pure stellar population (right panel), and of stellar continuum contaminated by nebular emission (left panel). The data we plot, in descending order, are the total color of each galaxy measured down to the Holmberg radius from Table~\ref{totlumtbl}, the estimated color of the burst from Table~\ref{burstclrtbl}, and the color of the host regions defined over two radial ranges corresponding to $\mu_B=24$--$26$ and $\mu_B=26$--$28$ mag arcsec${}^{-2}$ from Table~\ref{totclrtbl}  (see \S~\ref{colors} for details). The errorbars in the individual subplots in this figure are the same as in the corresponding tables. Note that for the burst colors all errorbars are underestimated, as they only contain the very small numerical error of integrating above the interpolated exponential disk, and the zero point uncertainties of the respective nights. \\

\noindent The models with nebular emission are from the \emph{Yddrasil} spectral synthesis code, as already described in \S~\ref{individ}. While these are suitable for the young population and for regions with nebular emission contribution, the analysis of NIR fluxes and colors is very sensitive to the treatment of thermally pulsing AGB stars. We therefore also present an independent set of models based on the~\citet{2008A&A...482..883M} isochrones, which feature one of the most up-to-date treatments of this evolutionary stage currently available. However, as these models lack nebular emission, they are only suitable for the outskirts of our target galaxies, where nebular emission tends to be subdominant. These model tracks assume a Salpeter initial mass function (IMF) and a star formation decay rate of $1$ Gyr, suitable for early--type galaxies. The star formation history (SFH) is introduced by weighting each consecutive point with the average star formation of the previous points, and is suitable for an old stellar population with no recent burst and very low SFR. The predictions of both SEMs tracks are shown for several metallicities ranging from very metal--poor ($Z=0.001$ or $Z=0.0004$) to solar ($Z=0.02$).\\

\noindent All SEMs tracks plotted in Figure~\ref{SEMs} assume zero redshift, which is a coarse approximation for our case since half of our sample is at a redshift of $\sim0.02$. The $B-V$ colors of the young component can vary by as much as $0.3$ mag from redshift $0.00$ to redshift $0.02$. Further, even though we have used ``standard'' $BVK$ filters in each observing run, filters over similar wavelength ranges from different observatories show small differences in transmission and shape. These differences can amount to e.g. $\sim0.15$ mag for the $B$ band. Summa summarum, the tracks plotted in Figure~\ref{SEMs} are only intended to guide the eye and give an overall estimate of the age and metallicity of the data. A much more careful and detailed analysis of all colors, with SEMs tracks tailored to each galaxy and filter, lies beyond the scope of this paper and will be provided in~\citet{Paper3}.\\

\noindent In the $B-V$ vs. $V-K$ plane, contribution from nebular emission to the stellar continuum shifts the predicted colors redwards in $B-V$ and bluewards in $V-K$, an effect which is much more drastic for ages younger than $1$ Gyr (left panel of Figure~\ref{SEMs}). In fact, ignoring nebular emission (right panel of Figure~\ref{SEMs}) leads to clearly irreconcilable differences between the observed and the predicted colors for the galaxy as a whole, for the burst, and also for the $\mu_B=24$--$26$ mag arcsec${}^{-2}$ region. Even though we present both sets of tracks, with and without nebular emission, it is clear that e.g. the burst colors can only be explained by the \emph{Yggdrasil} tracks because the young regions of the galaxies have larger amount of ionized gas and more young stars present to ionize the gas. In contrast, for the $\mu_B=26$--$28$ mag arcsec${}^{-2}$ region which contains predominantly old stars and much less gas, the colors are best reproduced by the stellar population tracks seen in the right panel. Not all targets have detectable $K$ band flux at these radial distances even given the large bin size of the integration, and the ones that do have quite large errorbars, but the trend towards an old stellar population is nevertheless clear. We note that the $\mu_B=24$--$26$ mag arcsec${}^{-2}$ region, which is at moderately large radii from the burst, still shows heavy nebular emission contamination. The colors here are on average similar to the total galaxy colors and much bluer in $V-K$ than in the $\mu_B\sim26$--$28$ mag arcsec${}^{-2}$ region. This indicates that there is still significant contribution from the young population to the light distribution of the galaxy. This region is therefore not to be relied upon for the measurement of structural parameters of the host such as the scale length $h_R$ and the central surface brightness $\mu_0$. Undoubtedly, for some individual galaxies the contamination may be negligible depending on the morphology, location and extent of the burst regions, and the distribution of the gas over the disk of the galaxy, but this would have to be examined on a case to case basis before any measurements over this region are considered reliable. This is not possible if the data are not deep enough to reach beyond the Holmberg radius with small errors. Our data allows this for the majority of the targets. We have therefore adopted the structural parameters of the host derived from the clearly host--dominated $\mu_B=26$--$28$ mag arcsec${}^{-2}$ region when analyzing possible correlations between the structural parameters and other galaxy properties. \\
\begin{center}
  \begin{table*}
    \begin{minipage}{150mm}
      \caption{Minimum Petrosian $A_{P}$ asymmetries measured in each filter over the area enclosed by the Petrosian radius $r[\eta(0.2)]$, here given in kpc for each galaxy and filter.}
      \protect\label{petrasymtbl}
      \begin{tabular}{@{}|l|r|r|r|r|r|r|r|r|r|r|r|r|r|r|r|@{}}
        \hline
        Galaxy&$A_U$&$r_U$&$A_B$&$r_B$&$A_V$&$r_V$&$A_R$&$r_R$&$A_I$&$r_I$&$A_H$&$r_H$&$A_K$&$r_K$\\\hline
        ESO185--13&$0.15$&$2.2$&$0.18$&$2.2$&$0.19$&$2.2$&$0.16$&$2.6$&$0.17$&$2.6$&$0.17$&$2.6$&$0.16$&$2.2$\\
        ESO249--31&&&$0.36$&$2.8$&$0.33$&$3.0$&$0.33$&$2.8$&$0.32$&$2.9$&$0.33$&$2.7$&$0.39$&$2.6$\\
        ESO338--04&&&$0.16$&$1.8$&$0.22$&$1.6$&$0.21$&$1.6$&$0.21$&$1.8$&$0.25$&$2.0$&$0.19$&$1.8$\\
        ESO400--43&&&&&$0.31$&$3.4$&&&$0.26$&$3.4$&$0.25$&$3.4$&$0.27$&$3.4$\\
        ESO421--02&$0.35$&$1.0$&$0.47$&$1.1$&$0.48$&$1.1$&$0.50$&$1.2$&$0.52$&$1.2$&$0.54$&$1.1$&$0.55$&$1.1$\\
        ESO462--20&$0.16$&$3.0$&$0.13$&$3.0$&$0.14$&$3.0$&&&&&$0.13$&$3.0$&$0.14$&$3.0$\\
        HE2--10&$0.28$&$0.3$&$0.26$&$0.4$&$0.24$&$0.7$&&&&&$0.12$&$0.6$&$0.12$&$0.6$\\
        HL293B&$0.25$&$0.3$&$0.31$&$0.3$&$0.29$&$0.3$&$0.32$&$0.4$&$0.27$&$0.5$&$0.35$&$0.5$&$0.39$&$0.5$\\
        IIZW40&$0.66$&$0.5$&$0.40$&$0.5$&$0.36$&$0.4$&&&$0.57$&$0.6$&$0.61$&$1.1$&$0.58$&$0.7$\\
        MK600&$0.23$&$0.5$&$0.15$&$0.5$&$0.17$&$0.5$&$0.13$&$0.6$&$0.11$&$0.7$&$0.27$&$0.7$&$0.31$&$0.7$\\
        MK900&$0.22$&$0.5$&$0.15$&$0.5$&$0.13$&$0.6$&$0.11$&$0.6$&$0.08$&$0.7$&$0.17$&$0.8$&$0.21$&$0.7$\\
        MK930&$0.34$&$2.9$&$0.36$&$2.9$&$0.37$&$2.9$&$0.33$&$2.9$&$0.33$&$2.9$&$0.34$&$2.9$&$0.34$&$2.9$\\
        MK996&$0.21$&$0.4$&$0.17$&$0.5$&$0.15$&$0.6$&$0.15$&$0.6$&$0.12$&$0.8$&$0.29$&$0.8$&$0.16$&$0.5$\\
        SBS0335--052E&&&$0.07$&$0.7$&&&&&$0.09$&$0.7$&$0.23$&$0.7$&$0.37$&$0.7$\\
        SBS0335--052W&&&$0.28$&$0.9$&&&&&$0.58$&$1.1$&$0.64$&$1.4$&$0.62$&$1.4$\\
        TOL0341--407&$0.36$&$3.6$&$0.35$&$3.6$&$0.38$&$3.6$&&&$0.34$&$3.6$&$0.37$&$3.6$&$0.43$&$3.6$\\
        TOL1457--262\emph{I}&&&$0.27$&$8.5$&$0.28$&$8.5$&&&&&&&$0.31$&$8.8$\\
        TOL1457--262\emph{II}&&&$0.40$&$4.3$&$0.42$&$4.3$&&&&&&&$0.35$&$4.3$\\
        UM133&$0.49$&$2.2$&$0.33$&$2.2$&$0.31$&$2.5$&$0.28$&$2.6$&$0.36$&$2.6$&$0.42$&$2.7$&$0.52$&$2.4$\\
        UM160&$0.54$&$5.5$&$0.47$&$6.1$&$0.49$&$5.5$&$0.45$&$5.4$&$0.49$&$5.3$&$0.39$&$5.2$&$0.50$&$4.8$\\
        UM238&$0.38$&$2.7$&$0.21$&$2.7$&$0.19$&$2.7$&$0.17$&$2.7$&$0.27$&$2.7$&$0.38$&$2.5$&$0.39$&$2.7$\\
        UM417&$0.53$&$0.6$&$0.39$&$0.6$&$0.28$&$0.5$&$0.34$&$1.7$&$0.33$&$1.8$&$0.58$&$2.1$&$0.65$&$2.0$\\
        UM448&&&$0.18$&$4.0$&$0.18$&$4.0$&$0.15$&$4.0$&$0.17$&$4.0$&&&$0.29$&$3.6$\\
        UM619&$0.36$&$4.9$&$0.18$&$4.9$&$0.15$&$4.9$&$0.14$&$4.9$&$0.19$&$4.6$&$0.23$&$4.3$&$0.23$&$4.3$\\
        \hline
      \end{tabular}
    \end{minipage}
  \end{table*}
\end{center}
\noindent In the $\mu_B=26$--$28$ mag arcsec${}^{-2}$ region there are a number of galaxies with measured $V-K$ which appear to be too red and indicate super--solar metallicities. The errorbars are large and all except the two most extreme points can be reconciled with the pure stellar population tracks with lower than solar metallicities, but we will nevertheless examine these on a case to case basis. Solar or near--solar metallicities are unusual for such large radii since we expect the (stellar) metallicity to drop with increased radius from the center~\citep{2008MNRAS.388L..10B}. The latter only holds, however, for regular morphology galaxies. The deviating galaxies, in order of decreasing $V-K$ excess, are UM619, UM238, UM160, and ESO421--02, which all cluster at the solar metallicity track. Statistically one would expect outliers like this in photometry measurements without this having to mean that there is something wrong with the measurements themselves. UM619 has nearly solar metallicity according to the literature~\citep{2001AJ....122.2318B}, and most of its star formation is distributed along the spiral arms at significant distances from the center. UM238 has all of its bright star--forming cluster complexes in the outskirts of the host, some of which fall inside the integration bin corresponding to the $\mu_B=26$--$28$ mag arcsec${}^{-2}$ region. Similarly, UM160 has highly irregular morphology, with many blue regions found inside of the relevant integration radial range. ESO421--02, on the other hand, has outskirts clearly dominated by a red host, but this galaxy shows nearly constant $V-K$ color for all physical regions we examined -- the difference between the total galaxy color and the color of the two outer regions is $\Delta(V-K)\lesssim0.06$ mag. Thus all of these measurements are either due to the outer regions being contaminated by starburst contribution, increasing its apparent metallicity, or the galaxy as a whole shows no metallicity gradient. The measurements contaminated by burst are therefore better compared to the \emph{Yggdrasil} tracks, which include nebular emission. Specifically the $Z=0.008$ track with an age $<1$ Gyr seems suitable for most of these data points. \\

\noindent Previous studies of BCGs have detected host $V-K$ colors which are difficult to reconcile with a normal stellar population of low metallicity~\citep{2002A&A...390..891B,2005mmgf.conf..355B,2006ApJ...650..812Z}. Some extreme cases, such as Haro 11, were impossible to reconcile with \emph{any} metallicity. However, judging by the $B-V$ vs. $V-K$ diagram, the case for an abnormally red host (a.k.a. the ``red halo'' phenomenon) in BCGs seems rather weak for this sample. Though there are only eight color measurements in the $\mu_B=26$--$28$ mag arcsec${}^{-2}$ region, and the errorbars are large, we see no clustering of the data points around high metallicity tracks or a general tendency towards abnormally red $V-K$ colors. We have also verified this with a $B-V$ vs. $V-I$ diagram (not presented here) using the same two models. In~\citet{2010MNRAS.405.1203M} we further refuted the extremely red color claimed for the host of Haro 11~\citep[][$V-K\sim4.2\pm0.8$]{2002A&A...390..891B,2005mmgf.conf..355B}, and measured instead the more modest color of $V-K\sim2.3\pm0.2$ over the same physical region. In other words, the present work casts doubt on whether any BCG truly has an abnormally red host, since Haro 11, previously the most extreme case of a red BCG host, is no longer extreme, and we find no clear indication of abnormally red colors in the present sample. Though there are currently no detections of an abnormally red host in individual BCGs left undisputed, here we cannot decisively conclude that the hosts of BCGs have normal colors. We will make such a determination in a future paper~\citet{Paper3}. Note that the lack of abnormal colors in BCG hosts has no direct bearing on the extremely red colors of halos around high- and low surface brightness stacked edge--on disks~\citep{2004MNRAS.347..556Z,2010MNRAS.405.2697B}.\\
\begin{figure}
  \begin{center}
    \includegraphics[width=8cm,height=10cm]{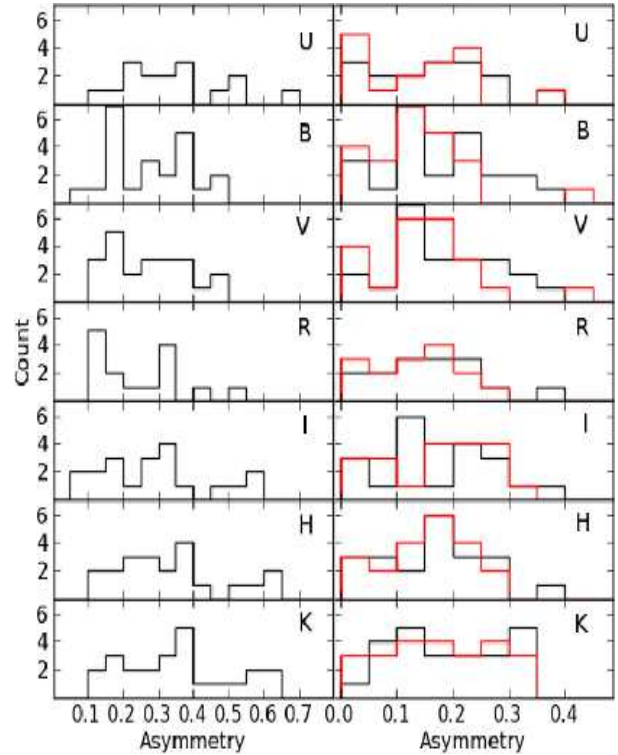}
    \caption{Histograms of $A_P$ Petrosian (left) and $A_H^\prime$ Holmberg (right, black) asymmetries. Overplotted in the right panel (red) is the dynamical $A_{dyn}$ asymmetry. Bin size of $0.05$ is used. The plotted data are from Tables~\ref{petrasymtbl} and~\ref{holmasymtbl}.}\protect\label{asymhist}
  \end{center}
\end{figure}

\noindent While we have noted that the $\mu_B=24-26$ mag arcsec${}^{-2}$ region for this sample contains a strong contribution from the star forming regions, from Figure~\ref{SEMs} we can still make the observation that all host galaxies in the sample do not appear to be younger than $5$ Gyr. The same conclusion, albeit with fewer available data points and larger errorbars, can be drawn from the $\mu_B=26-28$ mag arcsec${}^{-2}$ region in the same figure. The implied redshift at which the bulk of the star formation in the host must have occurred is thus $z\sim0.5$ ($8$ to $9$ Gyr after the Big Bang). While older stars could of course be present in all of our targets and can easily remain unseen in these data, it appears unlikely that the prototypical host for this sample could have been a red and dead elliptical galaxy since those have colors consistent with the Hubble age (compare to KIG732 in Appendix~\ref{kig}, with $B-V\sim1$, $V-K\sim3$ at large radii).\\
\subsection{Asymmetry and concentration}
\begin{figure}
  \begin{center}
    \includegraphics[width=6cm,height=6cm]{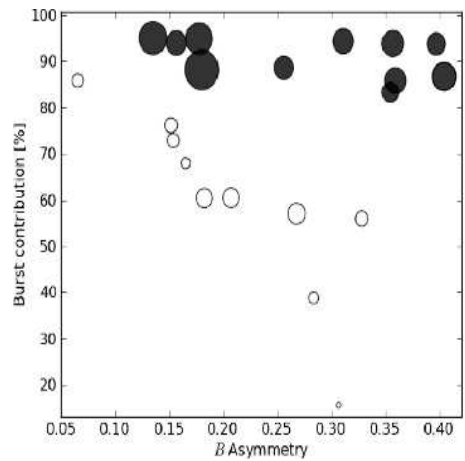}
    \caption{Relative burst contribution to the total luminosity in percent vs. the $B$ band Petrosian $A_P$ asymmetry for the sample. The filled black circles are the galaxies with true LSB hosts, $\mu_0\gtrsim23$ mag arcsec${}^{-2}$ and $h_r\gtrsim3$ kpc. The open circles are the remaining galaxies in the sample. The size of the markers reflects the respective scale length.}\protect\label{burstasym}
  \end{center}
\end{figure}

\protect\label{asymdiss}
\noindent The histograms in Figure~\ref{asymhist} show the distribution of Petrosian asymmetries (left panel), and ``Holmberg'' asymmetries (right panel) of the galaxies in the sample. In the case of the Petrosian asymmetry the size of the area over which it is measured is small (see Table~\ref{petrasymtbl}), and the faintest isophote included in that area is often as bright as $\mu\sim21$ mag arcsec${}^{-2}$. The Petrosian asymmetry is therefore mainly flocculent in nature -- it is a measure of the deviation from the symmetric ground state due to the presence of bright SF regions. As an example of how flocculent asymmetry can increase the total asymmetry of an otherwise symmetric galaxy, consider the case of HL293B. This galaxy has regular elliptical outer isophotes and should thus merit a rather small asymmetry, but instead has $A_B\sim0.3$ due to its off--center star formation region. One would therefore expect the composite asymmetry to correlate with relative burst contribution in the sense that as the burst contribution decreases the dominating asymmetry component changes from flocculent to dynamical, where the latter is a measure of low frequency structure. This does indeed seem to be the case for about half the sample as seen in Figure~\ref{burstasym}. As the contrast between star forming regions and faint dynamical features decreases with the burst contribution, the measured composite asymmetry increases as more weight is placed on the dynamical morphology. There appears to be an upper limit to the burst contribution at about $\gtrsim90\%$, above which the sampled asymmetry is purely flocculent. Perhaps not surprisingly, the galaxies that fall within this category are the BCGs consistent with true LSB hosts, as identified by Figure~\ref{papaderos} and discussed in \S~\ref{generaldiss}. The contrast between the bright burst and the LSB host is here very high, which would make the effect of dynamical asymmetry negligible since many tidal features or extended irregular structures are too faint to fall inside of the Petrosian $r[\eta(0.2)]$ radius.\\
\begin{figure}
  \begin{center}
    \includegraphics[width=8cm]{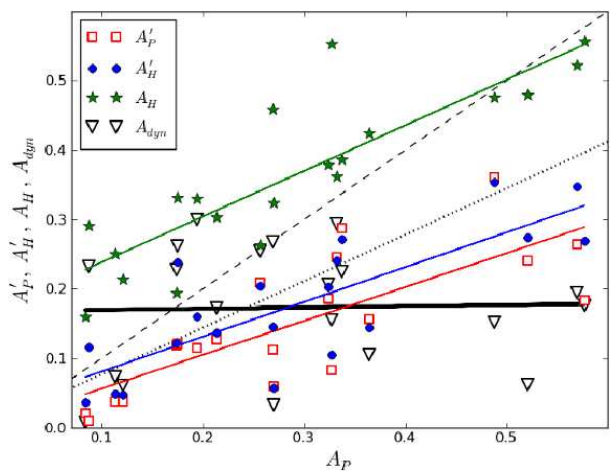}
    \caption{Petrosian asymmetry ($A_P$) vs. all alternative asymmetry measurements. $A_H$ is the Holmberg asymmetry without any smoothing of the images, while a $1\times1$ kpc smoothing box has been applied to $A_P^\prime$ (Petrosian), $A_H^\prime$ (Holmberg), and $A_{dyn}$ (dynamical). The uninterrupted lines are least square fits to the similarly colored data points. The dashed line has a slope of $1$. The dotted line is an extrapolation from the correlation of~\citet{2003ApJS..147....1C} for E, S0, Sa--b, Sc--d and Irr, which lie beyond $A_P\sim0.6$. All asymmetries here are from the $I$ band, in order to facilitate comparison with~\citet{2003ApJS..147....1C}.}\protect\label{allasymmetries}
  \end{center}
\end{figure}

\noindent The dynamical asymmetry should here be better represented by what we call the Holmberg asymmetry (Table~\ref{holmasymtbl}). The term is a bit misleading since the Holmberg radius is only defined in blue passbands, and we use it purely for abbreviation purposes. To reduce the effect of the bright, often irregular SF regions on the total asymmetry budget, and to compensate for the difference in distances to the galaxies in the sample, the images have been boxcar smoothed by a filter of size $1\times1$ kpc. This has the effect that predominantly large scale structures remain and the measured asymmetry should therefore be mainly dynamical, not flocculent. Our hope that the asymmetry obtained in such a fashion would be a reasonable proxy for the dynamical asymmetry was crushed by our findings in~\S~\ref{generaldiss} and \S~\ref{clrclrdiags}, where we established that for this sample of luminous BCGs there is still a strong contribution from the starburst down to the $\mu_B\sim26$ mag arcsec${}^{-2}$ isophotal level, which is nearly at the Holmberg radius. We will denote the Petrosian asymmetry with $A_P$, the unsmoothed Holmberg asymmetry with $A_H$, and the smoothed Holmberg asymmetry with $A_H^\prime$, where the prime indicates smoothing. In Figure~\ref{allasymmetries} we find a positive correlation between the Petrosian $A_P$ and Holmberg $A_H$, $A_H^\prime$ asymmetries in the sense that for a given galaxy large/small Petrosian nearly always implies large/small Holmberg asymmetries. This makes sense for starbursting luminous BCGs, as one would expect there to be a strong positive correlation between flocculent and dynamical asymmetries. A (wet) merger would lead to star formation, and the former would increase the dynamical component, while the latter would increase the flocculent component in the composite asymmetry. In our case, however, there are indications that the Petrosian and Holmberg asymmetries are measuring the same (flocculent) asymmetry component in over half the sample, with the numerically different Holmberg values being solely due to the applied smoothing, and not to the sampling of a dynamically dominated asymmetry. For a few cases (open circles in Figure~\ref{burstasym}) the Holmberg asymmetry successfully measures predominantly the dynamical component, as exemplified by HL293B. From a Petrosian asymmetry of $A_P(B)\sim0.3$ the value has now dropped to $\sim0.06$, which is much more suitable for a galaxy with such regular morphology.\\

\noindent To obtain a more universally successful measure of the dynamical asymmetry for our sample, we tweaked the Holmberg $A_H^\prime$ asymmetry to allow it to ignore star forming areas. All regions brighter than $\mu=25$ mag arcsec${}^{-2}$ in the optical, and $\mu=21$ mag arcsec${}^{-2}$ in the NIR were set to a constant value of $25$ and $21$ mag arcsec${}^{-2}$ respectively in the two wavelength regimes. This effectively circumvents the contribution of the flocculent asymmetry component to the total asymmetry budget, and it has the effect that more weight is given to the fainter structures. This modified asymmetry is a better estimate of the dynamical asymmetry component, and is given in Table~\ref{holmasymtbl}. We will refer to this asymmetry as simply dynamical asymmetry, $A_{dyn}$. For targets for which the Holmberg asymmetry ($A_H^\prime$) already measures the dynamical component, the $A_{dyn}$ values are very similar (e.g. HL293B, MK996, and other ellipticals). For targets for which the flocculent asymmetry dominated even the Holmberg values, $A_{dyn}$ varies from case to case depending on the morphology. ESO185--13, for example, now shows its more asymmetric nature, as the decreased flux contrast gives its faint tidal tail a chance to contribute significantly to the total asymmetry budget in the optical, resulting in $A_{dyn}(B)\sim0.2$ instead of $A_H^\prime(B)\sim0.1$. In the NIR the values stay the same, $A(NIR)\sim0.1$, since the tidal tail is invisible at these wavelengths.\\

\noindent Figure~\ref{allasymmetries} summarizes the behavior of all of these alternative asymmetry measurements versus the standard Petrosian asymmetry. In order to illustrate the effect of the resolution, both a smoothed ($A^\prime$) and unsmoothed ($A$) version of the Petrosian ($A_P,~A_P^\prime$) and Holmberg ($A_H,~A_H^\prime$) asymmetries are included in the figure. The correlation between Petrosian and Holmberg asymmetries (both smoothed and unsmoothed) is clear, with Pearson's $R=0.9;~0.8;~0.8$, and somewhat similar slopes for the fitted lines $\mathfrak{m}=0.66;~0.50;~0.49$ for $A_P$ vs. $A_H,~A_H^\prime,~A_P^\prime$, respectively.~\citet{2003ApJS..147....1C} finds $\mathfrak{m}=0.67$ for spheroids (E,S0), early and late type disks (Sa--b,Sc--d), and irregulars (Irr). From all of our asymmetry measurements, $A_H^\prime$ is best suited for direct comparison with Concelice's ``global'' $A_G(R)$ asymmetry. We get $A_H^\prime(I)=0.50\times A_P(I)+0.03$. This is consistent with Figure~\ref{burstasym} where for $\sim50\%$ of the targets the total asymmetry is completely dominated by the star formation.\\

\noindent In contrast to our previous comparison, in Figure~\ref{allasymmetries} we do not see any correlation between the Petrosian (flocculent, $A_P$) and the dynamical $A_{dyn}$ asymmetries. This makes sense because by substituting all star forming regions with a constant brightness we are effectively only sampling the dynamical asymmetry component of the host, and not of the composite galaxy. One does not expect any particular dependence between the flocculent asymmetry of the composite galaxy and the dynamical asymmetry of the host, except in some merger cases. We say \emph{some} merger cases, because there are also examples of very symmetric mergers, such as ESO421--02. \\

\noindent For a given galaxy, the Petrosian asymmetries do not show much variation from filter to filter, and preserve similar values even when going from the optical to the NIR regime (Table~\ref{petrasymtbl}). The same applies to the Holmberg asymmetries, which show similar negligible change with filter for the majority of the galaxies (Table~\ref{holmasymtbl}). The NIR asymmetry, regardless of how it is measured, should be more sensitive to the morphology of the old stellar population, and hence to the dynamical contribution to the total asymmetry. On the other hand, the optical asymmetry is more attuned to the flocculent asymmetry component since the young stellar population and the nebular gas would dominate the optical emission. The near constancy of the asymmetry across filters can then only imply one of two things -- that both the optical and NIR regimes are probing predominantly the flocculent asymmetry, which here dominates even in the NIR; alternatively, that the flocculent and dynamical asymmetries are equal and there is a perfect anti-correlation between the filter sensitivity towards the two asymmetries, in the sense that e.g. the $B$ band is very sensitive to the flocculent asymmetry and very insensitive to the dynamical asymmetry, and that this trend is reversed as one moves to the NIR. The latter is unlikely since it would require a remarkable degree of fine--tuning. Additionally, even if a specific filter is more sensitive to a particular population, it is not completely insensitive to the other population. Therefore when it comes to the asymmetry of the composite galaxy, i.e. either Petrosian or Holmberg, then both the optical and NIR asymmetries must be flocculent in nature. Note, however, that while the dominance of the flocculent asymmetry in both the optical and the NIR may seem intuitive and self-evident, we find large variations in asymmetry values between the two wavelength regimes for the volume--limited sample of emission line galaxies in~\citet{Paper2}, which correlate with morphological class. The constancy of the Petrosian and Holmberg asymmetry for this sample is due to the selection criterion which favors very luminous BCGs with intense star formation. These are predominantly burst dominated, as seen from Table~\ref{burstclrtbl} and Figure~\ref{burstasym}, with more than half the sample showing a relative burst contribution $\gtrsim70\%$. \\
\begin{figure*}
  \begin{center}
    \includegraphics[width=17cm,height=8cm]{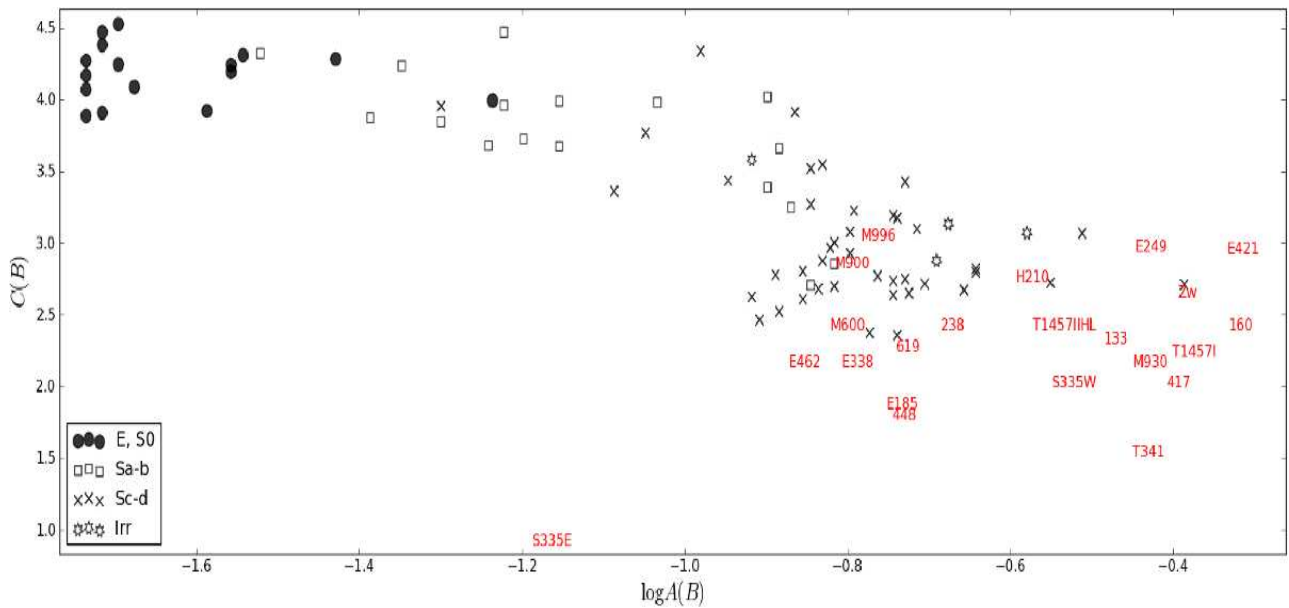}
    \caption{Concentration vs Petrosian $A_P$ Asymmetry for our sample (red text) compared to spheroids, early and late type disks, and irregulars taken from~\citet{2000ApJ...529..886C}. Abbreviations stand for: \emph{E185}=ESO185--13; \emph{E249}=ESO249--31; \emph{E338}=ESO338--04; \emph{E400}=ESO400--43; \emph{E421}=ESO421-02; \emph{E462}=ESO462-20; {H210}=HE2--10; \emph{HL}=HL293B; \emph{Zw}=IIZw40; \emph{M600}=MK600; \emph{M900}=MK900; \emph{M930}=MK930; \emph{M996}=MK996; \emph{S335E}=SBS0335--052E; \emph{S335W}=SBS0335--052W; \emph{T341}=TOL0341--407; \emph{T1457II}=TOL1457--262\emph{II}; \emph{T1457I}=TOL1457--262\emph{I}; \emph{133}=UM133; \emph{160}=UM160; \emph{238}=UM238; \emph{417}=UM417; \emph{448}=UM448; \emph{619}=UM619.}\protect\label{conselice}
  \end{center}
\end{figure*}

\noindent In Figure~\ref{conselice} we compare the concentration from Table~\ref{concentr} to the Petrosian $A_P$ asymmetry for our sample. Included in the plot are the spheroids (E and S0), early (Sa--b) and late type (Sc--d) disks, and irregular galaxies (Irr) from~\citet{2000ApJ...529..886C} and references therein. The normal galaxies exhibit a concentration--asymmetry correlation in the sense that the very symmetric galaxies are highly concentrated, with spheroids topping the scales in this correlation. Our sample of luminous BCGs occupies the opposite end of the extreme -- on average they display higher asymmetries than the other galaxy types, and are simultaneously much less concentrated. Although we have plotted $B$ band concentration and asymmetry, this trend is qualitatively preserved in the rest of the filters (Tables ~\ref{petrasymtbl} and~\ref{concentr}). The only outlier in this relation is SBS0335--052E. As already discussed in~\S \ref{sbsE} this galaxy has an extremely compact central region with fairly regular and bright outer isophotes, which would bring down its flocculent asymmetry, but is comparatively very extended in the optical, which would bring down its concentration.\\
\begin{table}
  \begin{minipage}{80mm}
    \caption{Minimum asymmetries measured in each filter. The two numbers per filter per galaxy are Holmberg $A_H^\prime$ asymmetries measured over the area enclosed by the Holmberg radius $r(\mu=26.5)$ in the optical and by $r(\mu=23)$ in the NIR (top value), and the dynamical $A_{dyn}$ asymmetries, with regions $\mu<25$ ($\mu<21$) set to $25$ ($21$) mag arcsec${}^{-2}$ in the optical (NIR) (bottom value). The images are pre--processed by a boxcar average of size $1\times1$ kpc. Dots indicate that the measured value was unreliable due to severe masking effects. }
    \protect\label{holmasymtbl}
    \tiny
    \begin{tabular}{@{}|l|l|l|l|l|l|l|l|l|@{}}
      \hline
      Galaxy&$A_U$&$A_B$&$A_V$&$A_R$&$A_I$&$A_H$&$A_K$\\\hline
      ESO185--13&$0.12$&$0.11$&$0.11$&$0.12$&$0.12$&$0.11$&$0.10$\\
      &$0.24$&$0.19$&$0.26$&$0.21$&$0.23$&$0.11$&$0.11$\\
      ESO249--31&&$0.20$&$0.20$&$0.19$&$0.20$&$0.18$&$0.19$\\
      &&$0.14$&$0.22$&$0.20$&$0.21$&$0.16$&$0.20$\\
      ESO338--04&&$0.14$&$0.13$&$0.17$&$0.14$&$0.17$&$0.14$\\
      &&$0.18$&$0.18$&$0.16$&$0.17$&$0.14$&$0.13$\\
      ESO400--43&&&$0.26$&&$0.20$&$0.18$&$0.20$\\
      &&&$0.13$&&$0.25$&$0.11$&$0.14$\\
      ESO421--02&$0.15$&$0.22$&$0.24$&$0.27$&$0.27$&$0.24$&$0.25$\\
      &$0.10$&$0.09$&$0.13$&$0.08$&$0.06$&$0.17$&$0.13$\\
      ESO462--20&$0.18$&$0.13$&$0.14$&&&$0.14$&$0.12$\\
      &$\dots$&$\dots$&$\dots$&&&$\dots$&$\dots$\\
      HE2--10&$0.09$&$0.12$&$0.13$&&&$0.06$&$0.06$\\
      &$\dots$&$\dots$&$\dots$&&&$\dots$&$\dots$\\
      HL293B&$0.05$&$0.06$&$0.06$&$0.05$&$0.06$&$0.03$&$0.06$\\
      &$0.02$&$0.02$&$0.02$&$0.02$&$0.03$&$0.03$&$0.05$\\
      IIZW40&$0.22$&$0.31$&$0.31$&&$0.35$&$0.35$&$0.33$\\
      &$0.22$&$0.19$&$0.19$&&$0.19$&$0.23$&$0.21$\\
      MK600&$0.03$&$0.04$&$0.12$&$0.07$&$0.05$&$0.04$&$0.10$\\
      &$0.04$&$0.06$&$0.12$&$0.06$&$0.07$&$0.04$&$0.08$\\
      MK900&$0.03$&$0.03$&$0.02$&$0.03$&$0.04$&$0.04$&$0.03$\\
      &$0.02$&$0.02$&$0.01$&$0.01$&$0.01$&$0.04$&$0.03$\\
      MK930&$0.23$&$0.24$&$0.25$&$0.24$&$0.24$&$0.21$&$0.26$\\
      &$0.21$&$0.18$&$0.21$&$0.19$&$0.29$&$0.18$&$0.25$\\
      MK996&$0.04$&$0.04$&$0.04$&$0.04$&$0.05$&$0.05$&$0.05$\\
      &$0.05$&$0.02$&$0.03$&$0.03$&$0.06$&$0.09$&$0.04$\\
      SBS0335--052E&&$0.11$&&&$0.12$&$0.09$&$0.09$\\
      &&$0.10$&&&$0.23$&$0.10$&$0.09$\\
      SBS0335--052W&&$0.21$&&&$0.27$&$0.22$&$0.17$\\
      &&$0.15$&&&$0.18$&$0.22$&$0.17$\\
      TOL0341--407&$0.21$&$0.25$&$0.24$&&$0.27$&$0.26$&$0.29$\\
      &$0.23$&$0.14$&$0.19$&&$0.23$&$0.16$&$0.16$\\
      TOL1457--262\emph{I}&&$0.31$&$0.34$&&&&$0.31$\\
      &&$0.18$&$0.18$&&&&$0.31$\\
      TOL1457--262\emph{II}&&$0.28$&$0.25$&&&&$0.28$\\
      &&$0.21$&$0.20$&&&&$0.26$\\
      UM133&$0.26$&$0.19$&$0.19$&$0.22$&$0.14$&$0.19$&$0.31$\\
      &$0.11$&$0.10$&$0.09$&$0.11$&$0.11$&$0.18$&$0.28$\\
      UM160&$0.37$&$0.37$&$0.37$&$0.37$&$0.35$&$0.28$&$0.31$\\
      &$0.16$&$0.13$&$0.14$&$0.13$&$0.15$&$0.24$&$0.27$\\
      UM238&$0.15$&$0.12$&$0.10$&$0.10$&$0.15$&$0.15$&$0.14$\\
      &$0.17$&$0.12$&$0.12$&$0.19$&$0.27$&$0.18$&$0.17$\\
      UM417&$0.28$&$0.21$&$0.29$&$0.17$&$0.11$&$0.30$&$0.34$\\
      &$0.11$&$0.09$&$0.11$&$0.10$&$0.16$&$0.30$&$0.33$\\
      UM448&&$0.18$&$0.19$&$0.21$&$0.24$&&$0.23$\\
      &&$0.22$&$0.21$&$0.24$&$0.26$&&$0.34$\\
      UM619&$0.19$&$0.12$&$0.11$&$0.11$&$0.16$&$0.16$&$0.17$\\
      &$0.22$&$0.11$&$0.16$&$0.26$&$0.30$&$0.24$&$0.23$\\
      \hline
    \end{tabular}
  \end{minipage}
\end{table}

\section[]{Conclusions}\protect\label{conclude}
\noindent We have obtained and analyzed very deep $UBVRIHKs$ broadband imaging data for a sample of 24 blue compact galaxies (BCGs). These are a part of a larger data set consisting of $46$ BCGs, which comprises the most extensive study to date of extremely deep optical and NIR surface photometry of this type of galaxies. The remaining galaxies are a volume limited sample and are therefore analyzed separately in~\citet{Paper2}. \\

\noindent In this work we have presented isophotal and elliptical integration surface brightness and color profiles, as well as extremely deep contour maps and RGB images. We have measured the total integrated colors of the targets, as well as over different regions of the host population. Where the morphology allows, we have estimated the burst luminosity and colors. All measurements were compared to stellar evolutionary models both with and without contribution by nebular emission. Structural parameters of the host population were derived from a least--squares fit of an exponential disk to the outskirts of the light distribution profiles. Our main results can be summarized as follows.
\begin{itemize}
\item For the targets lacking a morphological classification we have provided such a classification following~\citet{1986sfdg.conf...73L} based on an analysis of our contour and RGB plots. Such a classification is highly dependent on which faint surface brightness level one is looking at, since the structure in the outskirts can significantly change its physical properties beyond the Holmberg radius.
\item We find no significant difference in the shape and slope of isophotal and elliptical integration surface brightness and color profiles in the overlap region, except in the cases where the burst is not centrally located.
\item Our deep data reveal a previously undetected low surface brightness component beyond $\mu_B=26$ mag arcsec${}^{-2}$ for ESO185--13 and UM448. This new low surface brightness component has $\mu_0=24.7\pm0.12,~24.33\pm0.09$ mag arcsec${}^{-2}$, and $h_r=5.9\pm0.3,~9.3\pm0.4$ kpc, respectively. We note a similar low surface brightness structure for UM133 beyond $\mu_B=28$ mag arcsec${}^{-2}$. Even though this is at the edge of our reliability limit for the $B$ band, the deep contour images in the Appendix indicate that this feature could very well be real, but we have not dwelled on it further.
\item Regardless of the total luminosity of the BCG, the underlying host is consistent with being a dwarf with $M_B\gtrsim-18$, and in the majority of the cases it also qualifies as a true low surface brightness galaxy (LSBG) with $\mu_0\lesssim23$ mag arcsec${}^{-2}$. 
\item The most extended low surface brightness hosts have the brightest bursts in absolute terms, and the most dominant bursts in terms of percentage of total $B$ band luminosity.
\item We detect an optical bridge between the companion galaxies ESO400--43A\&B which we can trace down to the $\mu_V\sim28$th mag arcsec${}^{-2}$ isophotal level. In contrast, we detect no optical bridge between SBS0335--052E\&W at surface brightness levels brighter than $\mu_B\sim28$ mag arcsec${}^{-2}$.
\item The outer regions of a BCG do not always represent a region that is host--dominated, even if they have vaguely elliptical isophotes. Derivation of structural parameters therefore either needs to be performed on a case to case basis, which will complicate the analysis due to the inherent inconsistency of such a strategy, or the interpretation needs to allow for the obviously different structural properties of BCGs. Specifically, even in the cases of centrally--located star--forming regions we find the region defined by $\mu_B=24$--$26$ mag arcsec${}^{-2}$ to be on average strongly contaminated by nebular emission, which will affect the slope of the light distribution and hence both the derived scale length and the central surface brightness attributed to the host.
\item The fainter $\mu_B=26$--$28$ mag arcsec${}^{-2}$ region is on average free of nebular emission. Using this region for the derivation of the structural parameters, we find a significant difference between $\mu_0$ and $h_r$ for this region and the $\mu_B=24$--$26$ mag arcsec${}^{-2}$ region. With the parameters obtained from the fainter region we find no difference between the hosts of our luminous BCGs and other types of dwarfs like dEs and dIs. For the truly starbursting BCGs in the sample we find evidence suggestive of a tentative connection between low surface brightness galaxies and the BCG hosts, since they have comparable central surface brightness, scale lengths, and total luminosities.
\item We have achieved successful separation of the flocculent and dynamical asymmetry components for our galaxies using a very simple idea. There is no correlation between the dynamical and the flocculent asymmetry components (Pearson's $R=0.03$).
\item In contrast to spheroids (E,S0) and disks (Sa-b, Sc-d), luminous BCGs with LSBGs as hosts have total asymmetries completely dominated by the star formation event, i.e. by the flocculent component. This trend persists in both the optical and NIR regimes. 
\item In the concentration vs. asymmetry parameter space the BCGs are clearly distinguished from spheroids (E,S0) and disk  galaxies (Sa-b, Sc-d), occupying a region with high asymmetries and low concentrations, as expected.
\end{itemize}
\begin{table}
  \begin{minipage}{70mm}
    \caption{Concentration parameter for each filter as defined by Equation~\ref{eq:conc}.}
    \protect\label{concentr}
    \tiny
    \begin{tabular}{@{}lrrrrrrrr@{}}
      \hline
      Galaxy&$C_U$&$C_B$&$C_V$&$C_R$&$C_I$&$C_H$&$C_K$\\\hline
      ESO185--13&$1.8$&$1.8$&$1.8$&$1.8$&$2.1$&$1.8$&$1.8$\\
      ESO249--31&&$2.9$&$2.9$&$3.0$&$3.0$&$3.0$&$2.9$\\
      ESO338--04&&$2.1$&$2.1$&$2.1$&$2.4$&$2.4$&$2.4$\\
      ESO400--43&&&$1.8$&&$2.1$&$1.8$&$1.8$\\
      ESO421--02&$2.6$&$2.9$&$3.0$&$3.0$&$3.0$&$3.0$&$3.0$\\
      ESO462--20&$2.1$&$2.1$&$2.1$&&&$2.1$&$2.1$\\
      HE2--10&$2.0$&$2.7$&$2.8$&&&$2.6$&$2.6$\\
      HL293B&$2.0$&$2.4$&$2.4$&$2.4$&$2.7$&$2.1$&$2.1$\\
      IIZW40&$2.8$&$2.6$&$2.4$&&$3.0$&$3.3$&$3.0$\\
      MK600&$2.8$&$2.4$&$3.0$&$2.7$&$2.7$&$3.0$&$2.7$\\
      MK900&$2.6$&$2.8$&$3.0$&$2.6$&$2.9$&$3.0$&$2.9$\\
      MK930&$1.8$&$2.1$&$1.8$&$2.1$&$2.1$&$2.1$&$2.1$\\
      MK996&$2.4$&$3.0$&$2.4$&$2.4$&$2.8$&$2.8$&$3.0$\\
      SBS0335--052E&&$0.9$&&&$0.9$&$0.9$&$0.9$\\
      SBS0335--052W&&$2.0$&&&$2.4$&$2.1$&$2.7$\\
      TOL0341--407&$1.5$&$1.5$&$1.5$&&$1.7$&$1.7$&$1.7$\\
      TOL1457--262\emph{I}&&$2.2$&$2.2$&&&&$2.2$\\
      TOL1457--262\emph{II}&&$2.4$&$2.5$&&&&$2.2$\\
      UM133&$2.1$&$2.3$&$2.1$&$2.3$&$2.4$&$2.4$&$2.5$\\
      UM160&$2.3$&$2.4$&$2.5$&$2.4$&$2.6$&$2.9$&$2.7$\\
      UM238&$2.4$&$2.4$&$2.4$&$2.4$&$2.4$&$2.2$&$2.4$\\
      UM417&$2.0$&$2.0$&$1.5$&$2.8$&$2.2$&$2.6$&$3.0$\\
      UM448&&$1.8$&$1.8$&$1.8$&$2.4$&&$2.1$\\
      UM619&$2.1$&$2.2$&$2.7$&$2.2$&$2.7$&$2.6$&$2.7$\\
      \hline
    \end{tabular}
  \end{minipage}
\end{table}


\section*{Acknowledgments}
\noindent Part of the data presented in this paper are based on observations made with the Nordic Optical Telescope, operated on the island of La Palma jointly by Denmark, Finland, Iceland, Norway, and Sweden, in the Spanish Observatorio del Roque de los Muchachos of the Instituto de Astrofisica de Canarias. ALFOSC is provided by the Instituto de Astrofisica de Andalucia (IAA) under a joint agreement with the University of Copenhagen and NOTSA. \\

\noindent G.\"O. is a Royal Swedish Academy of Sciences Research Fellow supported by a grant from the Knut and Alice Wallenberg Foundation. G.\"O. acknowledges support from the Swedish research council (VR) and the Swedish National Space Board. E.Z. acknowledges research grants from the Swedish Research Council and the Swedish National Space Board. J.M. and I.M. acknowledge financial support from the Spanish grant AYA2010-15169 and from the Junta de Andalucia through TIC-114 and the Excellence Project P08-TIC-03531.\\

\noindent \textit{Based on observations collected at the European Organization for Astronomical Research in the Southern Hemisphere, Chile, proposal ID 07?.?-????}.\\

\noindent This work made use of the NASA/IPAC Extragalactic Database (NED) which is operated by the Jet Propulsion Laboratory, California Institute of Technology, under contract with the National Aeronautics and Space Administration.

%
\appendix

\section{KIG 732}\protect\label{kig}
\noindent This is a textbook pure elliptical galaxy. It is not a part of the sample, but it was nonetheless observed by us, and we present it here since the observations are somewhat deep and previously unpublished. The exposure times are $20$ minutes each for $BVI$, and $25$ minutes for $K$ and were obtained during $2005$ at the NOT. Note that there were only two $I$ band images taken during the corresponding observing night. We were therefore unable to perform a fringe correction. The effect of the lingering fringes is visible in the $V-I$ radial color profile beyond $r\sim40$ arcsec.
 \begin{figure*}
\begin{minipage}{150mm}
\centering
\includegraphics[width=15cm,height=18cm]{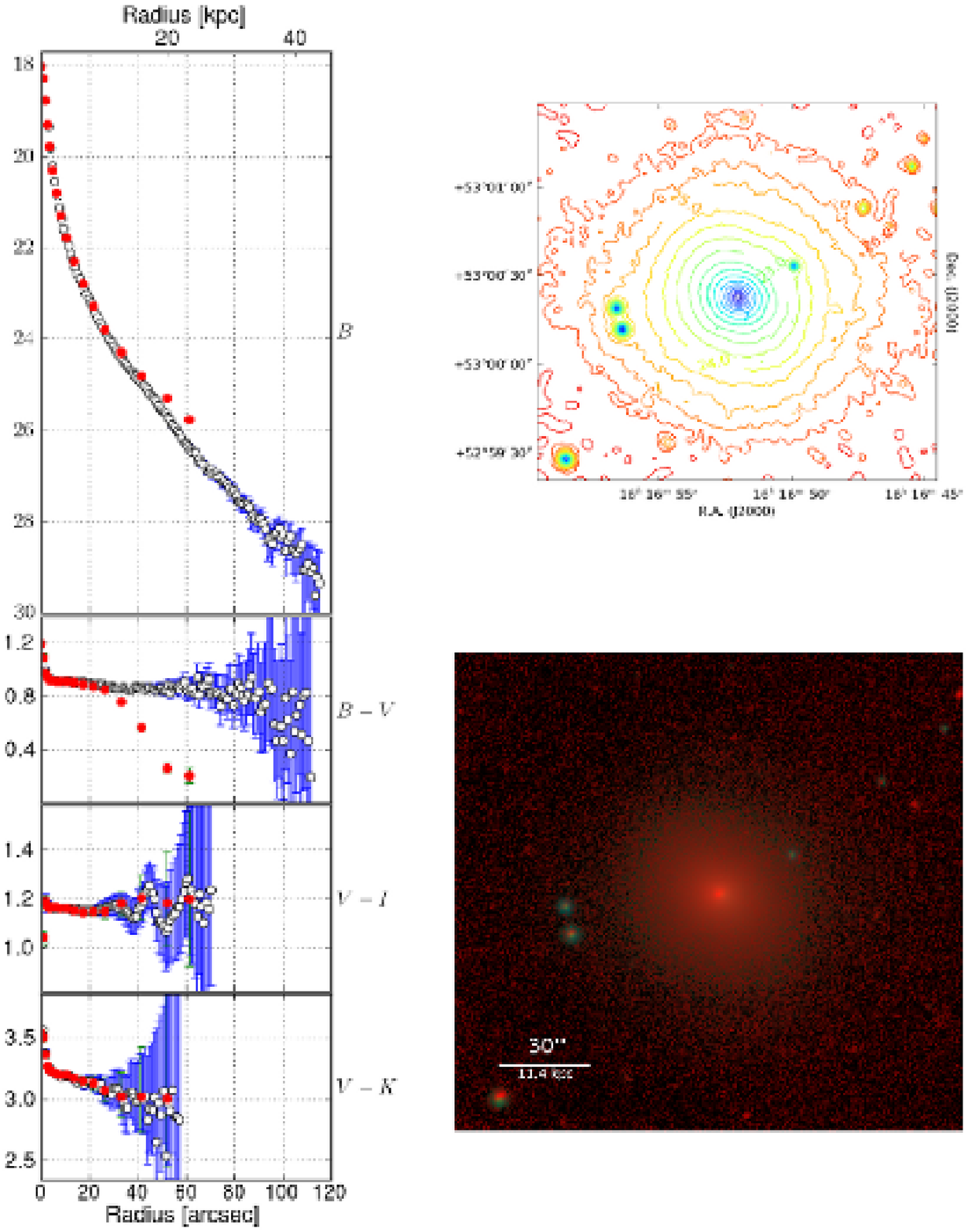}
\caption{\textbf{KIG732}. \textit{Left panel}: Surface brightness and color radial profiles for elliptical (open circles) and isophotal (red circles) integration. \textit{Upper right panel}: contour plot based on the $B$ band. Isophotes fainter than $22.0$, $24.5$, and $25.5$ are iteratively smoothed with a boxcar median filter of size $5$, $15$, and $21$ pixels respectively. \textit{Lower right panel}: A true color RGB composite image using the $U,B,I$ filters. Each channel has been corrected for Galactic extinction following \citet{1998ApJ...500..525S} and converted to the AB photometric system. The RGB composite was created by implementing the \citet{2004PASP..116..133L} algorithm.}
\protect\label{kigdatafig}
\end{minipage}
\end{figure*}
\section{Extremely faint contours}\protect\label{extremecontours}
\subsection{ESO185-13}
\begin{figure*}
\begin{center}
  \includegraphics[width=15cm,height=8cm]{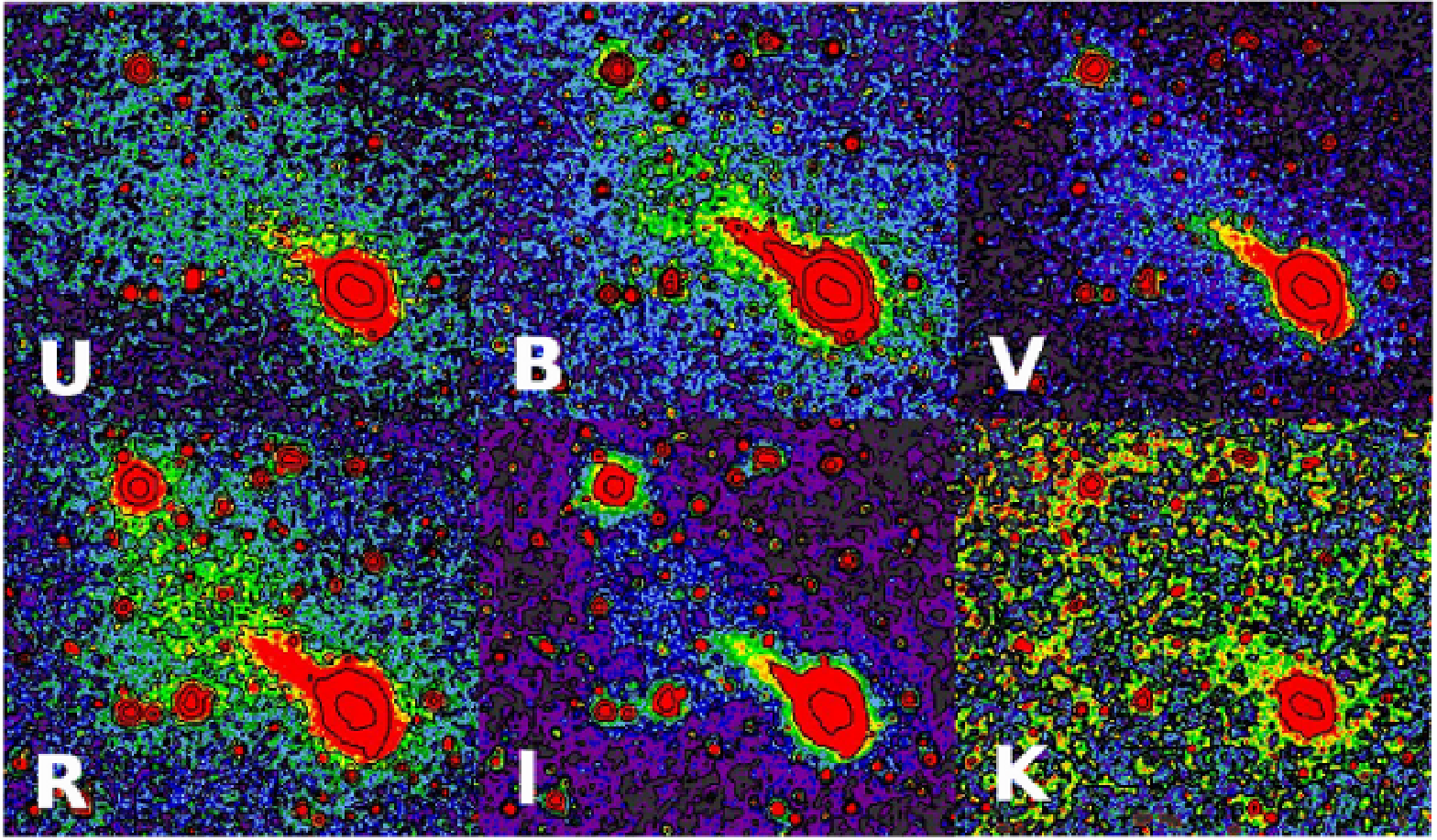}
\caption{\emph{ESO185-13} A faint structure is visible to the North--East along the tail in all optical filters. This feature is not seen in the NIR, neither in $H$ (not shown) or $K$. This could be tidal debris expelled from the main body of the galaxy, or a low luminosity dwarf which has passed through the galaxy. The average surface brightness in this feature is $\mu_U\sim23.6$, $\mu_B\sim29.2$, $\mu_V\sim28.9$, $\mu_R\sim27.7$, and $\mu_I\sim27.8$ mag arcsec${}^{-2}$. The contrast and the intensity cuts have been selected to enhance the faint features but are identical in all filters to facilitate direct comparison.}\protect\label{eso185plume}
\end{center}
\end{figure*}

\protect\label{eso185}
\noindent In Figure~\ref{eso185plume} we investigate the faint features of this galaxy. The faint structures are undetected in the NIR.
\subsection{UM448}
\begin{figure*}
\begin{center}
  \includegraphics[width=15cm,height=8cm]{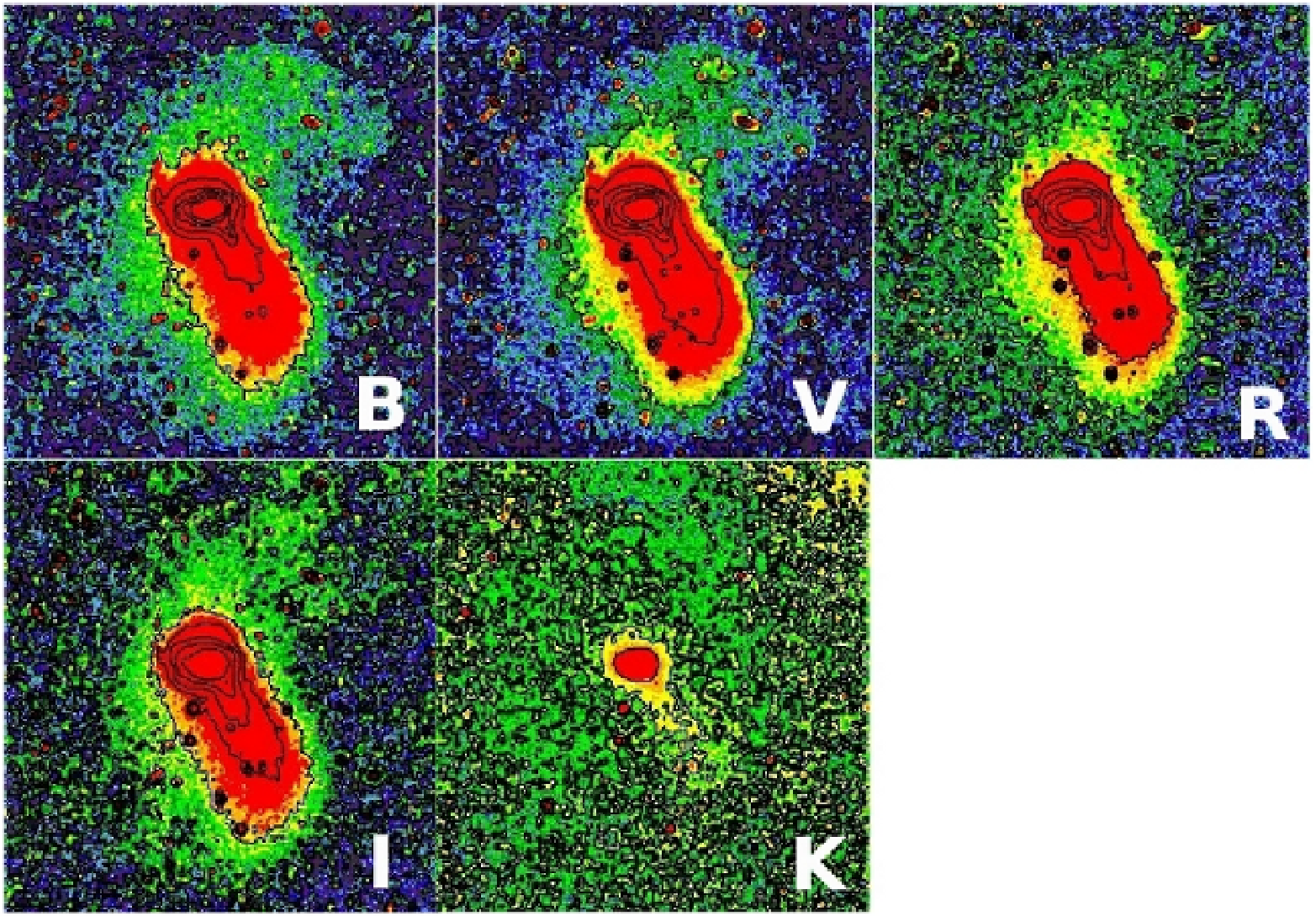}
\caption{\emph{UM448} A faint structure is visible in the optical, extending in a tidal arm to the North--West. Though not as clear, parts of this feature are also visible in the NIR, North--West and South--East of the main body of the galaxy. The average surface brightness measured over the tidal arm is $\mu_B\sim26.5$, $\mu_V\sim25.8$, $\mu_R\sim25.5$, and $\mu_I\sim24.5$ mag arcsec${}^{-2}$. The contrast and the intensity cuts are identical in all filters to facilitate direct comparison.}.\protect\label{um448plume}
\end{center}
\end{figure*}

\protect\label{um448}
\noindent In Figure~\ref{um448plume} we investigate the faint features of this galaxy in all filters. Note that the faint extended features are visible even in the $K$ band.
\subsection{Extreme contours}\protect\label{genappednix}
\noindent In Figure~\ref{deepcontours1} we show extremely deep contours for a selection of galaxies from our sample. The galaxies not shown either reveal no new information compared to the shallower contours in Figure~\ref{datafig} or deeper contours could not be reached. The images are all $B$ band, except for ESO400-43A\&B where we used the $V$ band. In units of mag arcsec${}^{-2}$ the contour levels are:
\begin{itemize}
\item ESO185-13: red -- 25.8, black -- 27.8
\item ESO338-04: red -- 25.7, black -- 27.7. Note the extended structure to the South--West.
\item ESO400-43: red -- 25.9, black -- 27.9. Note that there is an optical bridge between the two companions. ESO400-43B is not included in our sample since it was constantly at the edge of the frames.
\item ESO421-02: red -- 25.9, black -- 27.9. The regular shape of the outer contour is further indication that the two dwarf components involved in this merger could not have undergone a fly--through since there is no detectable tidal debris.
\item HE2-10: red -- 25.5, black -- 27.2
\item HL293B: red -- 25.7, black -- 27.7
\item IIZw40: red -- 23.5, black -- 26.0
\item MK600: red -- 25.8, black -- 27.8. The $25.8$th contour is the last one we can close around the galaxy. At fainter isophotes the contours merge with those from the bright surrounding stars.
\item MK900: red -- 25.8, black -- 27.8
\item MK930: red -- 25.5, black -- 27.5. The features to the North--East are possibly tidal debris from the recent merger.
\item MK996: red -- 25.8, black -- 27.8
\item SBS0335-052: red -- 25.8, black -- 27.8. Note the absence of an optical bridge between the East and West companions.
\item Tol0341-407: red -- 26.0, black -- 28.0. At faint isophotes the galaxy radically changes shape, looking more elliptical and regular.
\item UM133: red -- 27.9, black -- 29.1. These are the deepest contours for the sample and they are consistent with the feature we see in the surface brightness profile beyond $\mu_B\sim28$ mag arcsec${}^{-2}$. The bright star to the South--West of the galaxy goes below $B=29$ mag within $40$ arcsec, so it is not contributing to the structures visible to the East beyond the red contour.
\item UM160: red -- 26.0, black -- 28.0. The galaxy seemingly completely disappears beyond the $28$th isophote. Since it appears to be isolated, this could be a projection effect.
\item UM238: red -- 26.0, black -- 28.0. The structure to the South--West belongs to a bright star. It does not interfere with any measurements for this galaxy inwards of the $28$th isophote.
\item UM417: red -- 26.0, black -- 28.0
\item UM619: red -- 26.0, black -- 28.0
\end{itemize}
\begin{figure*}
\begin{center}
  \includegraphics[width=15cm,height=18cm]{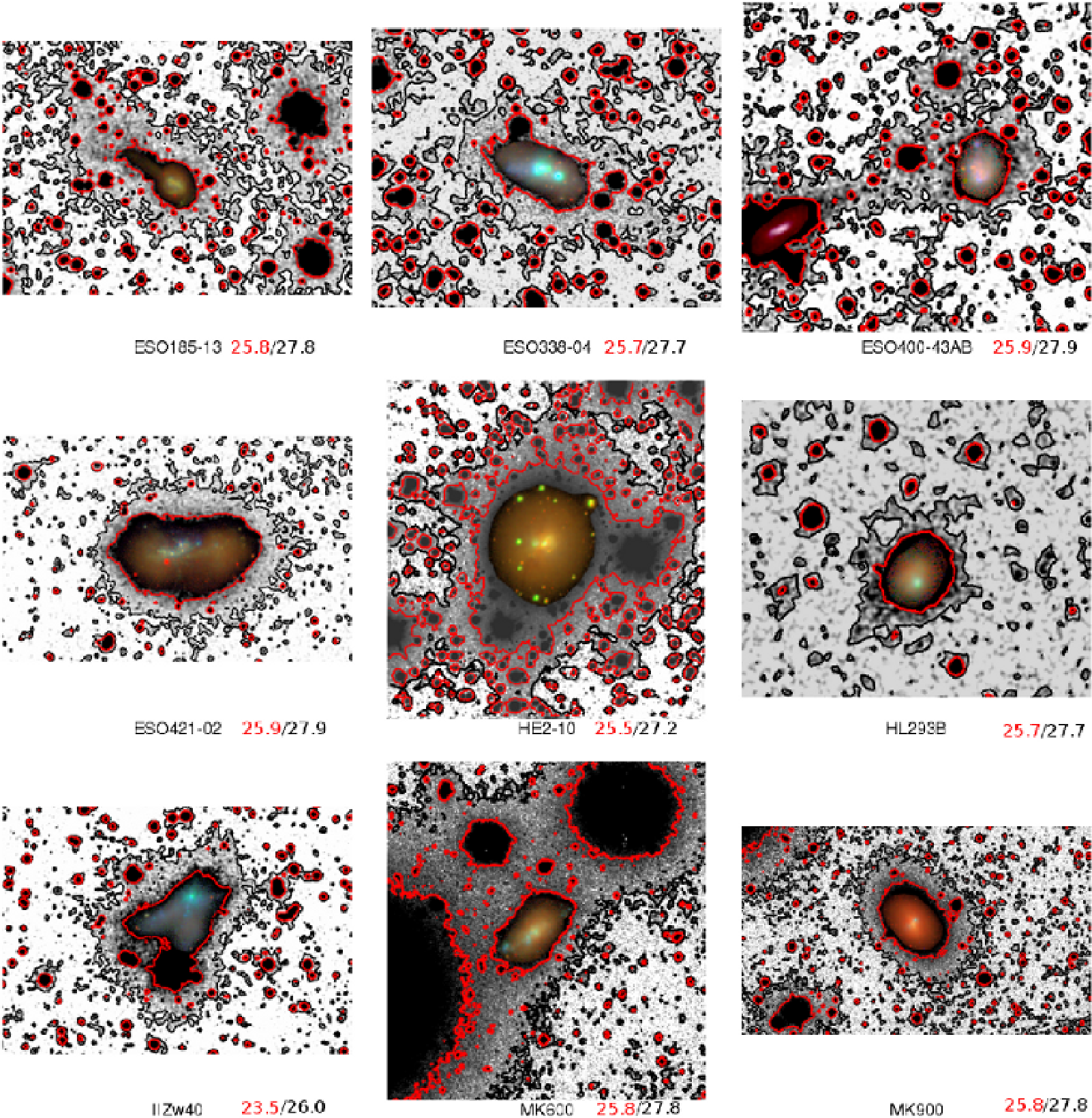}
\caption{Extreme contours. Numbers in red/black indicate the isophotal level of the red/black contour in mag arcsec${}^{-2}$.}\protect\label{deepcontours1}
\end{center}
\end{figure*}
\begin{figure*}
\begin{center}
  \includegraphics[width=15cm,height=18cm]{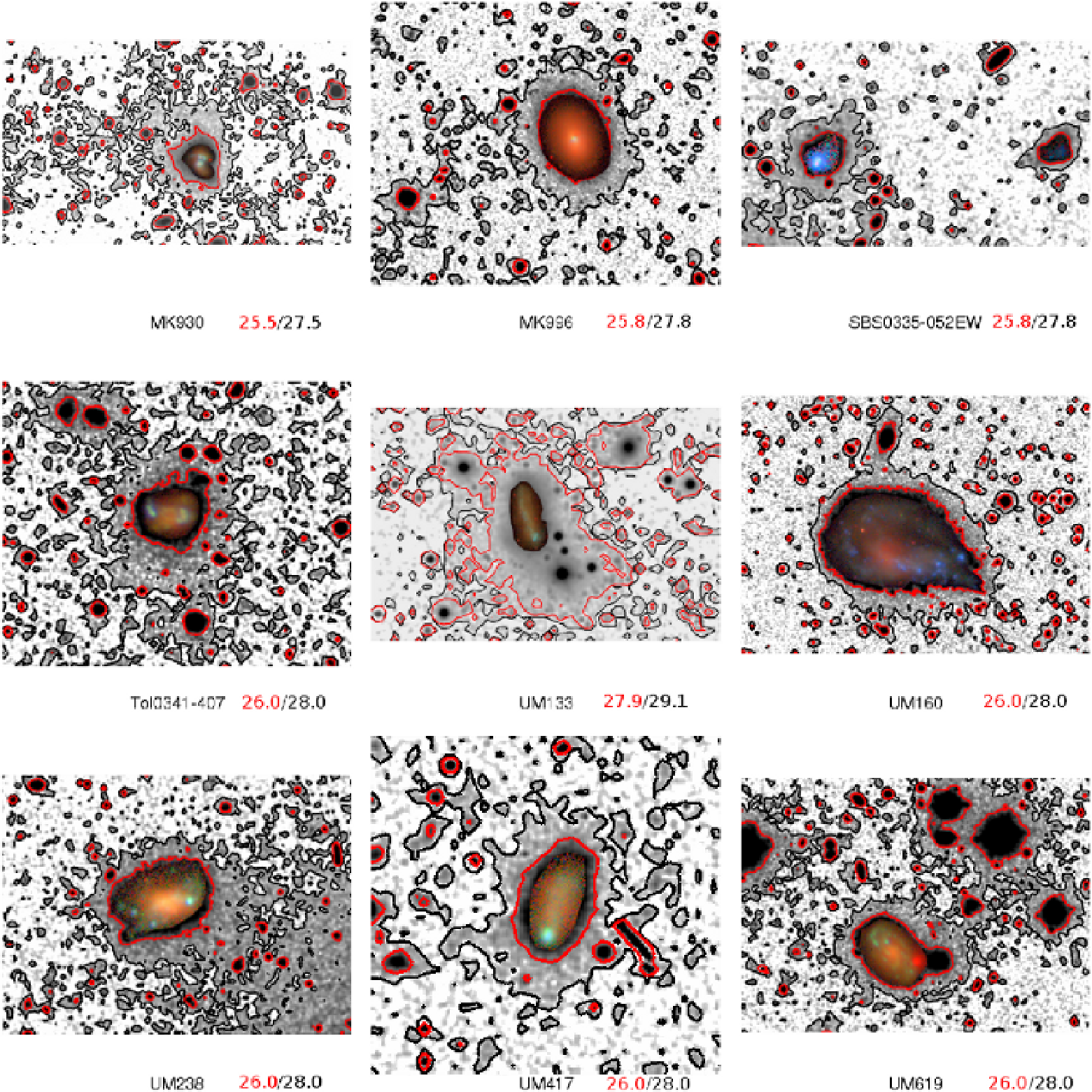}
\contcaption{ }
\end{center}
\end{figure*}
\bibliographystyle{mn2e}
\bibliography{micheva}

\bsp

\label{lastpage}

\end{document}